\documentclass[11pt]{elsarticle}
\usepackage{hyperref}

\usepackage{graphicx}

\usepackage{a4wide}
\usepackage{amssymb}
\usepackage{amsmath}
\usepackage{amsthm}
\usepackage{mathrsfs}
\usepackage{bm}
\usepackage{accents} 
\usepackage[utf8]{inputenc}
\usepackage[english]{babel}

\usepackage{multirow}
\usepackage{booktabs}
\usepackage{currfile}

\usepackage{macros-rohan}

\usepackage[numbers]{natbib}

\usepackage{color}

\newtheorem{theorem}{Theorem}[section]
\newtheorem{lemma}[theorem]{Lemma}
\newtheorem{e-proposition}[theorem]{Proposition}

\newtheorem{e-definition}[theorem]{Definition\rm}
\newtheorem{remark}{\it Remark\/ \rm}

\newcommand{\REMOVE}[1]{} 

\newcommand\blue[1]{\textcolor{blue}{#1}}
\newcommand\red[1]{\textcolor{red}{#1}}

\newcommand\ToDo[2]{\blue{#1}\red{\tt\small ({#2})}}
\newcommand\tododo[1]{#1}

\newcommand\RELEASE[1]{}

\newcommand{\correction}[1]{{{#1}}}
\newcommand{\correctionIJES}[1]{{{#1}}}

\newcommand\Cite[1]{\blue{\cite{#1}}}
\renewcommand\Citep[1]{\blue{\citep{#1}}}

\newcommand\Appx[1]{\ref{#1}}

\def\FT#1{{\underaccent{{\circ}}{{#1}}}} 
\def\LT#1{{\underaccent{{*}}{{#1}}}} 
\def\LTxt#1{\Lcal\{{#1}\}}  

\newcommand\HpdbO{\Hbm_{\#0}^1}

\newcommand\Average[2]{\left\langle{#1}\right\rangle_{#2}}

\newcommand\FEMapprox{\stackrel{\rm FEM}{\approx}}

\def\aYf#1#2{a_f \left ({#1},\,{#2}\right )}
\def\atYf#1#2{\tilde{a}_f \left ({#1},\,{#2}\right )}
\def\bYf#1#2#3{b_f \left ({#1};\,{#2},\,{#3}\right )}
\def\cYf#1#2#3{c_f \left ({#1};\,{#2},\,{#3}\right )}
\def\gYf#1#2#3{g_f \left ({#1};\,{#2},\,{#3}\right )}
\def\ipYf#1#2{\left \langle{#1},\,{#2}\right \rangle_{Y_f}}

\def\aYw#1#2{a_w \left ({#1},\,{#2}\right )}

\def\Dev#1{\wtilde{#1}}

\newcommand{\dlt}{\delta}

\newcommand\nillveps{{}} 

\newcommand{\cwto}{\rightharpoonup}
\newcommand{\csto}{\rightarrow}

\newcommand\dd{\mathrm{d}}

\newcommand\eeb[1]{\eb({#1})}

\newcommand\ggb[2]{\gb^{#2}({#1})}

\newcommand\zerobm{\bf{0}}

\newcommand\vop{{{\bf v} \kern-0.5em{\bf v}}}
\newcommand\Rop{{{\bf I} \kern-0.2em{\bf R}}}
\newcommand\Pop{{{\bf I} \kern-0.2em{\bf P}}}

\newcommand\rhs{{\rm right hand side~}}
\newcommand\lhs{{\rm left hand side~}}
\newcommand\wrt{{\rm with respect to~}}
\newcommand\ie{{\textit{i.e.~}}}
\newcommand\eg{{\textit{e.g.~}}}
\newcommand\cf{{\textit{cf.~}}}
\newcommand\ext{{\rm{ext}}}
\newcommand\phy{{\rm{phy}}}
\newcommand\Reynolds{{\rm{Re}\,}}
\newcommand\fsi{{\it{fs}}}

\newcommand\fbav{{\what{\bmi{f}}}}
\newcommand\trace{{\tt{tr}}}
\newcommand\traceG[2]{{\rm{trace}}_{#2}\left({#1}\right)}

\def\ENorm#1#2{|\kern-0.14em|\kern-0.14em |{#1} |\kern-0.14em |\kern-0.14em|_{#2}} 

\newcommand\Tuftxt{\Tcal_\veps\,}

\numberwithin{equation}{section}
\def\intY{\:\sim \kern-1.17em \int}
\def\intY{\sim \kern-1.2em \int}
\def\intYi#1{\sim \kern-1.2em \int_{Y_{#1}}}
\def\intOmY#1{\sim \kern-1.2em \int_{\Omega\times Y_{#1}}}
\def\intYsmall{\sim \kern-1.07em \int}
\def\intYs{- \kern-0.82em \int }

\begin{document}

\selectlanguage{english}

\begin{frontmatter}

\title{Modelling wave dispersion in fluid saturating periodic scaffolds}


\author[1]{Eduard Rohan}
\ead{rohan@kme.zcu.cz}
\author[1]{Robert Cimrman}
\ead{cimrman3@ntc.zcu.cz}



\address[1]{European Centre of Excellence, NTIS -- New Technologies for
Information Society, Faculty of Applied Sciences, University of West
Bohemia, Univerzitn\'{\i} 8, 30100 Pilsen, Czech Republic}


\begin{abstract}
  Acoustic waves in a slightly compressible  fluid saturating porous periodic structure are studied using two complementary approaches: 1) the periodic homogenization (PH) method provides effective model equations for a general dynamic problem imposed in a bounded medium, 2) harmonic acoustic waves are studied in an infinite medium using the Floquet-Bloch (FB) wave decomposition.
  In contrast with usual simplifications, the advection phenomenon of the Navier-Stokes equations is accounted for. For this, an acoustic approximation is applied to linearize the advection term. The homogenization results are based the periodic unfolding method combined with the asymptotic expansion technique providing a straight upscaling procedure which leads to the macroscopic model defined in terms of the effective model parameters. These are computed using the characteristic responses of the porous  microstructure.
  Using the FB theory, we derive dispersion equations for the scaffolds saturated by the inviscid, or the viscous, barotropic fluids, whereby the advection due to a permanent flow in the porous structures is respected. A computational study is performed for the numerical models obtained using the finite element discretization. For the FB methods-based dispersion analysis, quadratic eigenvalue problems must be solved. The numerical examples show influences of the microstructure size and of the advection generating an anisotropy of the acoustic waves dispersion.
\end{abstract}



\end{frontmatter}



\section{Introduction}\label{sec-intro}
Modelling of acoustic waves in fluid saturated porous media has been treated mostly using the homogenization theory \cite{Sanchez1980Book,Diaz-Alban-CPDE2014}, or using the phenomenological models based on the theory of porous media \cite{ThierryCoussyZinszner1987acoustics-porous,CarcioneBook2014}.
Although a large body of literature  devoted to this topic exists, some acoustic phenomena related to the nonlinearity arising from the flow model deserve a further attention. New challenges for the modelling of these phenomena emerge due to applications in biomedicine \cite{Raghavan-acoustic-streaming-2018}, chemistry and smart material design \cite{PhysRevB.99.134304}. As a step forward, we aim to explore models of the acoustic wave propagation in viscous and inviscid fluids saturating periodic rigid porous structures (scaffolds). Thus, we consider waves propagating only in the fluid flowing through channels with a periodic structure. To give an example of such structures, these can be created by sintered ceramic fibres, see \cite{kruisova2018}, where elastic waves in the solid were studied  without effects of the fluid. On the contrary, here we consider waves propagating in the fluid only, while  neglecting compliance of the solid.

It is well known that the standard Darcy law describing slow viscous flows in porous materials can be derived rigorously by the asymptotic homogenization of the steady Stokes flow \correctionIJES{\Citep{Sanchez1980Book}}, \cf \correctionIJES{\Cite{Zaki-2012}} where the unfolding  method of homogenization was employed.
Homogenization of the non-stationary incompressible Stokes flow was treated using the two-scale convergence method in \correction{\Cite{Allaire-nonsteady-NS-homog}}. The obtained Darcy permeability serves as the time-convolution kernel which, upon being transformed in the frequency domain, yields the so-called dynamic permeability. In this way, the viscodynamic operator \correctionIJES{\Citep{Norris1986}} can be introduced which comprises the tortuosity effects. Flows in periodic channels governed by the non-stationary Navier-Stokes (N-S) equations including the advection term were treated \correction{in \Cite{Mikel91}}, where the critical case of scaling the viscosity and the velocity \wrt the pressure was discovered. Using this critical scaling, the formal asymptotic expansion technique of homogenization was employed \correction{in \Cite{Zhengan-Hongxing-2008}} to derive the Forchheimer law from the N-S equations with the inertia terms. Upscaling of compressible, or incompressible flows governed by the \correctionIJES{N-S equations} was reported also \eg in \Cite{Mikel-Paoli99,Masmoudi-compressNS2002,Chen-homog-NS-Forchheimer2001,Kandel-long-wave-Forchheimer-2019}, computational aspects were studied in a number of papers \correctionIJES{\Citep{Laschet-2008-Forchheimer-homog,Peszynska2010-proc,Miroshnikova-2016}}.


In this paper, we consider acoustic wave propagation in periodic scaffolds saturated by  Newtonian, or inviscid, slightly compressible barotropic fluids. To respect the advection effects related to the permanent flow which is assumed to be independent of the acoustic perturbations, a linearization is employed to establish approximate models of acoustic waves. For the inviscid fluid, a reduced model involving the pressure only is derived.
We restrict the study  to the first order linearization of the fluid advection, but assume that the permanent flow can be quite important. In this context, to capture the acoustic streaming effect, where the flow is generated by the acoustic waves, higher order approximations are needed  \cite{Chafin2016WaveFlowIA,Raghavan-acoustic-streaming-2018,Wu-ac-stream-2018}.

Two approaches are examined to analyze the wave dispersion: the periodic homogenization (PH) and the Floquet-Bloch wave decomposition (FB), \cf~\cite{Collet2011}, enabling to analyze waves of lengths comparable with the periodicity size. Using both these approaches we  derive models of viscous and inviscid fluids in the rigid scaffolds, respecting the advection effect of the permanent fluid flow upon which perturbations induced by the wave propagation are superimposed. 

Pursuing the first approach for inviscid fluid, a macroscopic model governing the pressure fluctuations  is derived, which involves an advection term related to the time rate of the pressure. For the viscous fluid, the homogenization of the N-S equations in the rigid skeleton provides the dynamic permeability of the effective porous medium. The dispersion phenomenon is remarkable especially for low frequencies, whereas constant phase velocity characterizes the asymptotic behaviour for larger wave numbers. In both the cases, the PH approach captures the wave propagation for wave lengths significantly larger than the characteristic porosity size corresponding to one period of the lattice.  Using the second approach, the FB decomposition enables to capture the wave response for wave numbers within the whole first Brillouin zone. We derive the equations of the cell problems describing the local fluctuations of the wave polarization. The dispersion analysis leads to quadratic eigenvalue problems (QEP). In the case of viscous fluids, the QEP with rather complicated structure appears, that requires a suitable linearization demanding new variables to be introduced.  The computational analysis for the inviscid fluid model involving 3D scaffolds  with variable porosity is performed, showing the porosity influence on a frequency  band gap opening between the two lowest modes. However, the main emphasis is put on the illustration of the permanent flow advection influence on the sound speed, in both the inviscid and viscous fluids.




The plan of the paper is as follows. In Section~\ref{sec-flow-model} the model of fluid flow in porous structures and the acoustic waves are introduced using a decomposition of the model responses into the steady part and the acoustic fluctuations. For the inviscid and viscous fluid saturating periodic rigid scaffolds, the respective homogenized models are derived in Section~\ref{sec-homog}, where also the pressure plane wave propagation in the homogenized medium is described for the two fluids. In Section~\ref{sec-Bloch}, the Floquet-Bloch wave decomposition is applied to analyze the wave propagation in the considered porous structures; the generalized quadratic eigenvalue problems resulting from the finite-element discretization of the weak formulations are introduced and their linearization is explained.
Both the modelling approaches are illustrated on examples of 2D and 3D periodic scaffolds in Section~\ref{sec:examples}. There the dispersion analysis performed by solving the eigenvalue problems resulting from the Floquet-Bloch wave decomposition is compared with the corresponding approximation obtained by the homogenized model for viscous and inviscid fluids, whereby the size effects and advection phenomena are explored. Some technical results employed in the paper are postponed in the Appendix.

\paragraph{General notations}
Spatial position of a point is specified by its Cartesian coordinates, $x=(x_1, x_2, x_3) \in \RR^3$, where $\RR$ is the set of real numbers.  The boldface notation for vectors $\ab = (a_i)$ and second-order
tensors $\bb = (b_{ij})$ is used. \correction{The second-order identity tensor}
is denoted by $\Ib = (\delta_{ij})$.
The fourth-order elasticity tensor is denoted by $\Dop = (D_{ijkl})$. \correction{The
superposed dot denotes a derivative with respect to time.}
\correction{The gradient, divergence and Laplace operators are
 denoted by $\nabla, \nabla \cdot$ and $\nabla^2$, respectively.} \correctionIJES{When these} operators have
a subscript \correction{referring} to the space variable, it is for indicating that the operator
acts relatively at this space variable, for instance $\nabla_y = (\pd_i^y) = (\pd/\pd y_i)$.
The symbol dot `$\cdot$' denotes the scalar product between two vectors and the
symbol colon `$:$' stands for scalar (inner) product of two second-order
tensors, \eg $\Ab:\Bb = A_{ij}B_{ij} = \trace[\Ab^T\Bb] = A_{ki}B_{kj}\delta_{ij}$,
where $\trace[\star]$ is the trace of a tensor and superscript $T$ in $\star^T$ is the transposition operator.
Operator $\otimes$ designates the tensor product between two vectors, \eg $\ab\otimes\vb = (a_iv_j)$. Standard notations for functional spaces are adhered. Throughout the
paper, $x$ denotes the global (``macroscopic'') coordinates, while the
``local'' coordinates $y$ describe positions within the representative unit
cell $Y\subset\RR^3$.
The normal vectors on a boundary of domains $\Om_\alpha$ (or $Y_\alpha$) are denoted by $\nb^\alpha$, $\alpha = s,f$, to distinguish their orientation outward to $\Om_\alpha$ (or $Y_\alpha$) when dealing with the solid-fluid interfaces.
By $\eeb{\wb} = 1/2(\nabla\wb + (\nabla\wb)^T)$ we \correction{denote}  the strain of a vector field $\wb$ (displacements, or velocities).
The ``tilde''-notation can have various meanings which are explained \correctionIJES{through} the text and are clear within the particular context. The following
standard functional spaces are used: by $L^2(\Om)$ we refer to square integrable
functions defined in an open bounded domain $\Om$; by $H^1(\Om)$ we mean the Sobolev space
$W^{1,2}(\Om) \subset L^2(\Om)$ formed by square integrable functions
including their first generalized derivatives; space $C_0^\infty(\Om)$
is constituted by infinitely differentiable functions with the compact support, thus, with zero trace on $\pd \Om$; Bold
notation is used to denote spaces of vector-valued functions, e.g. $\Hdb(\Om)$;
by subscript $_\#$ we refer to the \emph{$Y$-periodic functions}.

\section{Model of fluid flow in porous structures and acoustic waves}\label{sec-flow-model}
We introduce formulations for acoustic waves in a fluid saturating  channels $\Om_f$ of a porous two-phase medium situated in an open bounded domain $\Om \subset \RR^3$. 
The solid skeleton occupies domain $\Om_s^\nillveps\subset\Om $, whereas
viscous, or inviscid slightly compressible fluids saturate the pores $\Om_f^\nillveps = \Om \setminus \ol{\Om_s^\nillveps} $ constituting a connected network of channels whose walls $\Gamma_\fsi^\nillveps = \ol{\Om_s^\nillveps} \cap  \ol{\Om_f^\nillveps}$ are impermeable for the fluid. Each of the two domains $\Om_k^\nillveps$, with $k = s,f$ are connected, hence also their interface $\Gamma_\fsi^\nillveps$ is {connected}.
 The fluid can flow through, or be loaded on the external part of $\pd\Om_f$ denoted by $\pd_\ext\Om_f^\nillveps = \pd\Om_f^\nillveps\setminus \Gamma_{fs}^\nillveps$.
Obviously, $\pd_\ext\Om_f^\nillveps \subset \pd\Om$ can be decomposed according to the specific boundary conditions.

We recall the Navier-Stokes equations;  the fluid velocity $\wb$ and pressure $p$ satisfy
\begin{equation}\label{eq-NS1}
\begin{split}
  \rho\left(\dt{}{t}\wb + \wb\cdot\nabla\wb \right)& = - \nabla p + \nabla\cdot\Dop\eeb{\wb}\;,\\
   \pdiff{\rho}{t} + \nabla\cdot(\rho\wb)  = 0\;,
\end{split}
\end{equation}
where $\rho$ is the fluid density, and
$\Dop\eeb{\wb}$ represents the viscous stress given by the velocity strain $\eeb{\wb} = 1/2(\nabla\wb + (\nabla\wb)^T)$ and by the viscosity tensor, $\Dop = (D_{ijkl})$ with $D_{ijkl} = \eta\delta_{ij}\delta_{kl} + \mu(\delta_{ik}\delta_{jl} + \delta_{il}\delta_{jk})$ depending on the 1st and the 2nd viscosity, $\mu$ and $\eta$, respectively.

\subsection{Assumptions for acoustic wave decomposition and linearization}\label{sec-flow-decompose}
We shall study acoustic waves propagating in a slightly compressible viscous, or inviscid fluid, while thermal effects are disregarded (the barotropic fluid).
The fluid flows through the rigid porous structure, whereby the following assumptions are made:

\begin{list}{}{}
\item (A1) The total fields $\wb$, $p$ and the mass density of the fluid, $\rho$,  are split into the ``stationary flow'' parts $\bar \wb$, $\bar \rho$ and $\bar p$ and the ``acoustic fluctuation'' parts $\tilde \wb$, $\tilde \rho$ and $\tilde p$, so that
\begin{equation}\label{eq-NS2}
\begin{split}
   \wb = \bar\wb + \tilde\wb\;,\quad p = \bar p + \tilde p\;,\quad \rho = \bar \rho + \tilde \rho\;.
\end{split}
\end{equation}

\item (A2) The fluid is assumed to be homogeneous and under the stationary flow described by $(\bar\wb,\bar p, \bar \rho)$ is considered as incompressible, thus
\begin{equation}\label{eq-NS2a}
\begin{split}
  \nabla\bar \rho & = 0\;, \quad \pdiff{}{t}\bar\wb = 0\;,\quad \pdiff{}{t}\bar\rho = 0\;.
\end{split}
\end{equation}

\item (A3) The acoustic perturbations introduced due to the split \eq{eq-NS2} are small, so that mutual multiplications of any two perturbations is neglected. {This assumption is important for the linearization.}

\item (A4)
  The acoustic response is barotropic: denoting by $p_0$ and $\rho_0$ reference state variables, it holds that $p-p_0 = c^2(\rho - \rho_0)$, where $c^2$ is the squared acoustic velocity $c = \sqrt{k_f/\rho_0}$ with the bulk stiffness $k_f = 1/\gamma_f$, thus,  $\gamma_f$ is the fluid compressibility for the reference state.
Since we consider the steady flow as incompressible, $\bar\rho = \bar\rho_0$, the reference pressure $p_0$ can be associated with the one of the steady flow, \ie  $\bar p \approx \bar p_0$.
  As a consequence, we have $\tilde p = c^2\tilde \rho$. In this paper we adhere to this linear approximation which is adequate in the context of the incompressibility of the stationary flow part, assumption A2). A nonlinear relationship between the pressure and the density, as treated \eg by \cite{Gilbert-Panchenko2004}, is required to handle some ``higher order effects'', such as the acoustic streaming, see \eg \cite{Wu-ac-stream-2018}.


\item (A5) The stationary flow is characterized by a periodic velocity field $\bar\wb$. This assumption is needed to analyze the acoustic waves using the Floquet-Bloch theory.

\end{list}

The above assumptions lead the following equations governing the acoustic fluctuations $(\tilde\ub,\tilde p)$ of the velocity and pressure fields (see \Appx{apx-ac-fl}),
\begin{equation}\label{eq-NS3}
\begin{split}
  \rho_0  \left(\dt{}{t}\tilde\ub + \bar\wb\cdot\nabla\tilde\ub + \tilde\ub\cdot\nabla\bar\wb\right)& = - \nabla \tilde p + \nabla\cdot\Dop\eeb{\tilde\ub}\;,\\
   \dt{}{t}\tilde p + \wb\cdot\nabla \tilde p & = - k_f \nabla\cdot\tilde\ub\;,
\end{split}
\end{equation}
while $(\bar\wb,\bar p)$ describing the stationary flow in the periodic scaffolds satisfy
\begin{equation}\label{eq-NS-flw}
\begin{split}
  \rho_0\bar\wb\cdot\nabla\bar\wb + \nabla \bar p -\mu\nabla^2{\bar \wb} & = \bar\fb\;,\\
 \nabla\cdot \bar\wb & = 0\;,
\end{split}
\end{equation}
where $\fb$ is the volume force.

Concerning Assumption (A5), in general, periodic heterogeneities in the flow can be caused by different aspects:
\begin{itemize}
\item presence of periodically distributed obstacles (or flow in a periodic porous structure) -- static fluid;
\item periodic heterogeneities in the fluid parameters; such a situation is relevant for a static fluid, \ie $\bar\wb \equiv \zerobm$, for instance two fluids, air bubbles...);
\item fluid flow through a periodic porous structure with non-negligible convection velocity which is periodically perturbed due to the porous structure.
\end{itemize}

\subsection{Inviscid fluid}
In this section we consider an inviscid compressible fluid characterized by constants $\rho_0$ and $k_f$ only, since the vanishing viscosity cancels the second \rhs term in \eq{eq-NS3}$_1$ involving $\Dop$.  

While for the viscous fluid model the advection velocity $\bar\wb$  is computed using \eq{eq-NS-flw}, for inviscid fluids the velocity field $\bar\wb$ is assumed to have vanishing vorticity. Hence, it is determined by  \eq{eq-NS-flw}$_2$  governing the steady incompressible flow: $\bar\wb = -\nabla \Psi$, where the potential $\Psi$  satisfies
\begin{equation}\label{eq-potflow}.
\begin{split}
  -\nabla^2 \Psi & = 0\quad \mbox{ in } \Om_f\;,\\
  \Psi & = \bar \Psi\quad \mbox{ on } \pd_\ext\Om_f\;,\\
  \nubf\cdot\nabla \Psi & = 0\quad\mbox{ on } \Gamma_\fsi\;.
\end{split}
\end{equation}
The incompressibility condition $\nabla\cdot \bar\wb = 0$ in $\Om_f$ and the impermeability of $\Gamma_\fsi$  are verified. 

For stationary fluids, \ie when $\bar\wb \equiv 0$, it is an easy exercise to eliminate velocity $\tilde\ub$ from \eq{eq-NS3}, recalling $\Dop$ is zero, so that the Helmholtz equation governing the pressure fluctuations $\tilde p$ is obtained. Alternatively, upon eliminating the pressure, the velocity must satisfy the wave equation supplemented by the zero vorticity constraint. 
The acoustic pressure waves in the free fluid propagate with the phase velocity $c_f = \sqrt{k_f/\rho_0}$, whereas the acoustic waves in the scaffolds are governed by either of the following two  hyperbolic equations,
\begin{equation}\label{eq-ivs2}
\begin{split}
\mbox{ pressure formulation: } \quad    \ddt{}{t}\tilde p & = c_f^2\nabla^2 \tilde p\;,\\
\mbox{ velocity formulation: } \quad    \ddt{}{t}\tilde\ub & = c_f^2\nabla(\nabla\cdot\tilde\ub)\quad \mbox{ and } \nabla\times \tilde\ub = \zerobm \;.
\end{split}
\end{equation}

When steady advection is respected, $\bar\wb \not \equiv \zerobm{}$, similar reduced formulations can be derived. However, it is not straightforward  to eliminate the velocity and to arrive at a pressure formulation. For this, divergence operator is applied to \eq{eq-NS3}$_2$, which yields,
\begin{equation}\label{eq-NSadv2}
\begin{split}
  \rho_0  \left(\nabla\cdot\dot{\tilde\ub} + 2 \pd_k \bar w_i \pd_i \tilde u_k +
  \bar\wb\cdot\nabla (\nabla\cdot\tilde\ub) + \tilde\ub\nabla(\nabla\cdot\bar\wb)
  \right) = -\nabla^2 \tilde p\;,
\end{split}
\end{equation}
where the last term on the left-hand side vanishes due to the incompressibility of $\bar\wb$.

The following lemma is employed to treat the second left-hand side term.

\begin{lemma}\label{lem1}
For any open bounded domain $Q \subset \Om_f$, it holds that
\begin{equation}\label{eq-lm1}
\begin{split}
  \int_Q \pd_k \bar w_i \pd_i \tilde u_k = \int_{\pd Q} \left(
  \bar w_i \pd_i\tilde u_k - \bar w_k \pd_i \tilde u_i\right) n_k
  + \int_Q \nabla\cdot\bar\wb \nabla\cdot\tilde\ub\;.
\end{split}
\end{equation}
\end{lemma}
The proof is a straightforward consequence of $\pd_i(\pd_k\tilde u_i) = \pd_k (\nabla\cdot\tilde\ub)$. 

As the consequence of Lemma~\ref{lem1} and the incompressibility $\nabla\cdot\bar\wb = 0$, the following approximation may be considered,
\begin{equation}\label{eq-lm2}
\begin{split}
 \Average{ \pd_i \bar w_k \pd_k \tilde u_i}{Q} & \approx |Q|^{-1}\int_{\pd Q} \left(\bar w_k \pd_k \tilde u_i- \bar w_i\Average{\nabla\cdot\tilde\ub}{Q}\right) n_i \\
 &\approx \bar w_k \pd_k \Average{\nabla\cdot\tilde\ub}{Q} -|Q|^{-1}\int_{\pd Q} \bar w_i n_i  \Average{\nabla\cdot\tilde\ub}{Q}\\
 & = \pd_w  \Average{\nabla\cdot\tilde\ub}{Q}\;,
\end{split}
\end{equation}
where $\Average{~}{Q} = |Q|^{-1}\int_Q$ is the average.
Therefore, we can employ the approximate relationship
\begin{equation}\label{eq-lm3}
\begin{split}
 \pd_i \bar w_k \pd_k \tilde u_i \approx \bar\wb\cdot\nabla(\nabla\cdot\tilde\ub)\;.
\end{split}
\end{equation}
Using \eq{eq-lm3} and upon substituting $\nabla\cdot\tilde\ub$ by \eq{eq-NS3}$_2$, the \lhs in \eq{eq-NSadv2} can be approximated, so that the acoustic waves in the pore fluid are governed by the following equation:

\begin{equation}\label{eq-NSadv3}
\begin{split}
  \left(\dt{}{t} + \zeta \bar\wb\cdot\nabla\right)\left(\dot{\tilde p} + \bar\wb\cdot\nabla{\tilde p}\right) & = c_f^2 \nabla^2 \tilde p\;,
\end{split}
\end{equation}
where $\zeta = 3$ and $c_f = \sqrt{k_f/\rho_0}$. Further we define $\theta = (1+\zeta)/2$,
thus, $\theta = 2$. However, in what follows, we keep the abstract notation $\zeta$ and $\theta$.







\section{Homogenization of fluid in periodic porous structures}\label{sec-homog}
We are confined to rigid scaffolds saturated by viscous, or inviscid fluids, so that, at the pore level, the acoustic waves are described by the models introduced in Section~\ref{sec-flow-model}.
The size of the porous microstructure is expressed by the scale parameter $\veps = \ell / L$ defined by the ratio of the micro- and macroscopic characteristic lengths, denoted by  $\ell$ and $L$, respectively. The homogenization procedure applied to derive an effective model of the rigid-porous medium consists in the asymptotic analysis $\veps\rightarrow 0$ of the micro-model presented above. 

The notation related to the geometry formerly introduced is adhered to, however, now the superscript ${}^\veps$ is appended to respect the dependence on the scale, \ie the fluid occupies pores $\Om_f^\veps = \Om \setminus \ol{\Om_s^\veps} $, bounded by the pore walls $\Gamma_\fsi^\veps = \ol{\Om_s^\veps} \cap  \ol{\Om_f^\veps}$, and by the external boundary of the pores $\pd_\ext\Om_f^\veps$.

\begin{figure}[t]
	\centering
	\includegraphics[width=0.7\linewidth]{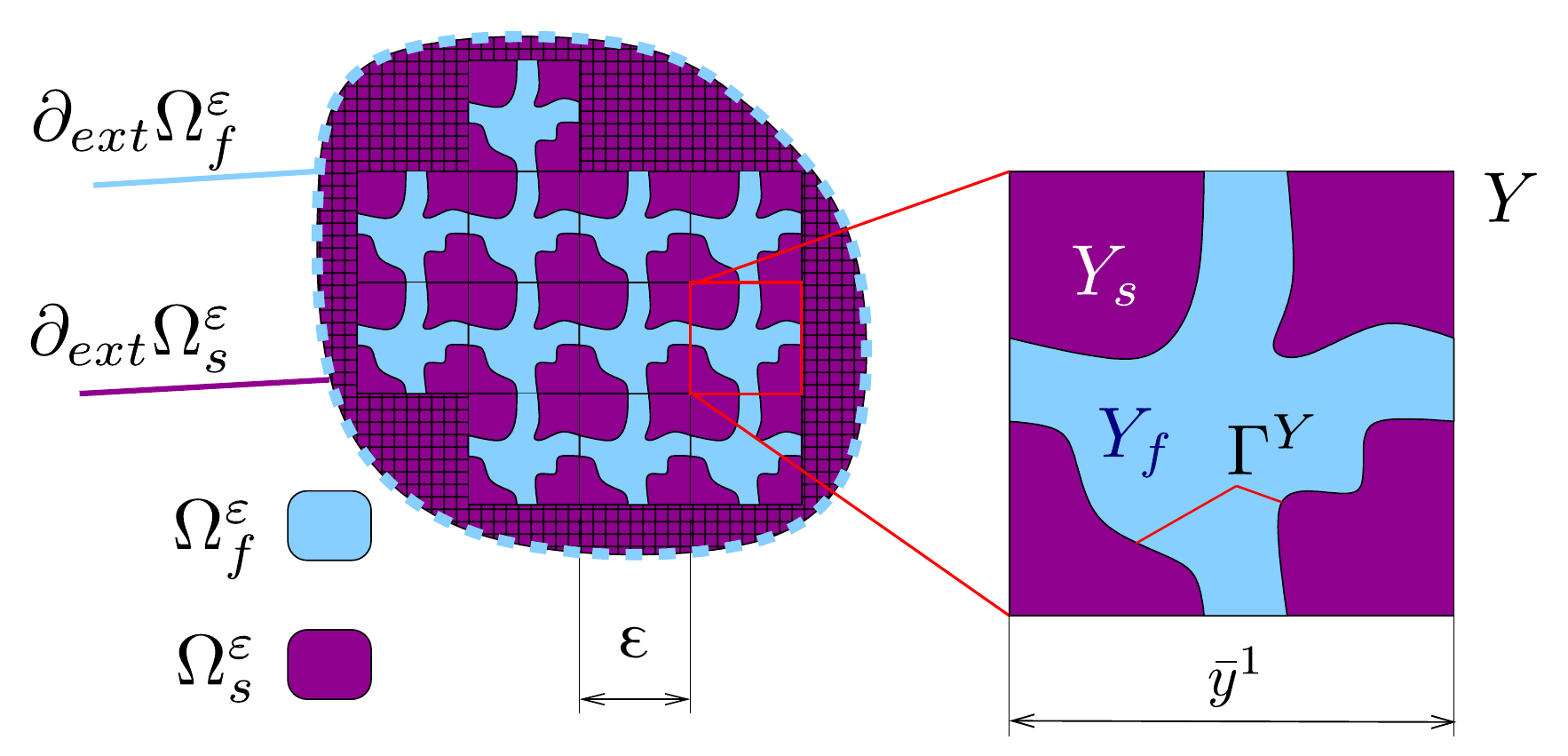}
        \caption{{A 2D scheme of the heterogeneous porous structure. (In 3D, both the solid and fluid subdomains $\Om_s^\veps$ and $\Om_f^\veps$ are connected.) Left: decomposition of the domain $\Om$ and its boundary $\pd\Om$ for a given scale $\veps$ into the solid and fluid parts. Right: the representative unit periodic cell $Y=\prod_{i=1}^3 ]0,\bar y^i[$.}}\label{fig-cell-Y}

\end{figure}

\subsection{Periodic microstructure}\label{sec-micro}
We consider a periodic structure of channels $\Om_f^\veps$ saturated by one homogeneous fluid. Due to the assumed periodicity of $\Om_f^\veps$, a representative periodic cell $\Zcal_f^\veps$ can be defined, which generates the fluid domain, see Fig.~\ref{fig-cell-Y}.
A periodic cell $\Zcal^\veps = \prod_{i=1}^3 ]0,\veps \bar y^i[$ can be introduced, such that $\Zcal^\veps = \Zcal_f^\veps \cup  \Zcal_s^\veps\cup\pd_{s}\Zcal_f^\veps$, where $\Zcal_s^\veps \subset \Zcal^\veps$ is the solid skeleton and $\pd_{s}\Zcal_f^\veps$ is the fluid-solid interface.
By $\pd_\# \Zcal_f^\veps = \pd \Zcal_f^\veps \setminus \pd_{s}\Zcal_f^\veps$ we denote the ``periodic part'' of the boundary. For the purpose of the homogenization we consider the ``unit periodic cell'' $Y = \veps^{-1}\Zcal^\veps$ which consists of the fluid and solid parts, $Y_f = \veps^{-1}\Zcal_f^\veps$ and $Y_s = \veps^{-1}\Zcal_s^\veps$ , respectively, accordingly the decomposition of $\Zcal^\veps$, thus, $Y = Y_f \cup Y_s \cup \Gamma$, where $\Gamma = \ol{Y_f} \cap \ol{Y_s}$ is the interface.

The homogenization procedure is presented below formally, without giving convergence proofs, however, the derivations of the limit equations can be followed. For this purpose, basics of the unfolding method are summarized in \Appx{apx-uf}.
We use the standard notation; for any $D\subset Y$ we abbreviate $\intYsmall_{D} = \frac{1}{|Y|}\int_{D}$; note that usually one may chose $\bar y^i$, $i=1,2,3$, such that $|Y|=1$.
Moreover, the following spaces are employed: $H_\#^1(Y_f) \subset H^1(Y_f)$ containing only $Y$-periodic functions. In analogy, a space of  $Y$-periodic vector-valued functions is denoted by  $\Hpdb(Y_f)$.


\subsection{Homogenization of an inviscid fluid}\label{sec-homog-ivs}
We recall the notation with the superscript ${}^\veps$ indicating dependence on the scale of pores $\Om_f^\veps$. The parameters $\theta = 2$, $\zeta = 3$ and $c_f$ are fixed. The homogenization procedure based on the periodic unfolding method \cite{Cioranescu2010} will be applied formally using the truncated asymptotic expansions of the pressure $p^\veps$ and the associated test functions $q^\veps$ involved in the weak formulation introduced below. It arises from  \eq{eq-NSadv3} supplemented by the Dirichlet boundary conditions given on $\pd_\ext \Om_f^\veps$ by $\tilde p^\pd$, and non-penetration conditions on $\Gamma_{fs}^\veps$, thus $\tilde\ub \cdot \nubf = 0$. The specific choice of the boundary conditions on $\pd_\ext \Om_f^\veps$ has no limiting consequences, since our aim is to analyze the wave dispersion in an unbounded media,

\subsubsection{Weak formulation for the acoustic waves with advection in an inviscid fluid}
 Recalling Assumption (A5), an advection velocity field $\bar\wb^\veps$ involved in \eq{eq-NSadv3} by virtue of the projected gradients $\pd_w^\veps:= \bar\wb^\veps\cdot\nabla$ must be introduced. For this we consider homogenization of the potential flow \eq{eq-potflow}; it is a classical academic problem of the homogenization in the perforated domains, see \eg \cite{Bensoussan1978book,Cioranescu-etal-2008}, nevertheless, a brief information on computing $\bar\wb^\veps$ in response to a given macroscopic velocity $\bar\wb^0$ is given in \Appx{appx-ivs-psi}.

We shall consider the following problem:
Find $\tilde p \in H^1(\Om_f)$ such that $\tilde p = \tilde p^\pd$ a.e. on $\pd_\ext \Om_f$, and
\begin{equation}\label{eq-NSadv5}
\begin{split}
  \int_{\Om_f^\veps}q^\veps\left(\ddot{\tilde p}^\veps + \theta\pd_w^\veps \dot{\tilde p}^\veps\right)
  - \int_{\Om_f^\veps}\pd_w^\veps q^\veps \left(\theta\dot{\tilde p}^\veps + \zeta\pd_w^\veps {\tilde p^\veps}\right)
  + c_f^2 \int_{\Om_f^\veps}\nabla{\tilde p^\veps}\cdot \nabla q^\veps = 0\;,
 \end{split}
\end{equation}
holds for a.a. $q^\veps \in H^1(\Om_f^\veps)$ such that $q^\veps = 0$ a.e. on $\pd_\ext \Om_f^\veps$.

 Note that \eq{eq-NSadv5} results from  \eq{eq-NSadv3} which is multiplied by $q$ and integrated in $\Om_f$, so that the obvious integration by parts leads to the following boundary integral,
\begin{equation}\label{eq-NSadv5a}
\begin{split}
 \Ical_{\pd\Om_f^\veps}(\tilde p^\veps,q^\veps) = \int_{\pd\Om_f}q^\veps \left(\nubf^f\cdot\nabla{\tilde p^\veps} - \zeta w_\nu^\veps \pd_w^\veps {\tilde p} - \theta w_\nu^\veps \dot{\tilde p}^\veps \right)\;,
\end{split}
\end{equation}
where  $w_\nu^\veps = \bar\wb^\veps\cdot\nubf^f$.
This integral vanishes on $\pd_\ext\Om_f$ due to the considered Dirichlet boundary conditions. On the walls $\Gamma_\fsi$, all the integrands in \eq{eq-NSadv5a} vanish as well due to the non-penetration condition which yields $\nubf^f\cdot \nabla  \tilde p^\veps = 0$ by virtue of \eq{eq-NS3}$_2$, since, in general, we assume $\nubf^f\cdot(\pd_w \tilde\ub + \tilde\ub\cdot\nabla\bar\wb) = 0$. For polygonal (polyhedral) interfaces this assumption can be verified, since $\pd_w \nubf^f:= \bar\wb\cdot\nabla \nubf^f = 0$.

\subsubsection{Limit problem}
We assume the advection velocity $\bar\wb^\veps$ to be of the order $o(\veps) = 1$, thus, using the unfolding operator, see \Appx{apx-uf},
$\Tuf{\bar\wb^\veps} = \bar\wb(x,y)$. Also the fluid phase velocity $c_f$ is assumed to be independent of $\veps$, so that the formal asymptotic expansion of $\tilde p^\veps$ is
\begin{equation}\label{eq-NSadv6}
\begin{split}
\Tuf{\tilde p^\veps} & = p^0(x,t) + \veps p^1(x,y,t)\;.
\end{split}
\end{equation}
The following sets are employed:
\begin{equation}\label{eq-NSadv6a}
\begin{split}
  W_0(\Om) & = H_0^1(\Om) = \{q \in H^1(\Om)\,|\; q = 0 \mbox{ on } \pd\Om\}\;, \\
  W_*(\Om) & = \{q \in H^1(\Om)\,|\; q = p^\pd \mbox{ on } \pd\Om\}\;,
\end{split}
\end{equation}
We may assume the Dirichlet boundary condition represented by $p^{\pd,\veps}$ converging with $\veps\rightarrow 0$ to $p^\pd$ defined on $\pd\Om$. Hence, by virtue of the convergence $\Tuf{\tilde p^\veps} \csto p^0$ 
we get $p^0 \in W_*(\Om)$.


\begin{remark}\label{rem-bc-p}
By virtue of the weak convergence of the unfolded pressure gradient,  \ie $\Tuf{p^\veps}(\cdot,t) \cwto \nabla_x p^0(\cdot,t) + \nabla_y p^1(\cdot,t)$ weakly in $L^2(\Om\times Y_y)$,
the trace of $p^0(\cdot,t) \in H^1(\Om)$ on $\pd_p \Om$ provides the limit trace  $p^0 =  p^\pd$ on  $\Om$.
\end{remark}

We define $\pd_w^x = \bar\wb\cdot\nabla_x$ and $\pd_w^y = \bar\wb\cdot\nabla_y$.
In the limit $\veps\rightarrow 0$, \eq{eq-NSadv5} yields the two-scale problem for $p^0$ and $p^1$ satisfying (recall \Appx{apx-uf} for the integration formula and $\intYsmall_{D} = \frac{1}{|Y|}\int_{D}$ for any $D\subset Y$)
\begin{equation}\label{eq-NSadv7}
\begin{split}
  \int_\Om q^0 \ddot p^0 + \theta\int_\Om q^0 \intY_{Y_f}\left (\pd_w^x \dot p^0 + \pd_w^y \dot p^1\right) - \theta\int_\Om\dot p^0 \intY_{Y_f}\left(\pd_w^x q^0 + \pd_w^y q^1\right) & \\
  - \zeta\int_\Om \intY_{Y_f}\left(\pd_w^x p^0 + \pd_w^y p^1\right)\left (\pd_w^x q^0 + \pd_w^y q^1\right) & \\
  + c_f^2 \int_\Om \intY_{Y_f}(\nabla_x p^0 + \nabla_y p^1)\cdot(\nabla_x q^0 + \nabla_y q^1) & = 0\;,
\end{split}
\end{equation}
for all $q^0 \in W_0(\Om)$ and $q^1 \in L^2(\Om; H_\#^1(Y_f))$.
Due to the incompressibility $\nabla\cdot\bar\wb = 0$ and since $\bar\wb(x,\cdot)$ is $Y$-periodic, for any $Y$-periodic function $\psi \in H_\#^1(Y_f)$
\begin{equation}\label{eq-NSadv7a}
\begin{split}
  \intY_{Y_f}\pd_w \psi = \intY_{Y_f}\bar\wb\cdot\nabla_y\psi = 0\;,
\end{split}
\end{equation}
so that, in the first line of \eq{eq-NSadv7}, both the integrals of $\pd_w^y \dot p^1$ and $\pd_w^y q^1$  vanish. The local problem is distinguished in \eq{eq-NSadv7} when putting $q^0 \equiv 0$.
By virtue of the linearity, the characteristic responses $\pi^k \in H_\#^1(Y_f)$, $k = 1,\dots,3$  are introduced, such that
\begin{equation}\label{eq-NSadv8}
\begin{split}
p^1 = \pi^k \pd_k^x p^0\;.
\end{split}
\end{equation}
Using the bilinear form (note that it involves $\zeta = 3$),
\begin{equation}\label{eq-NSadv9}
\begin{split}
  \aYw{p}{q}& = \intY_{Y_f}\left( c_f^2\nabla_y p\cdot \nabla_y q - \zeta\pd_w^y p \pd_w^y q \right) \\
  & = \intY_{Y_f} \left[\left(  c_f^2\Ib -\zeta \bar\wb\otimes\bar\wb \right) \nabla_y p \right]\cdot \nabla_y q \;,
\end{split}
\end{equation}
the autonomous problem for computing $\pi^k$ is established:
Find $\pi^k \in H_\#^1(Y_f)$, $k = 1,\dots,3$, such that
\begin{equation}\label{eq-NSadv10}
\begin{split}
  \aYw{\pi^k}{\psi} & = -\aYw{y_k}{\psi}\;,\quad \forall \psi \in H_\#^1(Y_f)\;.
\end{split}
\end{equation}

The macroscopic model of the acoustic waves which is presented below involves the  homogenized coefficients $\Acalbf = (\Acal_{ij})$ and the mean advection velocity $\wb^0$,  describing effective medium properties,
\begin{equation}\label{eq-NSadv11}
  \begin{split}
    \Acal_{ij} & =  \aYw{\pi^i + y_i}{y_j} = \aYw{\pi^i + y_i}{\pi^j + y_j}\;,\\
  \wb^0 & = \intY_{Y_f}\bar\wb\;,
\end{split}
\end{equation}
where the alternative symmetric expression for the \emph{anisotropic acoustic phase speed} $\Acalbf$  can be obtained due to problem \eq{eq-NSadv10}. Expression \eq{eq-NSadv11}$_2$ for the mean velocity $\wb^0$ is consistent with the definition of field $\bar\wb$ by virtue of the homogenization, see \eq{appx-ivs-psi}. In this context, when $\wb^0(x)$ varies with the macroscopic position, all expressions in \eq{eq-NSadv8}-\eq{eq-NSadv11} should be understood pointwise for a.a. $x\in \Om$.

\subsubsection{Macroscopic model of acoustic waves}
From the two-scale equation \eq{eq-NSadv7}, the macroscopic problem is obtained for $q^1 \equiv 0$ and the two-scale function $p^1$ is substituted by the split \eq{eq-NSadv8}, which yields
\begin{equation}\label{eq-NSadv12}
\begin{split}
  & \int_\Om q^0 \left(\phi_f\ddot p^0  + \theta\intY_{Y_f}\pd_w^x \dot p^0\right)
  - \int_\Om \dot p^0 \theta\intY_{Y_f} \pd_w^x q^0 \\
  & + \int_\Om  \zeta\intY_{Y_f}\left(\pd_w^x p^0 + \pd_w^y \pi^k \pd_k^x p^0 \right) \pd_w^x q^0
  +  c_f^2\int_\Om  \intY_{Y_f} \left(\nabla_x p^0 + \nabla_y \pi^k \pd_k^x p^0 \right) \cdot\nabla_x q^0 = 0\;,
\end{split}
\end{equation}
hence
\begin{equation}\label{eq-NSadv13}
\begin{split}
  \int_\Om \phi_f\ddot p^0 q^0
  +   \theta\int_\Om\left(\wb^0\cdot \nabla\dot p^0 q^0 -  \dot p^0 \wb^0\cdot \nabla q^0\right)
  + \int_\Om\aYw{\pi^k+ y_k}{y_l} \pd_k^x p^0 \pd_l^x q^0   = 0\;.
\end{split}
\end{equation}
In the last integral, the  expression involving the characteristic response $\pi^k$ is substituted using the homogenized $\Acalbf$, whereby the symmetry relationships follow due to the local micro-problem \eq{eq-NSadv10}.

\paragraph{Macroscopic problem}
Find $p^0(\cdot,t) \in W_*(\Om)$, such that
\begin{equation}\label{eq-NSadv15}
\begin{split}
  \int_\Om \phi_f  \ddot p^0 q^0 + \int_\Om (\Acalbf \nabla_x p^0)\cdot\nabla_xq^0
  +   \theta\int_\Om \wb^0\cdot(\nabla_x \dot p^0 q^0 - \nabla_xq^0\dot p^0)  = 0\;,
\end{split}
\end{equation}
for all {$q^0 \in W_0(\Om)$.}
It is worth to recall that the limit field $p^0$ satisfies the limit Dirichlet boundary condition given by $p^\pd$ on $\pd\Om$, as pointed above.

From \eq{eq-NSadv15}, the differential equation can be extracted upon integration by parts in $\Om$,
\begin{equation}\label{eq-NSadv15a}
\begin{split}
  \phi_f  \ddot p^0 - \nabla\cdot(\Acalbf \nabla_x p^0) + 2\theta \wb^0\cdot\nabla_x \dot p^0 = 0\;,
\end{split}
\end{equation}
whereby  the boundary integrals on $\pd\Om$ vanish due to $q^0\in W_0(\Om)$.

\begin{remark}\label{rem-1}
Besides the gradient projection $\wb^0\cdot\nabla_x \dot p^0$, the macroscopic stationary flow represented by $\wb^0$ influences the anisotropy of the wave propagation through  the coefficients $\Acalbf$ describing an anisotropic diffusion by virtue of \eq{eq-NSadv9} and \eq{eq-NSadv10}. Let us note that also for a static fluid,  $\bar\wb \equiv 0$, although coefficients $\Acalbf$ describe homogenized properties of an isotropic diffusion governed by the Laplace operator, the propagation is anisotropic in general due to the pore geometry given by $Y_f$.
\end{remark}

\subsubsection{Plane wave propagation in the homogenized medium}
Acoustic plane waves propagating in the homogenized infinite medium are expressed using the usual ansatz, defined in terms of
the (constant) amplitude $\bar p$, the wave direction $\nb$ and the wave number $\vkappa$ (the wave vector can be established, $\kappabf = \vkappa\nb$),
\begin{equation}\label{eq-NSadv19}
\begin{split}
  p^0(x,t) & := \bar p e^{\imu \om t} e^{-\imu \vkappa \nb\cdot\xb}\;.\\
\end{split}
\end{equation}
Substituted in \eq{eq-NSadv15a}, the following eigenvalue problem is obtained: For a given $\om \in \RR$ find $\vkappa \in \CC$, such that

\begin{equation}\label{eq-NSadv20}
\begin{split}
 \nb\otimes\nb : \Acalbf \vkappa^2  + 2\vkappa \om \theta \wb^0\cdot\nb -\phi_f \om^2 = 0\;.
\end{split}
\end{equation}
Using abbreviations,
\begin{equation}\label{eq-NSadv20a}
a_n =  \nb\otimes\nb : \Acalbf  \;,\quad
b_n = \theta\wb^0\cdot\nb\;,
\end{equation}
alternative relationships $\om\mapsto \vkappa_{1,2}$ and
$\vkappa\mapsto \om_{1,2}$
are established,
\begin{equation}\label{eq-NSadv20b}
\begin{split}
  \vkappa_{1,2} & = -\frac{\om}{a_n}(b_n \mp \sqrt{b_n^2 + a_n \phi_f})\;,\\
  \om_{1,2} & = \frac{\vkappa}{\phi}\left(b_n \mp \sqrt{b_n^2 + a_n \phi_f}\right)\;.
\end{split}
\end{equation}
Since $a_n >0$ due to the positive definiteness of $\Acalbf$, \eq{eq-NSadv20b} presents two waves of different sound speeds $c_{1,2} = \om/\vkappa_{1,2}$ (for a given $\om$), propagating in mutually opposite directions.
When $\bar\wb \equiv 0$, $b_n=0$ and a unique wave number (up to a sign) is obtained, $\vkappa = \om \sqrt{\phi_f/a_n} = \kappa_f c_f / \sqrt{a_n/\phi_f}$, where $\kappa_f = \om/c_f$ denotes the wave number of waves in a free stationary fluid. Then the effective sound speed of the static fluid in the rigid scaffolds is given by $c_* = \sqrt{(\nb\otimes\nb : \Acalbf)/\phi_f}$.


\subsection{Homogenization of a viscous fluid}

We consider homogenization of the linearized problem  arising from \eq{eq-NS3} which is now rewritten to respect its dependence on the scale parameter $\veps$. 
For any time $t>0$, assuming zero initial conditions, thus $\tilde\ub^\veps(x,t=0) = \bmi{0}$, and $\tilde p^\veps = 0$ in $\Om_f^\veps$,
the acoustic fluctuations $\tilde\ub^\veps$ and $\tilde p^\veps$ satisfy the following equations: 
\begin{equation}\label{eq-A1}
\begin{split}
  \rho_0\left( \pdiff{\tilde\ub^\veps}{t} + \bar\wb^\veps\cdot\nabla\tilde\ub^\veps + \tilde\ub^\veps\cdot\nabla\bar\wb^\veps\right) + \nabla \tilde p^\veps -\nabla\cdot\Dop^\veps\eeb{ \tilde\ub^\veps} & = \fb^\veps\;,\quad  \mbox{ in } \Om_f^\veps\;,\\
  \pdiff{\tilde p^\veps}{t} + \bar\wb^\veps\cdot\nabla\tilde p^\veps + k_f \cdot \nabla\cdot \ub^\veps & = 0\;,\quad  \mbox{ in } \Om_f^\veps\;,\\
  \tilde\ub^\veps & = \ub^{\pd,\veps}\;,\quad \mbox{ on } \pd_\ext\Om_f^\veps\;,\\
  \tilde\ub^\veps & = \bmi{0}\;,\quad \mbox{ on } \Gamma_\fsi^\veps\;,
\end{split}
\end{equation}
where $\ub^{\pd,\veps}(x,t)$ is a given function describing incident waves; we assume existence of an extension $\tilde\ub^{\pd,\veps}$ of $\ub^{\pd,\veps}$ to $\Om_f^\veps$, such that $\tilde\ub^{\pd,\veps} = 0$ on $\Gamma_{fs}^\veps$.
The following spaces and the set $\hat V^\veps(\Om_f^\veps,t)$ will be used:
\begin{equation}\label{eq-A1a}
\begin{split}
  V_0(\Om_f^\veps) & = \{\vb \in \Hdb(\Om_f^\veps)|\; \vb = \bmi{0}\;,\mbox{ on } \pd\Om_f^\veps\}\;,\\
  \hat V^\veps(\Om_f^\veps,t) & = \{\vb \in \Hdb(\Om_f^\veps)|\; \vb = \vb^{\pd,\veps}(\cdot,t)\;,\mbox{ on } \pd\Om_f^\veps\;,\quad \vb = \bmi{0}\;,\mbox{ on } \Gamma_{fs}^\veps\} \;,\\
  Q(\Om_f^\veps) & = L^2(\Om_f^\veps) \;,\\
  Q^1(\Om_f^\veps) & = H^1(\Om_f^\veps) \;.
\end{split}
\end{equation}
Alternatively $\hat V^\veps(\Om_f^\veps,t) = V_0(\Om_f^\veps) + \tilde\ub^{\pd,\veps}(\cdot,t)$. Below we suppress the time dependence in the notation, so that we use $\hat V^\veps(\Om_f^\veps)$.

The weak formulation of the acoustic problem arises from problem \eq{eq-A1} whereby the steady state flow $\bar \wb^\veps$ is assumed to be known.
The fluctuating velocity and pressure fields $(\wtilde\ub^\veps,\wtilde{p}^\veps)$, such that for a.a. $t > 0$,  $\wtilde\ub^\veps(t,\cdot) \in \hat V^\veps(\Om_f^\veps)$, and $\wtilde{p}^\veps(t,\cdot) \in Q^1(\Om_f^\veps)$, satisfy
\begin{equation}\label{eq-FS13}
  \begin{split}
    \rho_0\int_{\Om_f^\veps}\vb^\veps\cdot\left(\dot{\wtilde\ub}^\veps  + \bar\wb^\veps\cdot\nabla \wtilde\ub^\veps  + \wtilde\ub^\veps\cdot\nabla\bar\wb^\veps\right) & \\
    -  \int_{\Om_f^\veps} \wtilde{p}^\veps\nabla\cdot\vb^\veps +
    \int_{\Om_f^\veps}\Dop^{f,\veps} \eeb{ \wtilde\ub^\veps}:\eeb{\vb^\veps} & = \int_{\Om_f^\veps}\tilde\fb^f\cdot \vb^\veps\;, 
    \quad \forall \vb^\veps \in V_0^\veps(\Om_f^\veps)\;,\\
 \int_{\Om_f^\veps}q^\veps\left(
 \dot{\wtilde{p}}^\veps + \bar\wb^\veps \cdot \nabla\wtilde{p}^\veps  + k_f \nabla\cdot\wtilde\ub^\veps \right) & = 0\;,\quad \forall q^\veps \in Q^0(\Om_f^\veps)\;.
  \end{split}
\end{equation}

We pursue the formal procedure of the asymptotic analysis $\veps \rightarrow 0$ based on the formal unfolding method.
 The truncated asymptotic expansions and the viscosity scaling are adopted, however, in contrast with the steady state flow problem, see e.g. \cite{Chen-homog-NS-Forchheimer2001,Peszynska2010-proc}, a different scaling of the viscosity and of the advection velocity must be used,
\begin{equation}\label{eq-FS14a}
\begin{split}
\mu^\veps & = \veps^2 \bar\mu\;,\quad \Dop^{f,\veps} = \veps^2 2\bar\mu(\Iop - \frac{1}{3}\Ib\otimes\Ib)\;,\\
  \quad\quad  \Tuf{\bar\wb^\veps} & = \veps\bar\wb(x,y)\;,
\end{split}
\end{equation}
{where $\Iop$ is the fourth-order identity tensor and $\bar\wb(x,y)$ is the two-scale solution of steady state flow problem.}
The unfolded solutions $(\wtilde\ub^\veps,\wtilde\wb^\veps,\wtilde{p}^\veps)$ are represented by the truncated expansions
\begin{equation}\label{eq-FS14b}
\begin{split}
  \Tuf{\wtilde\ub^\veps(x,t)} & = \hat\ub(x,y,t)\;,\\
  \Tuf{\wtilde{p}^\veps(x,t)} & = p^0(x,t) + \veps p^1(x,y,t)\;.
\end{split}
\end{equation}
The test functions $\vb^\veps$ and $q^\veps$ are considered in the analogous form, thus
\begin{equation}\label{eq-FS14c}
\begin{split}
  \Tuf{\wtilde\vb^\veps(x)} & = \hat\vb(x,y)\;,\\
  \Tuf{q^\veps(x)} & = q^0(x) + \veps q^1(x,y)\;.
\end{split}
\end{equation}
All the two-scale functions are $Y$-periodic in the second variable $y$ and for almost all $t>0$
\begin{equation}\label{eq-FS14d}
\begin{split}
\hat\ub(\cdot,t),\hat\vb & \in L^2(\Om;\HpdbO(Y_f))\;,\\
p^1(\cdot,t),q^1 & \in L^2(\Om; H_\#^1(Y_f))\;,\quad p^0(\cdot,t) \in H^1(\Om)\;.
\end{split}
\end{equation}
Note that the non-slip condition of $\hat\ub$ on $\Gamma_\fsi$ is imposed due
to the space $\HpdbO(Y_f) \subset \Hpdb(Y_f)$ \tododo{introduced as follows}, see Section~\ref{sec-micro}, 
\begin{equation}\label{eq-Hper0}
\HpdbO(Y_f) = \{\vb \in \Hpdb(Y_f)|\;\vb = \bmi{0} \mbox{ on } \pd Y_f\setminus \pd Y\}\;.
\end{equation}

\begin{remark}\label{rem-u}
It is important to identify the limit boundary conditions for the homogenized flow.
To respect a prescribed velocity, in the limit, the Dirichlet condition of the micromodel {transforms} to a Neumann-type condition associated with the pressure. We assume the existence of $\ub^\pd \in L^2(\pd\Om)$ such that $\bar\phi_f\ub^\pd = \traceG{\phi_f\ub^0}{\Om}$, and
\begin{equation}\label{eq-FS29}
  \begin{split}
\int_{\pd\Om_f^\veps}\vphi \ub^{\pd,\veps}\cdot \nubf \rightarrow \int_{\pd\Om}\vphi\bar\phi_f\ub^\pd\cdot \nubf\;,\quad \forall \vphi \in L^2(\pd\Om)\;,
  \end{split}
\end{equation}
where $\bar\phi_f$ is the surface porosity while $\phi_f\in H^1(\Om)$ is the volume porosity. In \Appx{appx-bc-u}, we show $\ub^0\cdot\nubf = \ub^\pd\cdot \nubf$.   This identifies the desired  boundary condition for the limit macroscopic velocity $\ub^0$ which, however, is expressed in terms of the Darcy law involving $\nabla p^0$.
\end{remark}

The volume forces fluctuations $\tilde\fb^f$ are assumed to be independent of the scale $\veps$, such that the associated virtual powers converge, as follows,
\begin{equation}\label{eq-FS14f}
\begin{split}
  \int_{\Om_f^\veps} \tilde\fb^f\cdot \wtilde\vb^\veps & \rightarrow \int_\Om \fbav^f\cdot\intY_{Y_f}\hat\vb\;.
\end{split}
\end{equation}

\subsection{Limit problem}
The limit analysis of \eq{eq-A1} for $\veps\rightarrow 0 $ leads to the following 2-scale problem: For a.a. $t \in ]0,T]$, find
$(\hat\ub,p^0,p^1)$ such that $\hat\ub(t,\cdot) \in  L^2(\Om;\HpdbO(Y_f))$, $p^1 \in L^2(\Om\times Y_f)$, $p^0 \in H^1(\Om)$, and
\begin{equation}\label{eq-A8ex}
\begin{split}
  \rho_0\intY_{Y_f}\left(\pdiff{\hat\ub}{t} + \bar\wb\cdot \nabla_y \hat\ub + \hat\ub\cdot \nabla_y \bar\wb\right)\cdot\vb & \\
  + \intY_{Y_f}\bar\Dop\eeby{\hat\ub}:\eeby{\vb} + \intY_{Y_f}(\nabla_x p^0)\cdot\vb -\intY_{Y_f} p^1 \nabla_y\cdot\vb & = \fb\cdot \intY_{Y_f} \vb\;,\quad \forall \vb\in \HpdbO(Y_f)\;,\\
  \intY_{Y_f} q \nabla_y\cdot\hat\ub & = 0\;, \quad \forall q\in L^2(Y_f)\;,\\
  \phi_f\pdiff{p^0}{t}  + k_f \nabla_x\cdot\Ub  & = 0 \quad \mbox{ in }\Om\;,\\
  \quad\mbox{ where } \Ub & = \intY_{Y_f}\hat\ub\;,
\end{split}
\end{equation}
with initial and boundary conditions to be specified.

\paragraph{Characteristic responses}
For decoupling the scales,  the problem is transformed by the Laplace transformation $\LTxt{a}\mapsto \LT{a}$, so that the following split can be defined:
\begin{equation}\label{eq-A10a}
\begin{split}
  \LT{\hat\ub} & = \lam\LT{\chibf}^k\rho_0^{-1}(\LT{f}_k - \pd_k^x\LT{p^0})\;,\\
  \LT{p^1} & = \lam\LT{\pi}^k(\LT{f}_k - \pd_k^x\LT{p^0})\;.
\end{split}
\end{equation}
We shall employ the following notation:
\begin{equation}\label{eq-F7}
  \begin{split}
    \ipYf{\Ab}{\Bb} & = \intY_{Y_f} \Ab:\Bb\;,\\
  \aYf{\ub}{\vb} &  = \ipYf{\nabla_y \ub}{\nabla_y \vb}\;,\\
  \atYf{\ub}{\vb} &  = 2\ipYf{\eeby{\ub} - \frac{1}{3}\nabla_y\cdot\ub}{\eeby{\vb}- \frac{1}{3}\nabla_y\cdot\vb}\\
  & =  2\ipYf{\Dev{\eeby{\ub}}}{\Dev{\eeby{\vb}}}\;,\\
  \bYf{\bar\wb}{\ub}{\vb} & = 2\ipYf{\bar\wb\otimes\ub}{\eeby{\vb}}\;,\\
  \cYf{\bar\wb}{\ub}{\vb} & = \ipYf{\bar\wb\cdot\nabla_y \ub}{\vb}\;,\\
  \gYf{\bar\wb}{\ub}{\vb} & = \ipYf{\ub\cdot\nabla_y \bar\wb}{\vb}\;,
\end{split}
\end{equation}
where $\Dev{\eb} = \eb - (1/3) \eb:\Ib$ is the deviatoric part of the strain tensor.

Note that for $Y$-periodic fields $\ub,\vb$ vanishing on $\Gamma$ and satisfying $\nabla_y \cdot\ub = 0$ and  $\nabla_y\cdot \vb = 0$, it holds that $\aYf{\ub}{\vb} = \atYf{\ub}{\vb}$.
Using the notation introduced above,  \eq{eq-A8ex}$_1$ can be rewritten alternatively, as follows:
\begin{equation}\label{eq-A9}
  \begin{split}
    \ipYf{\pdiff{\hat\ub}{t}}{\vb} -  \bYf{\bar\wb}{\hat\ub}{\vb} + \frac{\bar\mu}{\rho_0}\atYf{\hat\ub}{\vb}
    - \frac{1}{\rho_0}\ipYf{p^1}{\nabla_y\cdot\vb} & =  \frac{1}{\rho_0}(f_k - \pd_k^x p^0) \ipYf{1}{v_k}\;,\\
    \ipYf{\pdiff{\hat\ub}{t}}{\vb} + \frac{1}{2}\left( \cYf{\bar\wb}{\hat\ub}{\vb} - \cYf{\bar\wb}{\vb}{\hat\ub}\right)
    + \gYf{\bar\wb}{\hat\ub}{\vb} &\\
    + \frac{\bar\mu}{\rho_0}\atYf{\hat\ub}{\vb}
    - \frac{1}{\rho_0}\ipYf{p^1}{\nabla_y\cdot\vb} & =  \frac{1}{\rho_0}(f_k - \pd_k^x p^0) \ipYf{1}{v_k}\;.
\end{split}
\end{equation}
Now, \eq{eq-A9} can be transformed and \eq{eq-A10a} substituted, which yields a problem for
$\LT{\chibf}^k \in \HpdbO(Y_f)$ and $\LT{\pi}^k \in L^2(Y_f)$, such that
\begin{equation}\label{eq-A10b}
  \begin{split}
    \lam\ipYf{\LT{\chibf}^k}{\vb} - \bYf{\bar\wb}{\LT{\chibf}^k}{\vb}
    + \frac{1}{\Reynolds}\atYf{\LT{\chibf}^k}{\vb} -\ipYf{\LT{\pi}^k}{\nabla_y\cdot\vb}  & = \frac{1}{\lam}\ipYf{1}{v_k}\quad \forall \vb \in \HpdbO(Y_f)\;,\\
    \ipYf{q}{\nabla_y\cdot\LT{\chibf}^k} & = 0\quad \forall q \in L^2(Y_f)\;,
\end{split}
\end{equation}
where the $\frac{1}{\Reynolds} = \bar\mu/\rho_0$ is the Reynolds number; note that, for a given scale parameter $\veps_0 > 0, $$\frac{1}{\Reynolds} = \mu^\phy/(\veps_0^2 \rho_0)$ can be expressed by the physical viscosity.

By virtue of \eq{eq-A8ex}$_{3,4}$ which are the macroscopic equations, the effective velocity is computed using the permeability which is defined by (note $\bar\wb(x,y)$ is a two-scale function, in general)
\begin{equation}\label{eq-A11}
\begin{split}
  \LT{\Kcal}_{ij}(\lam,x) & = \frac{\lam}{\rho_0}\intY_{Y_f}\LT{w}_i^j =  \frac{\lam}{\rho_0}\ipYf{1}{\LT{w}_i^j}\\
    & =  \frac{1}{\rho_0}\left(\lam^3\ipYf{\LT{\chibf}^i}{\LT{\chibf}^j} -\lam^2\bYf{\bar\wb}{\LT{\chibf}^i}{\LT{\chibf}^j}
    + \lam\frac{1}{\Reynolds}\atYf{\LT{\chibf}^i}{{\chibf}^j}\right)\;.
\end{split}
\end{equation}
It is possible to define
\begin{equation}\label{eq-A11a}
\begin{split}
  \LT{\what\Kcal}_{ij}(\lam,x) & = \lam^2\intY_{Y_f}\LT{w}_i^j = \lam \rho_0 \LT{\Kcal}_{ij}(\lam,x)\\
  & = \lam^4\ipYf{\LT{\chibf}^i}{\LT{\chibf}^j} -\lam^3\bYf{\bar\wb}{\LT{\chibf}^i}{\LT{\chibf}^j}
    + \lam^2\frac{1}{\Reynolds}\atYf{\LT{\chibf}^i}{{\chibf}^j}\;,
\end{split}
\end{equation}
which will be used in the the plane wave propagation analysis.

It is possible compute $\chibf^k$ by solving problem \eq{eq-A10b} transformed in the time domain: for a.a. $t\in ]0,\infty[$ let $\chibf^k(t,\cdot) \in \HpdbO(Y_f)$ and $\pi^k(t,\cdot) \in L^2(Y_f)$ satisfy
\begin{equation}\label{eq-A10c}
  \begin{split}
    \pdiff{}{t}\ipYf{\chibf^k}{\vb} - \bYf{\bar\wb}{{\chibf}^k}{\vb}
    + \frac{1}{\Reynolds}\atYf{{\chibf}^k}{\vb} -\ipYf{{\pi}^k}{\nabla_y\cdot\vb}  & = H(t) \ipYf{1}{v_k}\quad \forall \vb \in \HpdbO(Y_f)\;,\\
    \ipYf{q}{\nabla_y\cdot{\chibf}^k} & = 0\quad \forall q \in L^2(Y_f)\;,\\
    \chibf^k(0,y) & = \bmi{0}\quad \mbox{ for a.a. } y \in Y_f\;,
\end{split}
\end{equation}
where $H(t)$ is the Heaviside function (since $t>0$, $H(t) = 1$ in the \rhs term).
 We assume zero initial conditions $\hat\ub(t=0,x,y) = \bmi{0}$ in $\Om\times Y_f$. Therefore, the macroscopic velocity is given by
\begin{equation}\label{eq-A10d}
\begin{split}
  \Ub(t,x) & = \int_0^t \dt{}{t} \left(\intY_{Y_f} \chibf^k(t-s,y) \dd y \right) (f_k(s,x) - \pd_k^x p^0(s,x))\dd s\;.
\end{split}
\end{equation}
Then, the permeability can be evaluated, as follows:
\begin{equation}\label{eq-A11b}
\begin{split}
  {\Kcal}_{ij}(t) & =  \frac{1}{\rho_0}\dt{}{t}\intY_{Y_f} \chi_i^j(t,y)\dd y\;.
\end{split}
\end{equation}
Note the physical dimensions: $[w_i^j] = \rm{s}$, hence $[{\Kcal}_{ij}] = [\rho_0^{-1}][w_i^j] = \frac{\rm{m}^3 \cdot \rm{s}}{\rm{kg}} = \frac{\rm{m}^2}{\rm{Pa} \cdot  \rm{s}^2}$.

\subsection{Macroscopic model --- time domain}
As announced earlier, the macroscopic model is given by \eq{eq-A8ex}$_{3,4}$. Using the Darcy law
\begin{equation}\label{eq-A12}
\begin{split}
  \LT{\Ub}(\lam,x) & =  \LT{\Kcalbf}(\LT{\fb} - \nabla_x\LT{p^0})\;,\\
  \Ub(t,x) & = \int_0^t \Kcalbf(t-s)(\fb(s) - \nabla_x p^0(s))\dd s\;,
\end{split}
\end{equation}
the macroscopic acoustic problems reads: Given $p^0(t=0,\cdot)$ by the initial condition and $\fb(t,\cdot)$ in $\Om$, for a.a. $t > 0$ find $p^0(t,x)$, such that
\begin{equation}\label{eq-A13}
\begin{split}
  \phi_f\pdiff{p^0}{t}  + k_f \nabla_x\cdot\int_0^t\Kcalbf(t-s)(\fb(s) - \nabla_x p^0(s))\dd s & = 0\quad \mbox{ in } \Om\;,\\
  \nubf\cdot\Ub & = \nubf\cdot\ub^\pd\quad \mbox{ on }\pd\Om\;,
\end{split}
\end{equation}
see Remark~\ref{rem-u}, where $\Ub$ is expressed by \eq{eq-A12}$_2$.

\paragraph{Macroscopic model --- frequency domain}
In the frequency domain ($\lam=\imu\om$), \eq{eq-A8ex}$_{3,4}$ transform to
\begin{equation}\label{eq-A14}
  \begin{split}
   \imu\om \phi_f \FT{p^0} + \frac{k_f}{\imu\om\rho_0} \nabla_x\cdot\FT{\what\Kcalbf}(\imu\om) (\FT{\fb} - \nabla_x\FT{p^0}) & = 0\quad \mbox{ in } \Om\;.
\end{split}
\end{equation}
Recalling the notation formerly introduced, $\kappa_f = \om / c_f$ designating the wave number associated with the phase velocity of wave propagation $c_f = \sqrt{k_f/\rho_0}$ in the unconfined (free) fluid, the ansatz \eq{eq-NSadv19} for the pressure plane wave yields the wave number of the pressure wave in the porous medium,
\begin{equation}\label{eq-w3A}
  \begin{split}
    \vkappa & = \kappa_f\sqrt{\frac{\phi_f}{\FT{\what\Kcalbf}(\imu\om):\nb\otimes\nb}}
    = \sqrt{\frac{-\imu\om\phi_f}{k_f \FT{\Kcalbf}(\imu\om):\nb\otimes\nb}}\;,
\end{split}
\end{equation}
where the alternative expressions correspond to the permeabilities introduced in \eq{eq-A11} and \eq{eq-A11a}, respectively.


\section{Floquet-Bloch wave decomposition for analysis of fluid acoustics}\label{sec-Bloch}

We shall consider plane waves propagating in an infinite porous medium, such that $\Om_f^\veps$ is unbounded, being generated as a periodic structure by the representative scaffold cell $\Zcal_f^\veps\subset\Zcal^\veps$; we recall that $\Zcal^\veps = \veps Y$ is the real size of the periodic cell $Y$. Since in the following analysis the scale is fixed, $\veps> 0$, for the sake of brevity,  we shall drop the superscript $\veps$ which is related to a given scale of the structure, thus, $\Zcal_f$ is used to refer to $\Zcal_f^\veps$, etc.

By virtue of the Floquet-Bloch theory, the wave response of the flow models introduced in Section~\ref{sec-flow-model} is represented by functions $\tilde\ub$ and $\tilde p$ expressed in the decomposed form,
\begin{equation}\label{eq-bloch-1}
\begin{split}
\tilde\ub(x,t) & = \ub(x)e^{-\imu\kappabf\cdot\xb}e^{\imu\om t}\;,\\
\tilde p(x,t) & = p(x)e^{-\imu\kappabf\cdot\xb}e^{\imu\om t}\;,
\end{split}
\end{equation}
where $\kappabf = \vkappa\nb$ is the wave vector given by the  direction of wave propagation $\nb$, and the wave number $\vkappa$.
Functions $\ub$ and $p$ are $\Zcal^\veps$-periodic  and, on the channel walls $\pd_s\Zcal_f^\veps$, satisfy the boundary conditions according to the flow model.

To employ the weak formulations for derivation of the dispersion relationships, it is convenient to consider the test functions, such that they describe waves propagating in the opposite directions with respect to those of \eq{eq-bloch-1}, thus
\begin{equation}\label{eq-bloch-1a}
\begin{split}
\tilde\vb(x,t) & = \vb(x)e^{\imu\kappabf\cdot\xb}e^{\imu\om t}\;,\\
\tilde q(x,t) & = q(x)e^{\imu\kappabf\cdot\xb}e^{\imu\om t}\;,
\end{split}
\end{equation}
where $\vb,q$ are $\Zcal$-periodic.
As a consequence, the strain tensor applied to $\tilde\ub$, or to another function of the same form, is expressed in terms of the wave-strain $\ggb{\tilde\ub}{\nb}$, thus,
\begin{equation}\label{eq-uwp-B8}
\begin{split}
\eeb{\tilde\ub} = \left(\eeb{\ub} - \imu\vkappa \ggb{\ub}{\nb}\right)e^{-\imu\kappabf\cdot\xb}e^{\imu\om t}\;,\quad \mbox{ where }
\ggb{\ub}{\nb} = \frac{1}{2}(\ub\otimes\nb + \nb\otimes\ub)\;.
\end{split}
\end{equation}
In analogy, $\eeb{\tilde\vb} = \left(\eeb{\vb} + \imu\vkappa \ggb{\vb}{\nb}\right)e^{\imu\kappabf\cdot\xb}e^{\imu\om t}$.



\subsection{Floquet-Bloch wave decomposition for inviscid flows}\label{sec-FBivs}


The plane wave propagation in the porous medium is governed by \eq{eq-NSadv3}, involving the two parameters  $\theta = 2$ and $\zeta = 3$. In analogy with \eq{eq-NSadv5}, the weak formulation is obtained which describes the wave propagation in the representative cell $\Zcal_f$,
\begin{equation}\label{eq-NSadv16}
\begin{split}
  \int_{\Zcal_f}\tilde q (\ddot{\tilde{p}} + \theta\pd_w \dot{\tilde{p}}) - \int_{\Zcal_f}\pd_w\tilde q(\theta\dot{\tilde{p}} + \zeta\pd_w {\tilde{p}})
  + c_f^2 \int_{\Zcal_f}\nabla \tilde p \cdot \nabla \tilde q  & = \Ical_{\pd\Zcal_f}(\tilde p,q)\;,
\end{split}
\end{equation}
where $\Ical_{\pd\Zcal_f}(\tilde p,q)$ is defined according to \eq{eq-NSadv5a}.
For the wave ansatz \eq{eq-bloch-1}$_2$ and \eq{eq-bloch-1a}$_2$ substituted in \eq{eq-NSadv16}, there the right-hand side integral $\Ical_{\pd\Zcal_f}(\tilde p,q)$ vanishes. The velocity field is confined by the non-penetration and the free-slip conditions, so that  $\ub\cdot\nubf^f = 0$ is prescribed on $\pd_s\Zcal_f$. As a consequence, for almost any time $t$, the admissible acoustic pressure $p(\cdot,t)$ must satisfy
\begin{equation}\label{eq-ivs3a}
\begin{split}
  p(\cdot,t) \in P_\#(\Zcal_f) := \{q \in  H_\#^1(\Zcal_f)| q \mbox{ is $\Zcal$-periodic, } \nabla (q e^{-\imu\kappabf\cdot\xb}) \cdot\nubf^f = 0 \mbox{ on }  \pd_s\Zcal_f\}\;.
\end{split}
\end{equation}
Note that, in the definition of $P_\#(\Zcal_f^\veps) $, the zero normal-projected gradient concerns $\tilde p$ involving the exponential part and not only $p$, see \eq{eq-bloch-1}.

To find nontrivial solutions to \eq{eq-NSadv16}, the following dispersion problem must be solved: Find $p \in P_\#(\Zcal_f)$ and $\vkappa \in \CC$, such that
\begin{equation}\label{eq-NSadv18}
  \begin{split}
    -\om^2\int_{\Zcal_f} p q + \imu \om \theta\left(\int_{\Zcal_f}\bar\wb\cdot[\nabla p - \imu\vkappa \nb p]q
    - \int_{\Zcal_f}\bar\wb\cdot[\nabla q + \imu\vkappa \nb q]p \right) \\
    -  \zeta\int_{\Zcal_f}\bar\wb\cdot[\nabla p - \imu\vkappa \nb p][\nabla q + \imu\vkappa \nb q]\cdot\bar\wb
    + c_f^2  \int_{\Zcal_f}[\nabla p - \imu\vkappa \nb p]\cdot[\nabla q + \imu\vkappa \nb q] = 0\;,
\end{split}
\end{equation}
for all $q \in P_\#(\Zcal_f)$. Alternatively, for a given $\kappa \in \RR$, frequency $\om \in \CC$ can be computed

\paragraph{FEM -- discretization of the Bloch wave analysis} Problem
\eq{eq-NSadv18} is solved numerically using the finite element method (FEM).
{The matrices resulting from the conforming FEM discretization of
  integral forms involved in \eq{eq-NSadv18} are introduced below, employing a
  self-explaining notation}:

\begin{equation}\label{eq-NSadv21}
  \begin{split}
 \int_{\Zcal_f}\nabla p\cdot \nabla q & \FEMapprox = \qbm^T\Cbm\pbm\;,\\  
 \int_{\Zcal_f}\nabla p\cdot \nb q & \FEMapprox   = \qbm^T\Ybm \pbm\;,\\  
  \int_{\Zcal_f}p q & \FEMapprox \qbm^T\Mbm\pbm\;,  \\
   \zeta\int_{\Zcal_f}\bar\wb\cdot\nabla p \bar\wb\cdot\nabla q & \FEMapprox   \qbm^T\Gbm_w\pbm\;,\\
  \zeta\int_{\Zcal_f}\bar\wb\cdot\nabla p (\bar\wb \cdot \nb) q & \FEMapprox  \qbm^T\Rbm_w \pbm\;,\\
  \zeta\int_{\Zcal_f} (\bar\wb \cdot \nb) p (\bar\wb \cdot \nb) q & \FEMapprox \qbm^T\Wbm_w\pbm\;,\\
 \theta\int_{\Zcal_f} (\bar\wb \cdot \nb) p q & \FEMapprox \qbm^T\Vbm_w\pbm\;,\\
 \theta\int_{\Zcal_f}q (\bar\wb\cdot\nabla p) & \FEMapprox \qbm^T\Nbm_w\pbm\;.
\end{split}
\end{equation}
Above the column matrices $\pbm$ and $\qbm$ represent all degrees of freedom (DOFs) arising from the discretization of $p$ and $q$, respectively, considering $\Zcal_f$, whereas the DOFs reduction due to the periodic conditions on the external channel surfaces $\pd_\# \Zcal_f = \pd\Zcal \cap \pd\Zcal_f$ is applied.
Using this notation, identity \eq{eq-NSadv18} is approximated by the following equation,
\begin{equation}\label{eq-NSadv22}
\begin{split}
  \left[c_f^2 \Cbm + (c_f^2\vkappa^2 - \om^2) \Mbm + \vkappa^2 \Wbm_w + 2\vkappa\om \Vbm_w - \Gbm_w \right]\pbm \\
  + \imu\left[ \om(\Nbm_w - \Nbm_w^T) - \vkappa(\Rbm_w - \Rbm_w^T) - \vkappa c_f^2(\Ybm - \Ybm^T) \right]\pbm = 0\;.
\end{split}
\end{equation}
Two alternative eigenvalue problems can be considered:
a) for a given real frequency $\om$, find a complex $\vkappa \in \CC$ which satisfies \eq{eq-NSadv22}, or b) for a given real $\vkappa \in \RR$, find a complex frequency $\om \in \CC$ satisfying \eq{eq-NSadv22}. Both these formulations lead to a quadratic eigenvalue problem (QEP). To transform it into a standard eigenvalue problem, we employ the following notations,
\begin{equation}\label{eq-NSadv23}
\begin{split}
  \Qbm & := \Rbm_w + c_f^2 \Ybm \;,\\
  \Hbm & := \Gbm_w - c_f^2 \Cbm\;,\\
  \Ubm & := \Wbm_w + c_f^2 \Mbm\;.
\end{split}
\end{equation}
Then, using the Choleski decomposition, auxiliary variables $\sbm$ and $\rbm$ can be introduced, such that
\begin{equation}\label{eq-NSadv23a}
  \begin{split}
    \Ubm & =    \Tbm_U^T \Tbm_U\;,\quad  \vkappa^2 \Ubm \pbm  = \vkappa \Tbm_U^T \sbm\;,\quad  \vkappa \Tbm_U\pbm   = \sbm\;,\\
    \Mbm & = \Tbm_M^T\Tbm_M\;,\quad \om^2 \Mbm \pbm  = \om \Tbm_M^T \rbm\;,\quad  \Tbm_M\pbm = \rbm\;.
 \end{split}
\end{equation}
Let us first consider the problem for computing $\vkappa$ for a given $\om$.
Using the decomposition in \eq{eq-NSadv23a}$_1$, the identity \eq{eq-NSadv22} can be transformed into the following standard eigenvalue problem for computing $\lam_\vkappa := 1/\vkappa$ which satisfy,
\begin{equation}\label{eq-NSadv24}
\begin{split}
  \left[
    \begin{array}{ll}
\Abm_\vkappa(\om), & \Tbm_U^T \\ \Tbm_U, & \zerobm
      \end{array}
    \right]
  \left[
    \begin{array}{c} \pbm \\ \sbm
      \end{array}
    \right] =
  \lam_\kappa
  \left[
    \begin{array}{ll}
      \Bbm_\vkappa(\om), & \zerobm \\
      \zerobm, & \Ibm
      \end{array}
    \right]
  \left[
    \begin{array}{c} \pbm \\ \sbm
      \end{array}
    \right]\;, \\
  \mbox{ where } \Abm_\vkappa(\om) = 2\om \Vbm_w -\imu(\Qbm - \Qbm^T)\;,\quad \Bbm_\vkappa(\om) = \om^2 \Mbm + \Hbm + \imu\om (\Nbm_w^T - \Nbm_w)\;.
\end{split}
\end{equation}
The reason for computing $\lam_\vkappa$ (which is proportional to the wavelength) rather than $\vkappa$ arises form the regularity of the \rhs block matrix of this generalized eigenvalue problem, since the \lhs matrix may become nearly singular. This may happen because of vanishing $\Vbm_w$, when $\bar\wb \cdot \nb\approx 0$, thus, when the advection flow field $\bar\wb$ is almost orthogonal to the wave direction $\nb$.

Alternatively, $\om$ can be computed in response to $\vkappa$. For this, in analogy, we employ the decomposition in \eq{eq-NSadv23a}$_2$. Using the notation \eq{eq-NSadv23}, the following generalized eigenvalue problem can be established which yields a real $\lam_\om := 1/\om$,
\begin{equation}\label{eq-NSadv25}
\begin{split}
  \left[
    \begin{array}{ll}
 \Abm_\om(\vkappa), & -\Tbm_M^T \\ -\Tbm_M, & \zerobm
      \end{array}
    \right]
  \left[
    \begin{array}{c} \pbm \\ \rbm
      \end{array}
    \right] =
  \lam_\om
  \left[
    \begin{array}{ll}
     \Bbm_\om(\vkappa), & \zerobm \\
      \zerobm, & -\Ibm
      \end{array}
    \right]
  \left[
    \begin{array}{c} \pbm \\ \rbm
      \end{array}
    \right] \;, \\
  \mbox{ where } \Abm_\om(\vkappa) = 2\vkappa \Vbm_w + \imu (\Nbm_w - \Nbm_w^T)\;,\quad  \Bbm_\om(\vkappa) =  \Hbm - \vkappa^2 \Ubm + \imu{\vkappa}(\Qbm - \Qbm^T)\;.
\end{split}
\end{equation}
As in the case of \eq{eq-NSadv24}, the \rhs matrix is regular even for vanishing $\bar\wb$, whereas the \lhs matrix may be not.

\begin{remark}\label{rem-qep}
  All matrices defined in \eq{eq-NSadv21} are real, whereby $\Cbm,\Mbm,\Gbm,\Vbm$ and $\Wbm$are symmetric, thereby both the block matrices $\Abm$ and $\Bbm$ in the QEPs problems \eq{eq-NSadv24} and \eq{eq-NSadv25} are Hermitian. Therefore, their eigenvalues $\lam$ are real, or come in complex-conjugate pairs, see \eg \cite{Tisseur2001TheQE}. The definiteness of the diagonal blocks and, therefore, of the whole block matrices in both the QEP depend on the respective parameters, \ie on $\omega$ in \eq{eq-NSadv24}, and on $\vkappa$ in  \eq{eq-NSadv25}. Moreover, all matrices labelled by subscript $_w$ depend on the advection and matrices $\Rbm_w$, $\Vbm_w$ and $\Wbm_w$ may become nearly singular,  when $\bar\wb \cdot \nb\approx 0$.
  However, as we demonstrate in the numerical examples reported in Section~\ref{sec:examples}, up to rather high advection velocities $w^0 = |\wb^0|$, the pressure modes corresponding to predictions obtained by the homogenization approach are featured by real eigenvalues $\lam\in\RR$, as expected by virtue of the inviscid fluid properties.
\end{remark}


\subsection{Floquet-Bloch wave decomposition for viscous fluids}
In analogy with the Bloch wave analysis for the model of inviscid fluids in an infinite periodic porous structure, due to the wave decomposition ansatz \eq{eq-bloch-1}, we consider the reduced problem imposed in the periodic cell $\Zcal_f$.
We consider a steady flow represented by velocity field $\bar\wb$ in the scaffold periodic cell $\Zcal_f$. The weak formulation for $(\tilde\ub(x),\tilde p(x))$ given by the ansatz \eq{eq-bloch-1} is derived from \eq{eq-NS3} upon multiplying \eq{eq-NS3}$_1$ by $\tilde\vb$ and integrating over $\Zcal_f$; this yields
\begin{equation}\label{eq-vsc0}
\begin{split}
  \int_{\Zcal_f}\rho_0  \left(\dt{}{t}\tilde\ub + \bar\wb\cdot\nabla\tilde\ub + \tilde\ub\cdot\nabla\bar\wb\right)\cdot\tilde\vb & \\
+ \int_{\Zcal_f}\Dop\eeb{\tilde\ub}:\eeb{\tilde\vb} -  \int_{\Zcal_f}\tilde p \nabla\cdot\tilde\vb & =
\int_{\pd\Zcal_f}(\Dop(\eeb{\tilde\ub} - \tilde p \Ib):(\nubf^f\otimes \tilde\vb)\;,\\
\int_{\Zcal_f}\tilde q\left(\dt{}{t}\tilde p + \bar\wb\cdot\nabla \tilde p + k_f \nabla\cdot\tilde\ub\right) & = 0\;,
\end{split}
\end{equation}
to hold for all test functions $(\tilde\vb,\tilde q)$ in the form \eq{eq-bloch-1a} which describes waves running in the opposite direction \wrt the waves $(\tilde\ub,\tilde p)$.
Above, the \rhs boundary integral vanishes due to the $\Zcal$-periodicity and vanishing $\ub,\vb$ on the solid-fluid interface.
\begin{remark}\label{rem-symex}
Due  to $\nabla\cdot\bar\wb = 0$, the following integration by parts can be employed in \eq{eq-vsc0},
\begin{equation}\label{eq-vsc1a}
  \begin{split}
    \int_{\Zcal_f}q \bar\wb\cdot\nabla \tilde p & =
    \frac{1}{2}\int_{\Zcal_f}\left(q \bar\wb\cdot\nabla \tilde p - \tilde p \bar\wb\cdot\nabla q\right)\;,\\
    \int_{\Zcal_f} \bar\wb\cdot\nabla\tilde\ub\cdot\vb & = \frac{1}{2}\int_{\Zcal_f}\left(
    \bar\wb\cdot\nabla\tilde\ub\cdot\vb - \bar\wb\cdot\nabla\vb\cdot\tilde\ub
    \right)\;.
\end{split}
\end{equation}
\end{remark}

Now the wave ansatz \eq{eq-bloch-1} is substituted in \eq{eq-vsc0}, so that \eq{eq-vsc0} leads to the following dispersion eigenvalue problem:
Given $\vkappa \in \RR$, find a complex $\om  \in \CC$, 
such that
\begin{equation}\label{eq-vsc10}
\begin{split}
  \rho_0\int_{\Zcal_f}\left(\ub\cdot\nabla\bar\wb + \bar\wb\cdot\nabla\ub -\imu\vkappa(\nb\cdot\bar\wb)\ub\right)\cdot\vb - \int_{\Zcal_f}( \nabla\cdot\vb+\imu\vkappa \nb \cdot\vb) p & \\
  +   \int_{\Zcal_f}\left(\Dop \eeb{\ub}: \eeb{\vb} + \vkappa^2 \Dop\ggb{\ub}{\nb}:\ggb{\vb}{\nb}\right)& \\
  + \imu\vkappa \int_{\Zcal_f}\left(\Dop \eeb{\ub}:\ggb{\vb}{\nb} -  \Dop\ggb{\ub}{\nb} : \eeb{\vb}\right) & = - \imu\om \rho_0 \int_{\Zcal_f}\ub\cdot\vb\;,\quad \\
   \gamma\int_{\Zcal_f}(\bar\wb\cdot\nabla p - \imu\vkappa\nb\cdot\bar\wb p)q + \int_{\Zcal_f}q(\nabla\cdot\ub- \imu\vkappa\nb\cdot\ub) & = - \imu\om \gamma \int_{\Zcal_f} p q \;,
\end{split}
\end{equation}
for all $\vb \in \HpdbO(\Zcal_f)$ and for all $q \in H_\#^1(\Zcal_f)$.

\paragraph{FEM approximation and matrix notation.}



{In analogy with the inviscid case, the FEM model is obtained upon discretizing problem \eq{eq-vsc10}. The non-slip condition for the velocity is applied on $\pd_s\Zcal_f$ along with the periodicity conditions on $\pd_\#\Zcal_f$. In addition to column matrices $\pbm$ and $\qbm$ representing the pressure field, column matrices $\ubm$ and $\vbm$ involve all degrees of freedom (DOFs) arising form the discretization of $\ub$ and $\vb$, respectively, respecting the boundary conditions.}

The matrices obtained by the FE discretization of \eq{eq-vsc10} are defined below in \eq{eq-vsc6} and \eq{eq-vsc6a} --- some of the matrix symbols are reused with different meanings with respect those introduced above in Section~\ref{sec-FBivs}.


The first group of matrices arises from the discretized integrals which do not depend on the advection velocity,
\begin{equation}\label{eq-vsc6}
\begin{split}
  \int_{\Zcal_f}\Dop\eeb{\ub}:\eeb{\vb}    & \FEMapprox \vbm^T\Abm\ubm\;,\\ 
  \int_{\Zcal_f}\Dop\ggb{\ub}{\nb}:\ggb{\vb}{\nb}  & \FEMapprox\vbm^T\Sbm\ubm\;,\\ 
  \int_{\Zcal_f}\Dop\eeb{\ub}:\ggb{\vb}{\nb}  & \FEMapprox \vbm^T\Xbm\ubm\;,\\ 
  \int_{\Zcal_f}pq &\FEMapprox \qbm^T\Qbm\pbm\;,\\
  \int_{\Zcal_f}\ub\cdot\vb &\FEMapprox \vbm^T\Mbm\ubm\;,\\
  \int_{\Zcal_f}q \nabla\cdot\ub &\FEMapprox \qbm^T\Bbm\ubm\;,\\
  \int_{\Zcal_f}q \nb\cdot\ub &\FEMapprox \qbm^T\Nbm_n\ubm\;,\\
\end{split}
\end{equation}
whereas the second group of matrices depend on $\bar\wb$; the respective matrices vanish for the static fluid, when $\bar\wb\equiv 0$,
\begin{equation}\label{eq-vsc6a}
\begin{split}
  \int_{\Zcal_f}(\nb\cdot\bar\wb) pq &\FEMapprox \qbm^T\Qbm_w\pbm\;,\\
  \int_{\Zcal_f} (\nb\cdot\bar\wb) \ub\cdot\vb &\FEMapprox \vbm^T\Mbm_w\ubm\;,\\
  \int_{\Zcal_f}q \bar\wb\cdot\nabla p &\FEMapprox \frac{1}{2}\qbm^T(\Ybm_w - \Ybm_w^T)\pbm\;,\\  
  \int_{\Zcal_f} (\bar\wb\cdot\nabla \ub)\cdot\vb &\FEMapprox \frac{1}{2}\vbm^T(\Cbm_w - \Cbm_w^T)\ubm \;,\\ 
  \int_{\Zcal_f} (\ub\cdot\nabla \bar\wb) \cdot\vb &\FEMapprox \vbm^T\Gbm_w\ubm\;, 
\end{split}
\end{equation}
where matrices $\Ybm_w$ and $\Cbm_w$ were introduced by virtue of the expressions \eq{eq-vsc1a}.
Note that $\Gbm_w$ is non-symmetric.
Using this notation, the discretized form of \eq{eq-vsc10} can be written, as
\begin{equation}\label{eq-vsc11}
  \begin{split}
    \rho_0[ \imu\om \Mbm - \imu\vkappa \Mbm_w + \frac{1}{2}(\Cbm_w - \Cbm_w^T) + \Gbm_w]\ubm - (\Bbm^T + \imu\vkappa\Nbm_n^T)\pbm & \\
    + \left[\Abm  + \imu\vkappa (\Xbm - \Xbm^T) + \vkappa^2 \Sbm  \right] \ubm= \textbf{0}\;,\\
[\imu\om \Qbm + \frac{1}{2}(\Ybm_w - \Ybm_w^T) - \imu\vkappa \Qbm_w]\pbm + k_f(\Bbm - \imu\vkappa\Nbm_n)\cdot\ubm & = \textbf{0}\;.
\end{split}
\end{equation}

\begin{remark}
  \label{rem1}
  Formally one can consider the degenerate case of an inviscid fluid, $\mu = 0$, so that matrices $\Abm, \Sbm$ and $\Xbm$ vanish, however, the irrotational velocity constraint $\nabla\times\tilde\ub = 0$ should be imposed to suppress spurious modes, and also the free-slip, non-penetration conditions must be applied, \ie $\nubf^f\cdot\ub = 0$ on $\pd_s\Zcal_f$.
\end{remark}

Two alternative formulations of the dispersion problem arising from \eq{eq-vsc11} can be considered:
1) For a given real frequency $\om \in \RR$, compute a complex wave number $\vkappa \in \CC$, such that \eq{eq-vsc11} holds with nonvanishing eigenvectors, $\ubm,\pbm$; 2) For a given real wave number $\vkappa \in \RR$, compute a complex frequency $\om \in \CC$ and nonvanishing $\ubm,\pbm$;



In this paper we use the first formulation, although it leads to the quadratic eigenvalue problem (QEP) which requires to introduce an additional variable and to use a substitution enabling  to derive a standard generalized eigenvalue problem. The second alternative seems to be advantageous, since no such substitutions are needed, however, it appears to be quite cumbersome to distinguish the correct pressure modes among all the dispersion curves. Nevertheless, even this second formulation is outlined briefly in Section~\ref{sec-k2o}. 


\subsubsection{Formulation $\om \mapsto \vkappa$}\label{sec-o2k}
We consider the problem: For a given $\om \in \RR$, find $\vkappa \in \CC$, such that \eq{eq-vsc11} admits a non-trivial solution $(\ubm, \pbm)$. It is advisable to transformed this QEP into a problem involving $\vkappa$ in the linear form. To do so, the Choleski decomposition of $\Sbm = \Vbm^T\Vbm$ is applied, which yields the substitution $\zbm:= \imu\vkappa\Vbm\ubm$ and $\vkappa^2\Sbm\ubm = -\imu\vkappa\Vbm^T\zbm$.

The following block  matrices can be established, where $\gamma = 1/k_f$,
\begin{equation}\label{eq-vsc12b}
  \begin{split}
    \Pop_w := \left (
    \begin{array}{lll}
      \Abm + \rho_0[ \frac{1}{2}(\Cbm_w - \Cbm_w^T)+ \imu\om\Mbm + \Gbm_w]\;, & -\Bbm^T \;, &  \textbf{0}\\
      \Bbm\;,& \imu\om \gamma\Qbm + \frac{\gamma}{2}(\Ybm_w - \Ybm_w^T)\;, &  \textbf{0}\\
      \textbf{0}\;, &  \textbf{0}\;, & \Ibm
    \end{array}\right) \;,\\
    \Nop_w := \left (
    \begin{array}{lll}
      \rho_0 \Mbm_w + (\Xbm^T - \Xbm)\;, & \Nbm_n^T\;, &  \Vbm^T\\
      \Nbm_n\;,  & \gamma\Qbm_w \;, &\textbf{0} \\
      \Vbm\;, & \textbf{0}\;, &\textbf{0}
    \end{array}\right) \;,\quad\quad
    \vop = \left (
    \begin{array}{l}
      \ubm \\ \pbm \\ \zbm
    \end{array}\right)\;.
\end{split}
\end{equation}
Now \eq{eq-vsc11} can be rewritten,
\begin{equation}\label{eq-vsc13b}
  \begin{split}
    \frac{1}{\imu\vkappa}\Pop_w\vop = \Nop_w \vop\;,
\end{split}
\end{equation}
where   $\Pop_w$ is regular. Unfortunately, because of the combination of the viscosity and advection effects associated with velocity $\bar\wb$, no special properties can be identified concerning matrices $\Pop_w$ and $\Nop_w$. Nevertheless, this formulation of the generalized eigenvalue problem has been implemented to analyze the dispersion $\omega \mapsto \vkappa$, as reported in Section~\ref{sec:examples-viscous}. The eigenmodes of interest are those with the smallest imaginary part $\vkappa_I$ (for a given real $\omega$).


\begin{remark}
\label{rem3}
  \textbf{(Balancing the equations)}
Since $\rho_0/\gamma \approx 10^{10}$, the system should be balanced to improve conditioning of matrix $\Nop$. For this we set,
\begin{equation}\label{eq-vsc14}
\begin{split}
  \tilde\ubm:=\rho_0^{1/2} \ubm\;,\quad \tilde\pbm:=\gamma^{1/2} \pbm\;,\\
  \tilde\Bbm:=(\rho_0\gamma)^{-1/2} \Bbm\;,\quad \tilde\Nbm_n:=(\rho_0\gamma)^{-1/2} \Nbm_n\;,\\
  \tilde\Abm:=\Abm/\rho_0\;,\quad\tilde\Sbm:=\Sbm/\rho_0\;,\quad \tilde\Xbm:=\Xbm/\rho_0\;.
\end{split}
\end{equation}
Using these substitutions, matrices \eq{eq-vsc12b} are replaced by
\begin{equation}\label{eq-vsc14c}
  \begin{split}
    \tilde\Pop := \left (
    \begin{array}{lll}
      \tilde\Abm +  \imu\om\Mbm + \frac{1}{2}(\Cbm_w - \Cbm_w^T) + \Gbm_w\;, & -\tilde\Bbm^T \;, &  \textbf{0}\\
      \tilde\Bbm\;,& \imu\om \Qbm + \frac{1}{2}(\Ybm_w - \Ybm_w^T)\;, &  \textbf{0}\\
      \textbf{0}\;, &  \textbf{0}\;, & \Ibm
    \end{array}\right) \;,\\
    \tilde\Nop := \left (
    \begin{array}{lll}
      \Mbm_w + \tilde\Xbm^T - \tilde\Xbm\;, & \tilde\Nbm_n^T\;, &  \tilde\Vbm^T\\
      \tilde\Nbm_n\;,  & \Qbm_w \;, &\textbf{0} \\
      \tilde\Vbm\;, & \textbf{0}\;, &\textbf{0}
    \end{array}\right) \;,\quad\quad
    \tilde\vop = \left (
    \begin{array}{l}
      \tilde\ubm \\ \tilde\pbm \\ \tilde\zbm
    \end{array}\right)\;,
\end{split}
\end{equation}
where $\tilde\Sbm = \tilde\Vbm^T\tilde\Vbm$, hence $\tilde\Vbm:=\Vbm \rho_0^{-1/2}$. 

\end{remark}

\subsubsection{Formulation $\vkappa \mapsto \om$}\label{sec-k2o}
Alternatively, one may consider a given real wave~number $\vkappa \in \RR$ and find an $\om \in \CC$, such that \eq{eq-vsc11} admits a nontrivial solution $(\om,(\ubm,\pbm))$. In this case, \eq{eq-vsc11} presents a linear eigenvalue problem for the complex wave frequency $\om$, whose
imaginary part signifies the wave attenuation.
The following matrices can be established, recalling $\gamma = 1/k_f$,
\begin{equation}\label{eq-vsc12a}
  \begin{split}
    \Rop_w := \left (
    \begin{array}{ll}
      \Abm + \vkappa^2 \Sbm  +  \imu\vkappa (\Xbm - \Xbm^T)\;,&
      \textbf{0}  \\
       \textbf{0}\;, &  \textbf{0}
    \end{array}\right)\;,\\
    \Kop_w := \left (
    \begin{array}{ll}
      \rho_0[ \frac{1}{2}(\Cbm_w - \Cbm_w^T)- \imu\vkappa \Mbm_w + \Gbm_w]\;, & -(\Bbm^T+\imu\vkappa\Nbm_n^T) \\
      \Bbm -\imu\vkappa\Nbm_n & -\imu\vkappa \gamma\Qbm_w + \frac{\gamma}{2}(\Ybm_w - \Ybm_w^T)
    \end{array}\right) \;,\\
    \Mop := \left (
    \begin{array}{ll}
      \rho_0 \Mbm\;, &  \textbf{0}\\
      \textbf{0} & \gamma \Qbm
    \end{array}\right) \;,\quad\quad
    \vop = \left (
    \begin{array}{l}
      \ubm \\ \pbm
    \end{array}\right)\;.
\end{split}
\end{equation}
Now \eq{eq-vsc11} can be rewritten,
\begin{equation}\label{eq-vsc13a}
  \begin{split}
    [\Kop_w + \Rop_w]\vop = - \imu\om \Mop \vop\;,
\end{split}
\end{equation}
where  $\Mop$ is real symmetric positive definite and $\Rop_w$ being associated with the fluid viscosity, is Hermitean, but singular.
Matrix $\imu\Kop_w$ is general, but becomes Hermitean for the case of a static fluid, \ie when $\bar\wb = 0$ so that the nonsymmetric matrix $\Gbm_w$ becomes zero.

\begin{remark}
  \label{rem-iv-k2o}
As the consequence of the above statements, for static and inviscid fluids, \eq{eq-vsc13a} yields real frequencies, thus, $\RR\ni\vkappa\mapsto\om \in \RR$.
In contrast,  when $\bar\wb \not \equiv 0$, even for inviscid fluids when $\Rop_w$ vanishes, \eq{eq-vsc13a} yields complex frequencies in general, as discussed in Remark~\ref{rem-qep}. However, in such a case, by virtue of Remark~\ref{rem1}, free slip boundary conditions should be respected (thus, the displacement-associated matrices modified) and the irrotationality constraint related to the velocity field should be involved to reduce spurious oscillation modes.
\end{remark}

\def\reg{\textsuperscript{\textregistered}}
\def\nb{\bm{n}}
\def\veps{\varepsilon}
\def\bbw{\bar{\bm{w}}}
\def\bbwM{\bar{\bm{w}}^0 }
\def\mbw{\bar{w}^0}


\section{Numerical examples}
\label{sec:examples}

Properties and mutual correspondence of the models introduced in the preceding
sections we be illustrated using examples of 2D and 3D geometries representing
the periodic scaffolds saturated by viscous, or inviscid fluids. For this we
compute dispersion curves and phase velocities characterizing the wave
propagation in such media. Besides comparison of the responses provided by the
two modelling approaches, \ie the periodic homogenization (PH) and the
Floquet-Bloch wave decomposition (FB), we aim to explore the following
phenomena related to the wave propagation:
\begin{itemize}
\item the flow advection -- the influence of the permanent fluid flow given by
  the macroscopic velocity $\bbwM$ (its size $\mbw = |\bbwM|$ and
  orientation \wrt the wave direction $\nb$;
\item the scale effects -- the influence of the characteristic size of the
  microstructure, as described by the scale parameter $\veps^0$.
\end{itemize}

Numerical solutions of the characteristic problems associated with the
homogenized models, and of the eigenvalue problems, arising from the FB
analysis, were computed using the finite element method implemented in the
software SfePy \cite{sfepy2019}. In all the examples reported below, meshes
with a sufficient resolution have been used, such that any further uniform
refinement did not change the wave dispersion results for long wave lengths
near the origin of the Brillouin zone. In all figures, results of the numerical
experiments are displayed in the basic SI units, unless specified otherwise.

\subsection{Inviscid fluid}
\label{sec:examples-inviscid}

\begin{figure}[htp!]
  \centering
  \begin{tabular}{ccc}
    & $d=1.5 \cdot 10^{-4}$~m & $d=2.3 \cdot 10^{-4}$~m \\
      \includegraphics[width=0.3\linewidth]{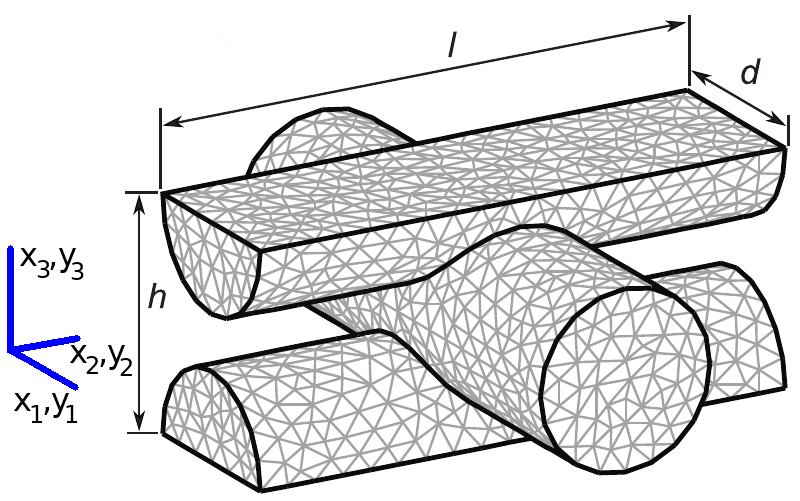}
    &
      \includegraphics[width=0.3\linewidth]{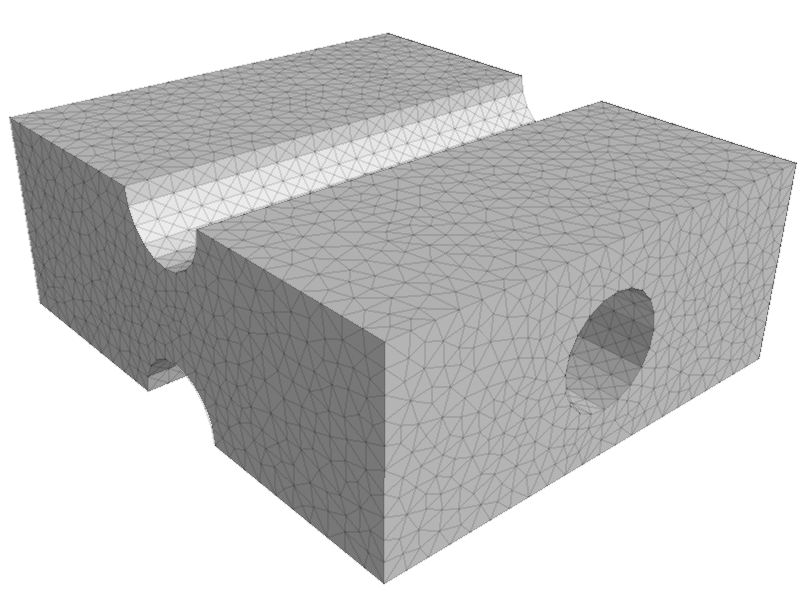}
    &
      \includegraphics[width=0.3\linewidth]{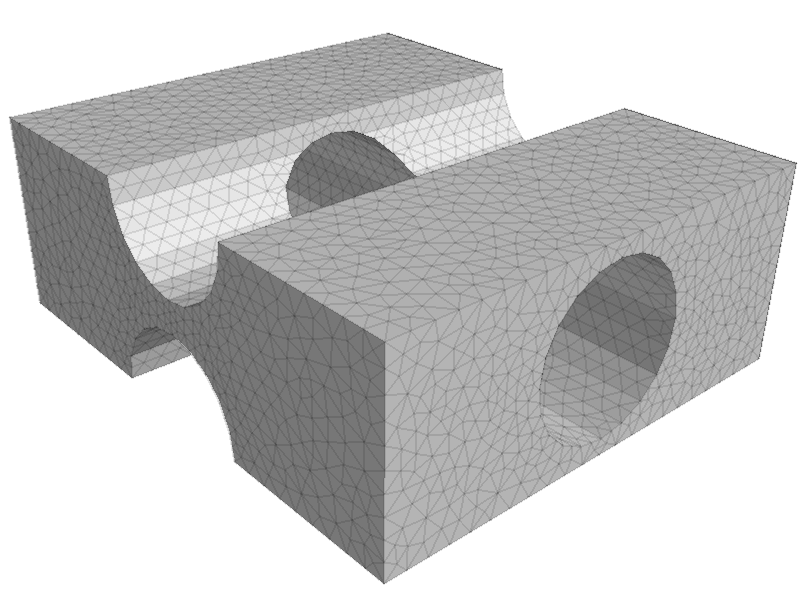}\\
      solid fibres $\Zcal_s$ & fluid in $\Zcal_f$ & fluid in $\Zcal_f$
  \end{tabular}
  \caption{The periodic cell of solid fibres (left, from \cite{kruisova2018})
    and corresponding finite element meshes ($l = 6.5 \cdot 10^{-4}$~m,
    $h = 2.75 \cdot 10^{-4}$~m) with two radii of the fibres for inviscid flow
    experiments.}
  \label{fig:mesh-3d-inviscid}
\end{figure}

The inviscid fluid parameters were those of water
$\gamma = 5 \cdot 10^{-10}$~Pa$^{-1}$, $\rho_0 = 1000$~kg/m$^3$. We considered
3D robocast structures, \cite{kruisova2018}, produced by sinterization of
ceramic fibres. The representative cell $\Zcal$ is illustrated in
Fig.~\ref{fig:mesh-3d-inviscid} (left). This domain generates a lattice of
periodic 3D scaffolds. The largest edge of the cell $\Zcal$
($7.5 \cdot 10^{-4}$~m in the $x_1$ direction) limits the shortest wavelengths
for the FB analysis, hence, in the reciprocal lattice, the corresponding 1st
irreducible Brillouin zone spans $8378$~m$^{-1}$. The P1 finite elements on
tetrahedrons were used for the discretization of both the FB and PH analyses.

\subsubsection{Static fluid, $w^0 = 0$}
\label{sec:example-inviscid-now}

Below we study the influence of the solid fibres diameters on the dispersion
properties using \eq{eq-NSadv22} for the Bloch wave analysis (the frequencies
$\omega$ were computed for given wave numbers $\vkappa$) and
(\ref{eq-NSadv20b}) for the homogenization-based analysis, simplified for the
case of zero convective velocity, $\bar\wb \equiv \zerobm$. Specifically,
(\ref{eq-NSadv22}) then presents a generalized eigenvalue problem for
$\lam:=\om^2$, which satisfies
\begin{equation}\label{eq-NSadv22a}
  \Abm_0\pbm = \om^2 \Mbm\pbm\;\quad \mbox{ with }\Abm_0 := c_f^2\left[\Cbm + \vkappa^2 \Mbm - \imu \vkappa (\Ybm - \Ybm^T) \right]\;.
\end{equation}
Due to the positive definiteness of both $\Abm_0$ and $\Mbm$, real positive
$\om^2$ can be computed for given $\vkappa \in \RR$. Examples of the finite
element (FE) meshes of the 3D domains are shown for two different fibre
diameters in Fig.~\ref{fig:mesh-3d-inviscid} (middle, right). The eigenvalues
were computed using the ARPACK solver (implicitly restarted Lanczos method
\cite{arpack1998}) through the SciPy package (function \texttt{eigsh()}
\cite{scipy2019}). The shift-invert mode was used to accelerate the calculation
of the smallest eigenvalues.

\begin{figure}[htp!]
  \centering
  \begin{tabular}{cc}
    \multicolumn{2}{c}{$d = 0.1 \cdot 10^{-4}$~m}
    \\
      \includegraphics[width=0.5\linewidth]{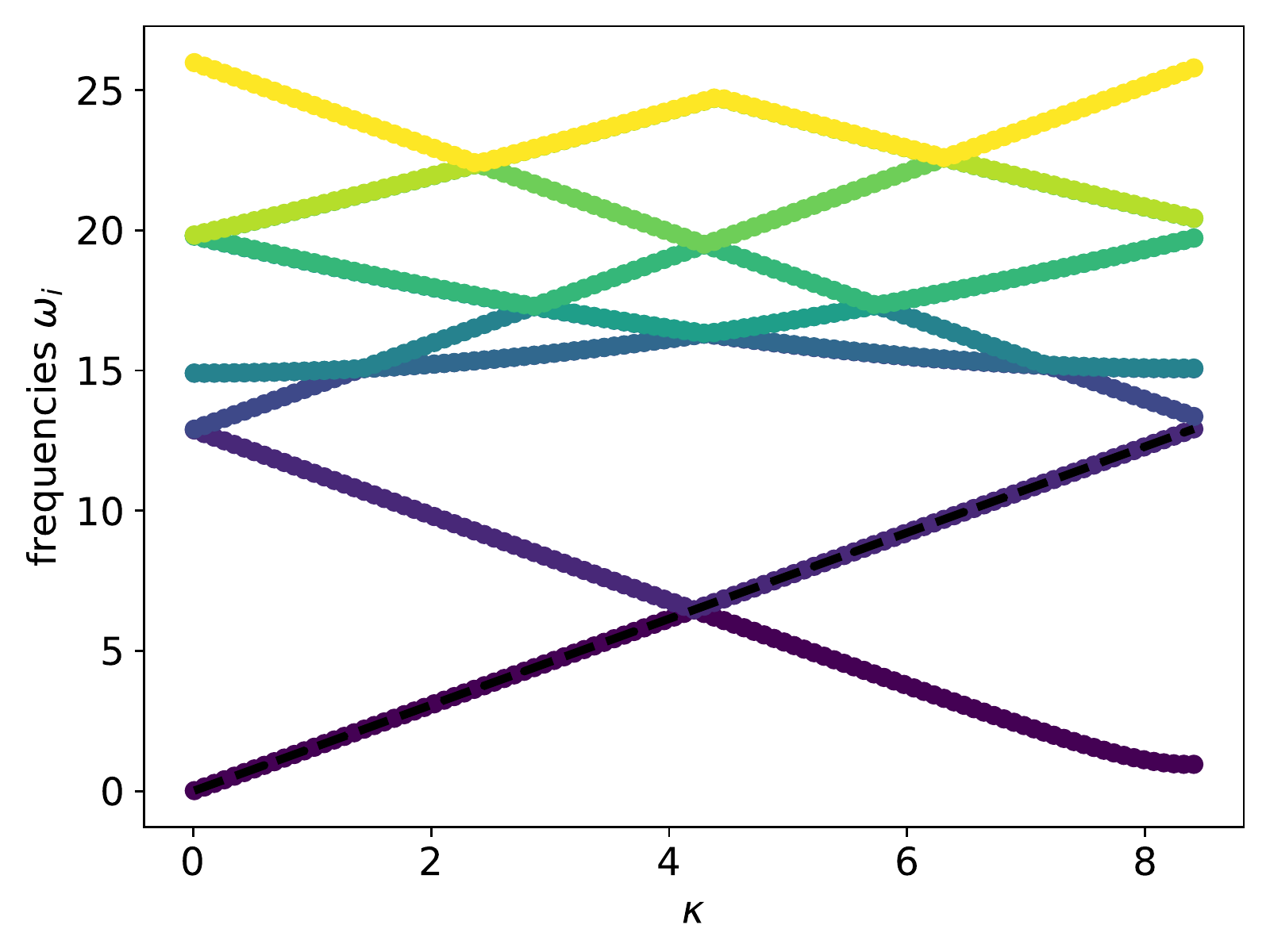}
    &
      \includegraphics[width=0.5\linewidth]{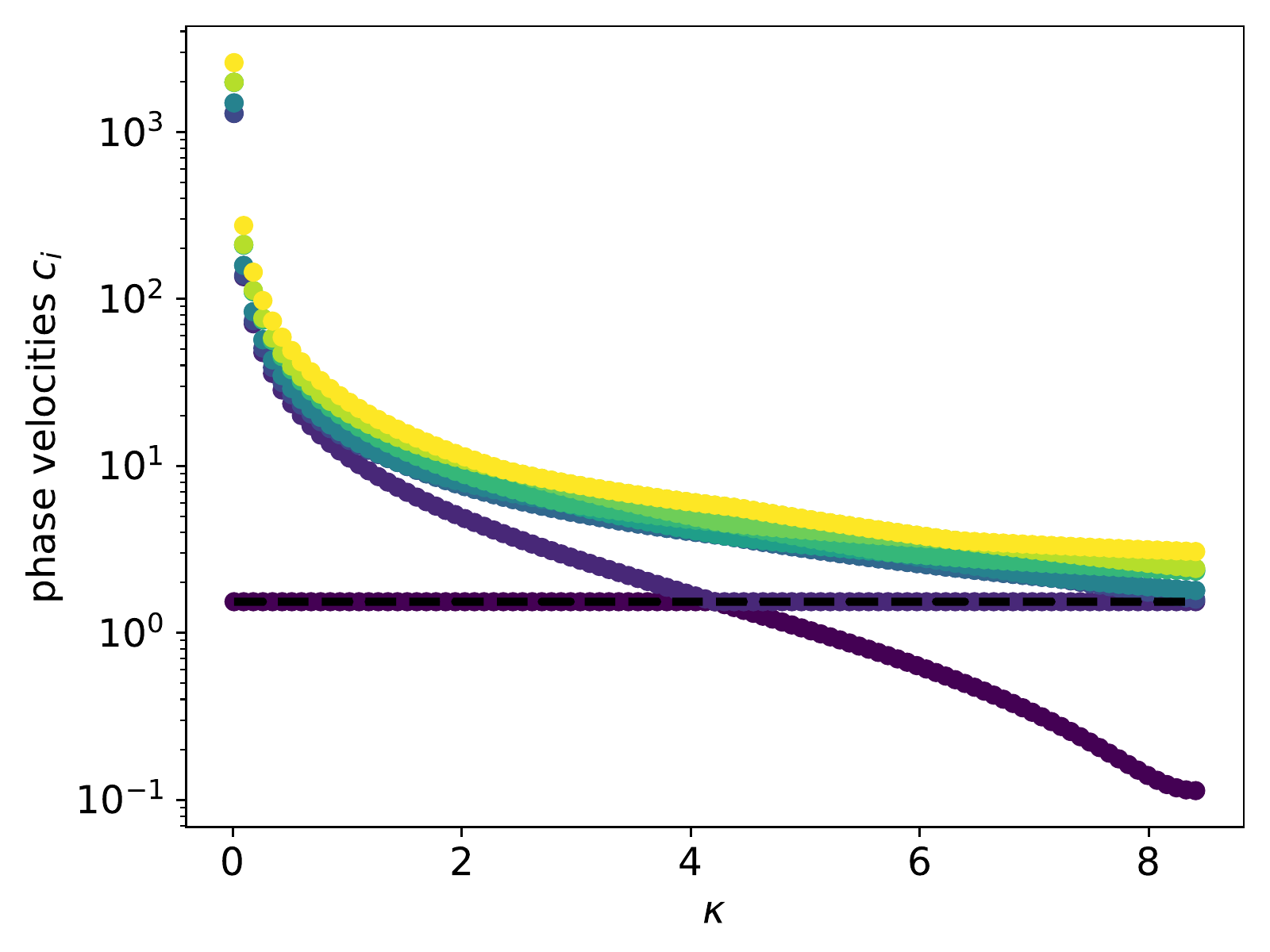}
    \\
    \multicolumn{2}{c}{$d = 1.5 \cdot 10^{-4}$~m}
    \\
      \includegraphics[width=0.5\linewidth]{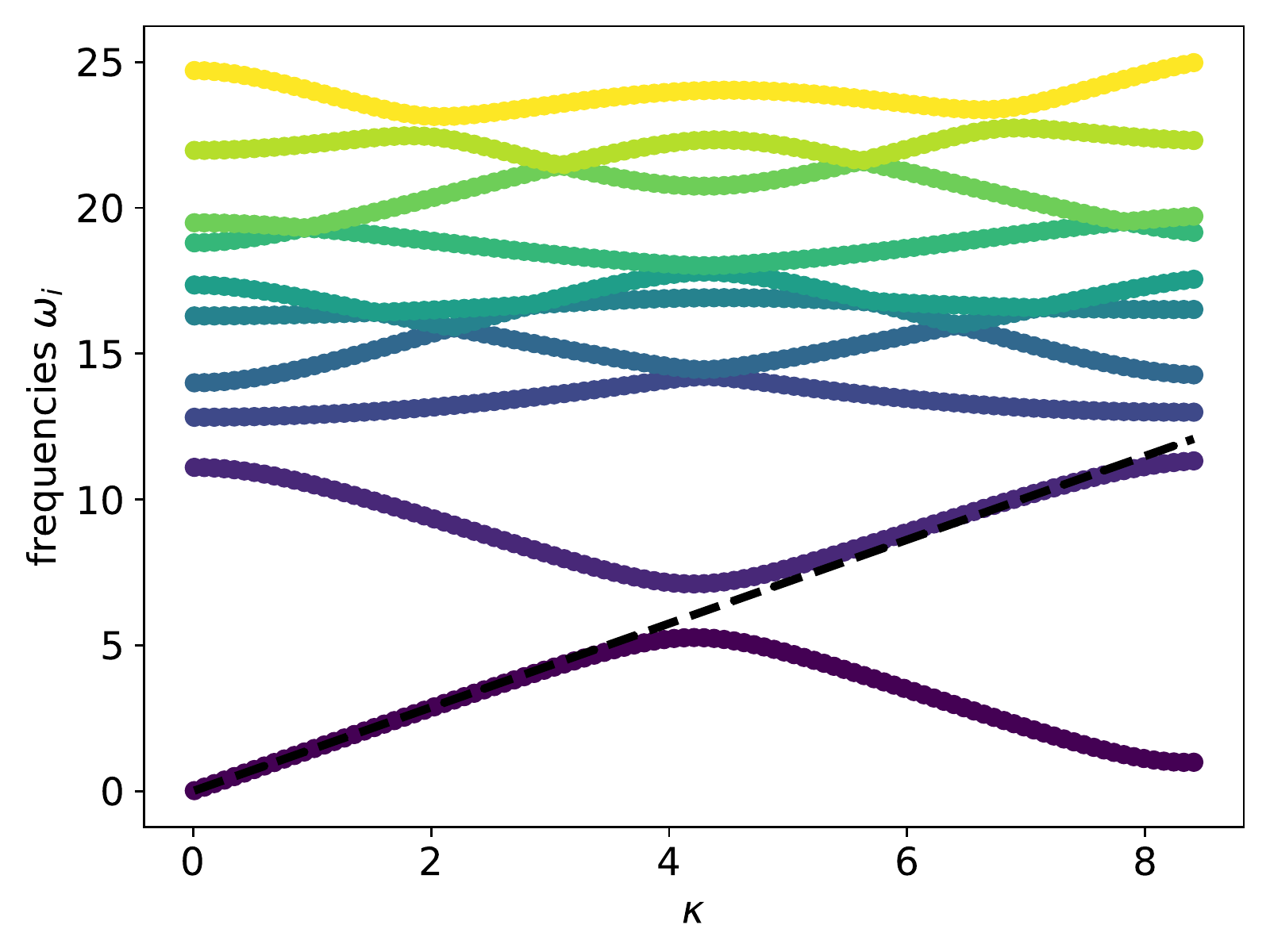}
    &
      \includegraphics[width=0.5\linewidth]{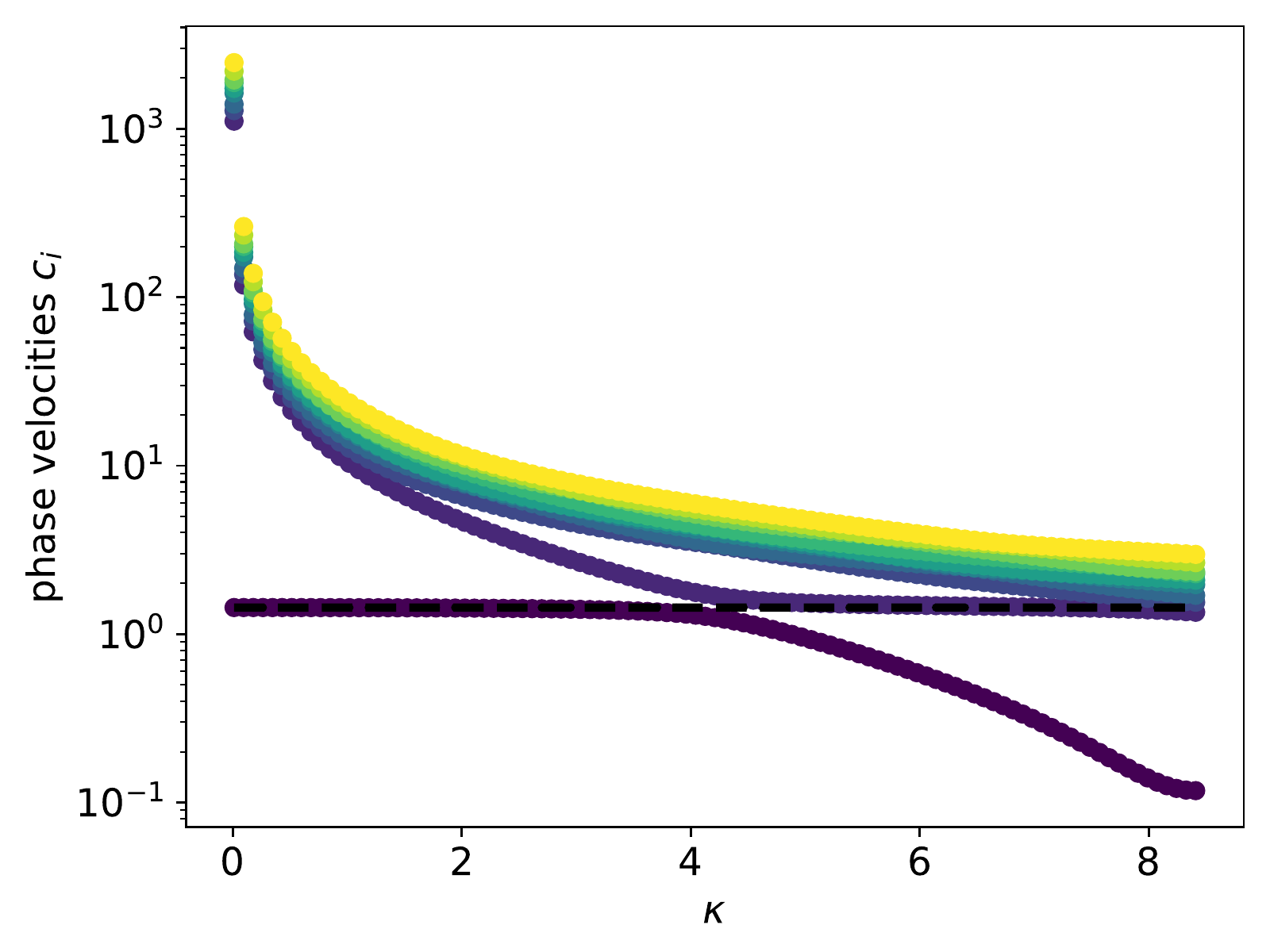}
    \\
    \multicolumn{2}{c}{$d = 2.3 \cdot 10^{-4}$~m}
    \\
      \includegraphics[width=0.5\linewidth]{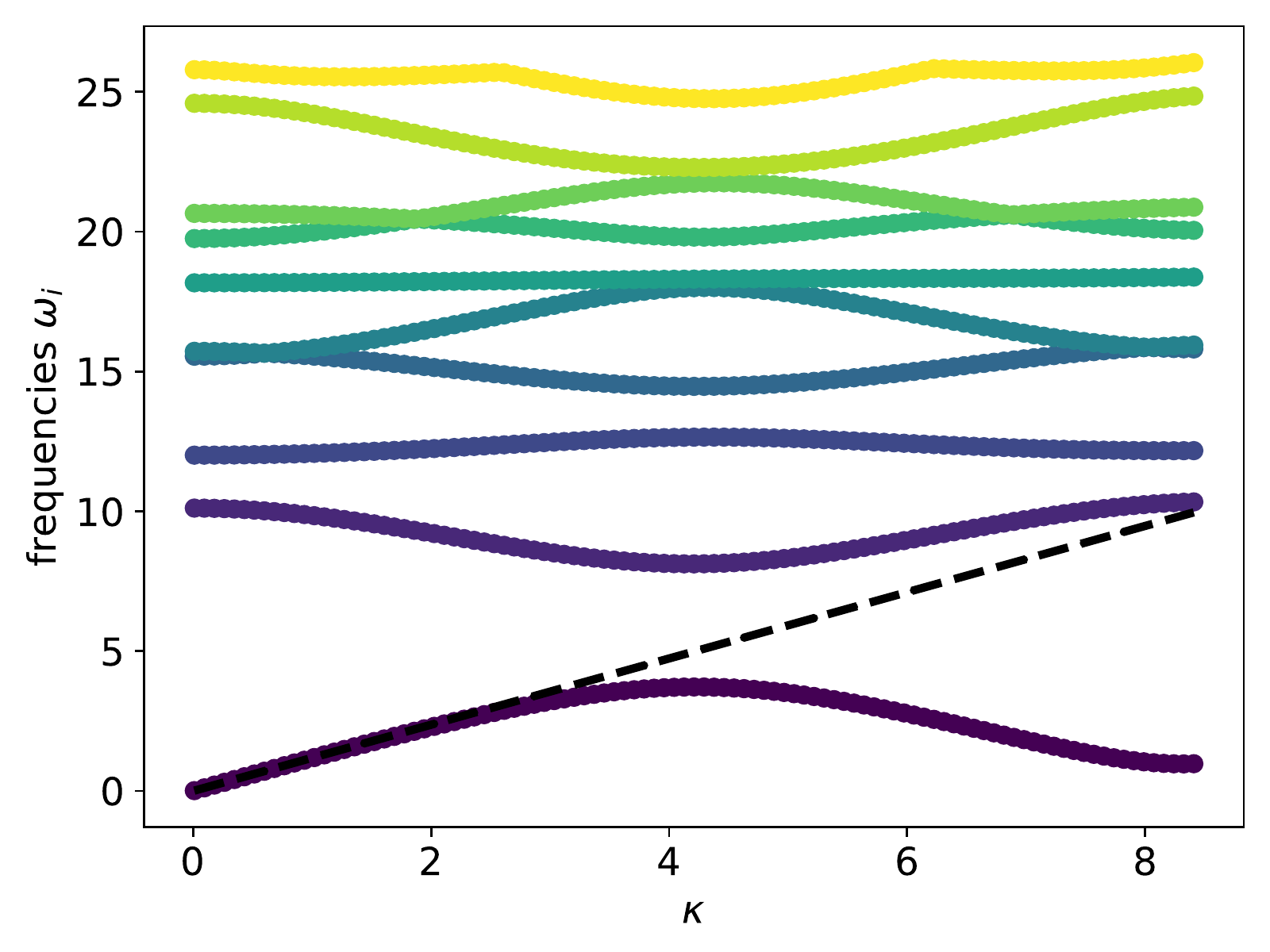}
    &
      \includegraphics[width=0.5\linewidth]{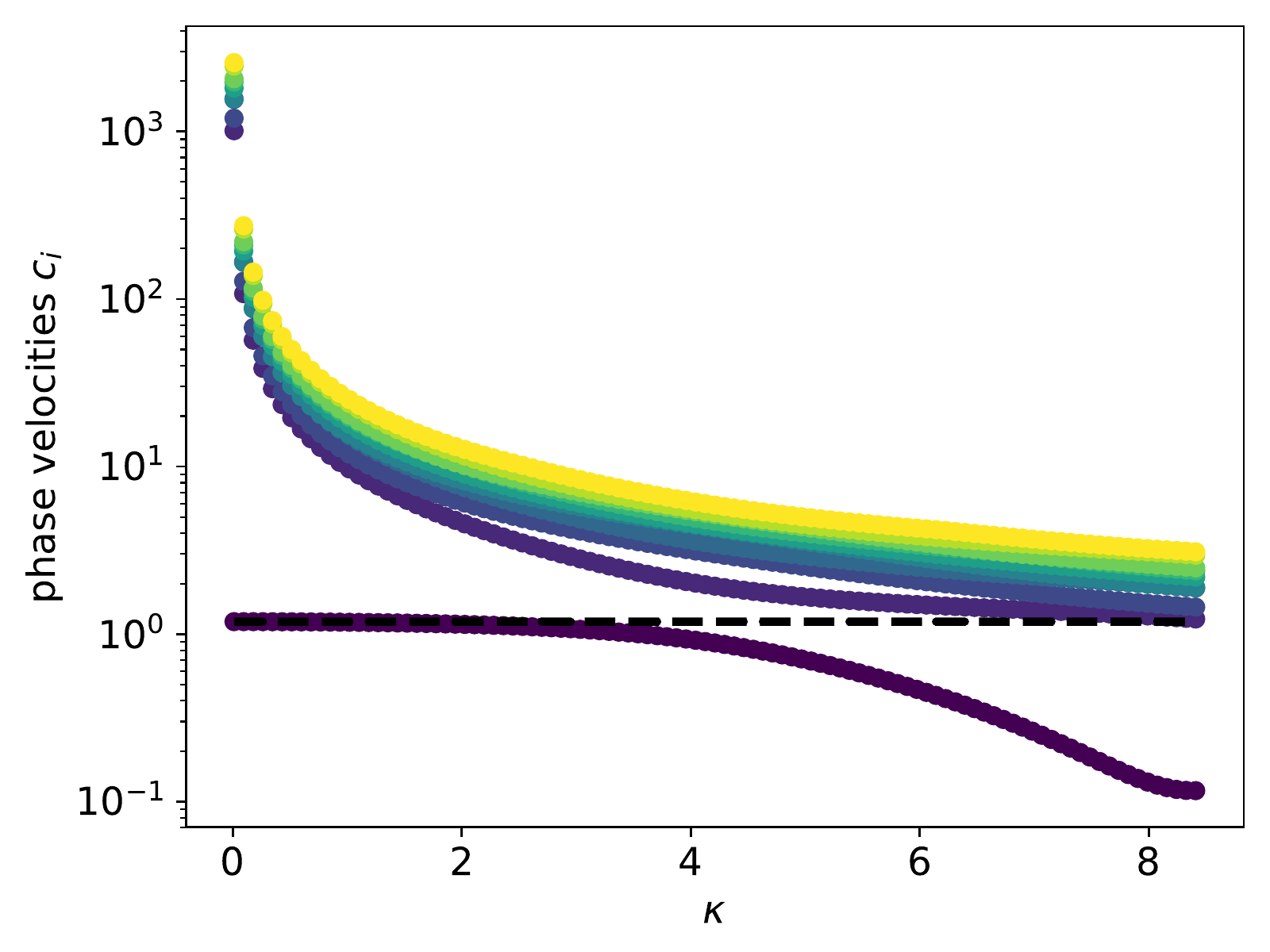}
  \end{tabular}
  \caption{The influence of the solid fibres diameters $d$ on the dispersion
    properties. As the radius of the fibres increases, a band gap appears and
    increases. The Bloch-based (dots) and homogenization-based
    (lines) dispersion results are shown using the dispersion curves (left),
    and the phase velocity curves (right). Units: $\vkappa$: 1/mm, $\om$: MHz.}
  \label{fig:bg-3d-inviscid}
\end{figure}

We considered incident waves in the direction $\nb = (1, 0, 0)^T$ and examined
influence of the porosity. While increasing fibres radii, thus reducing the
fluid volume fraction, a partial band gap appears and increases, as illustrated
in Fig.~\ref{fig:bg-3d-inviscid} in terms of the dispersion curves
($\om(\vkappa)$ dependence) and phase velocity plots. The dashed black lines
correspond to the homogenization-based asymptotes, see \eq{eq-NSadv20b}, while
the thick color lines depict results of the FB analysis, obtained by solving
\eq{eq-NSadv22a}.

\begin{figure}[htp!]
  \centering
  \begin{tabular}{cc}
    \includegraphics[width=0.5\linewidth]{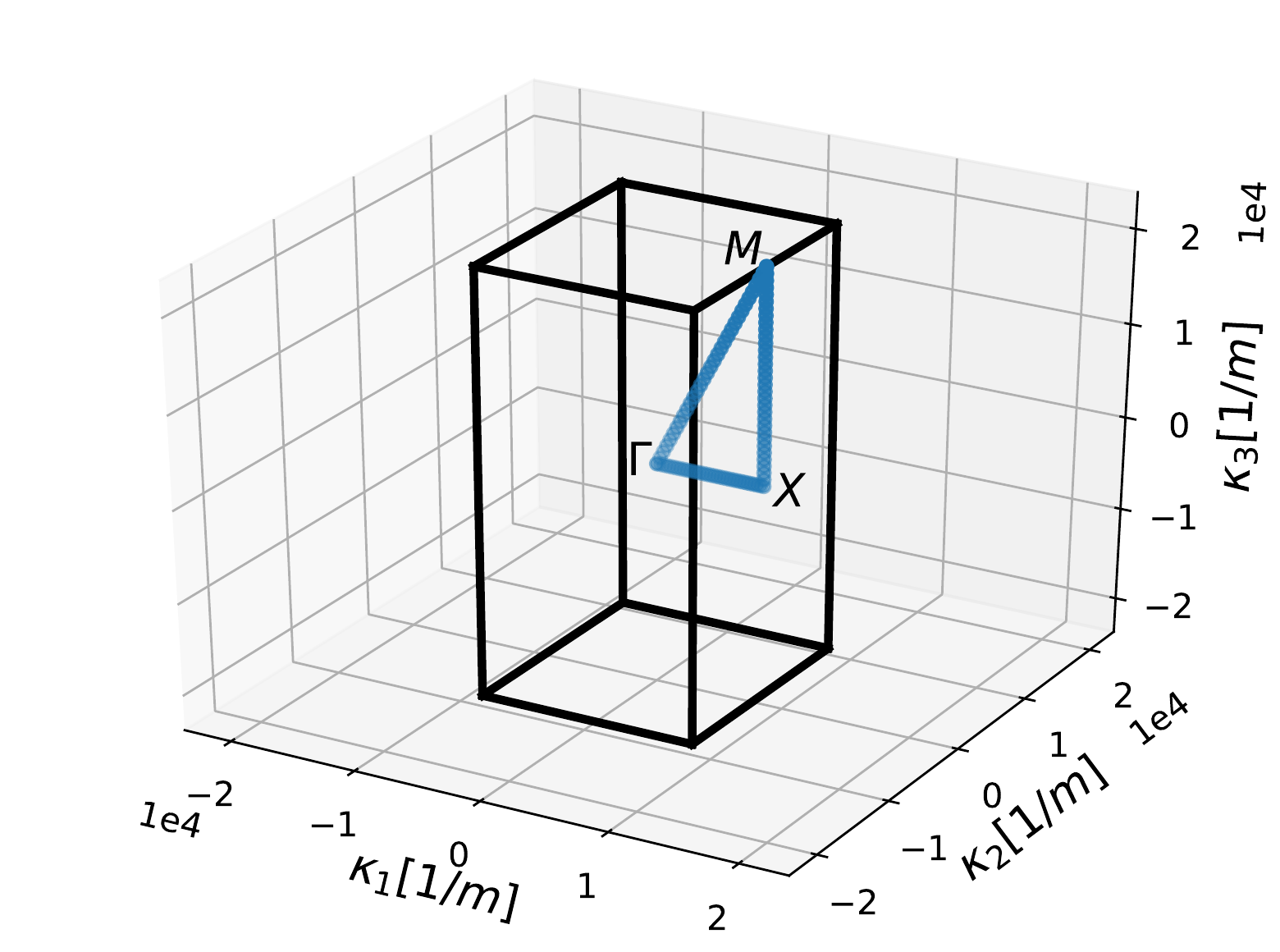}
    &
    \includegraphics[width=0.5\linewidth]{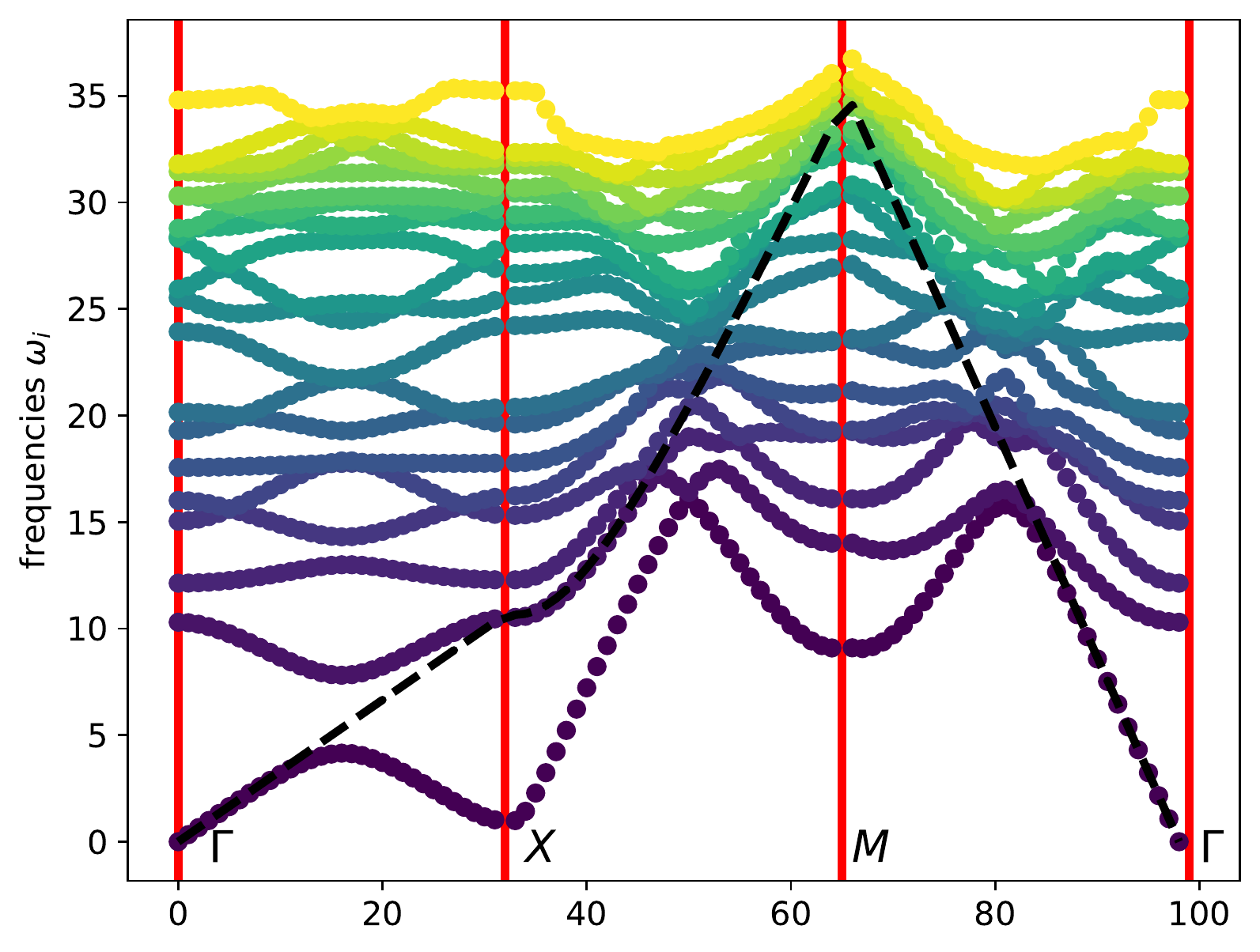}
  \end{tabular}
  \caption{Left: a path of the wave vectors $\vkappa \nb$ on the Bragg planes,
    bounding the Brillouin zone: $\vkappa_2 = 0$ and $\vkappa_1$, $\vkappa_3$
    vary. Right: the dispersion curves computed by the FB analysis (dotted
    color lines), and the corresponding homogenization-based predictions (the
    black dashed lines). Units: [$\vkappa$] = 1/mm, [$\om$] = MHz.}
  \label{fig:brillouin-3d-inviscid}
\end{figure}

Further, using the mesh in Fig.~\ref{fig:flow-3d-inviscid} (left), fixing the
fiber diameter $d=2.1 \cdot 10^{-4}$~m, the wave vector $\kappabf = \vkappa \nb$
was modified on a path in the reciprocal lattice; we traced a part of the
boundary of the Brillouin zone, while fixing $\vkappa_2 = 0$ and varying
$\vkappa_1$, $\vkappa_3$, see Fig.~\ref{fig:brillouin-3d-inviscid} (left). The
dispersion curves (the color lines) resulting from the FB analysis are
displayed in Fig.~\ref{fig:brillouin-3d-inviscid} (right), where the dashed
line represents the homogenization-based asymptotes.

\subsubsection{Nonstationary inviscid fluid -- effects of the advection}
\label{sec:example-inviscid-w}

We report the influence of the advection magnitude $\mbw$ (the macroscopic flow
velocity) and the incident wave direction $\nb$ on the dispersion properties.
The dispersion analysis was performed for $\mbw \in (0, 300)$ using
(\ref{eq-NSadv25}) for the FB analysis (the frequencies $\omega$ were computed
for given wave numbers $\vkappa$) and (\ref{eq-NSadv20b}) for the
homogenization-based analysis. The eigenvalues were computed using the
MATLAB\reg{} function \textrm{eigs()} \cite{stewart2002, arpack1998} via the
MATLAB Engine API for Python.

\begin{figure}[htp!]
  \centering
  \begin{tabular}{cc}
    $d = 2.1 \cdot 10^{-4}$~m & \\
      \includegraphics[width=0.4\linewidth]{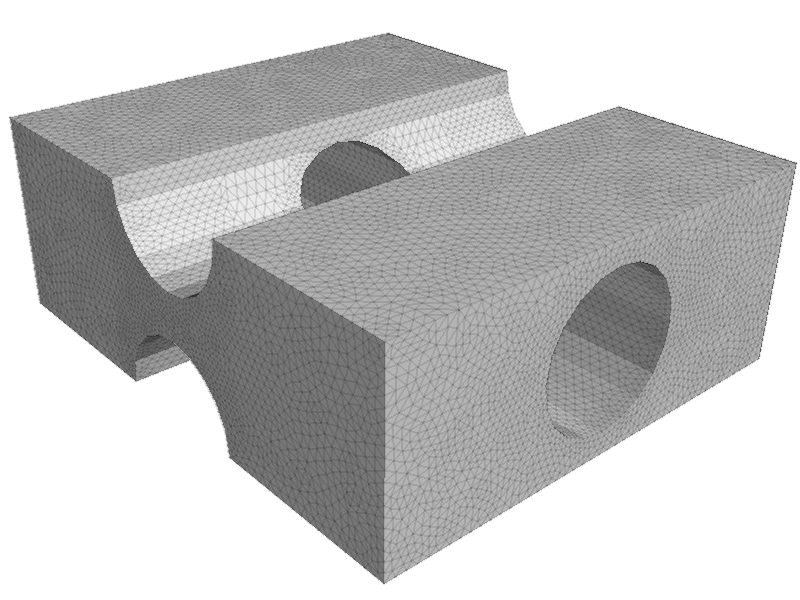}
    &
      \includegraphics[width=0.5\linewidth]{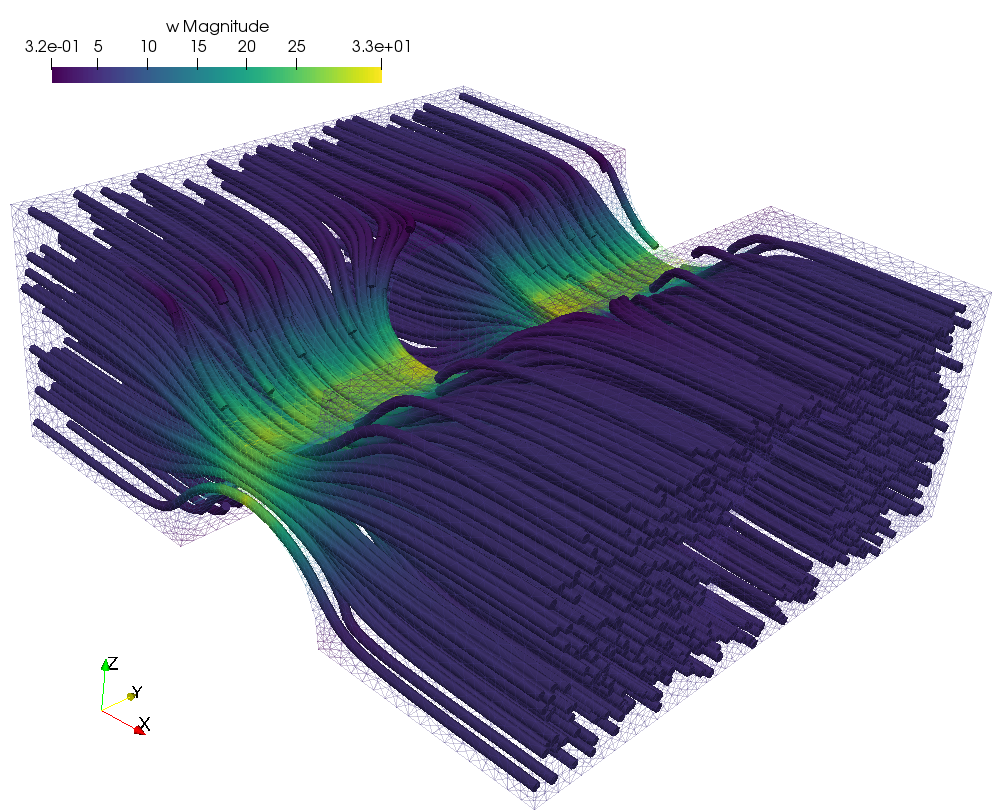}
  \end{tabular}
  \caption{The finite element mesh (left) and inviscid flow streamlines
    (right).}
  \label{fig:flow-3d-inviscid}
\end{figure}

The macroscopic velocity $\bbwM = (\mbw,0,0)$ had the overall direction along
the $x_1$-axis and was specified using its magnitude $\mbw$. The local velocity
field $\bar\wb$ was established by virtue of the homogenization result, see
\Appx{appx-ivs-psi}. Field $\bar\wb$ is illustrated in
Fig.~\ref{fig:flow-3d-inviscid}, in terms of the streamlines distributed in the
3D computational domain.

Two cases studies were performed for two different directions $\nb$ of the
incident waves:
\begin{itemize}
\item Study I-A: $\nb = (0, 1, 0)^T$,
  i.e. perpendicular to the macroscopic flow, thus $\nb\perp\bbwM$.
\item Study I-B: $\nb = (1, 0, 0)^T$,
  i.e. in the direction of the macroscopic flow, thus $\nb\parallel\bbwM$.
\end{itemize}

\begin{figure}[htp!]
  \centering
  \begin{tabular}{cc}
    \multicolumn{2}{c}{$\mbw = 1$} \\
    \includegraphics[width=0.5\linewidth]{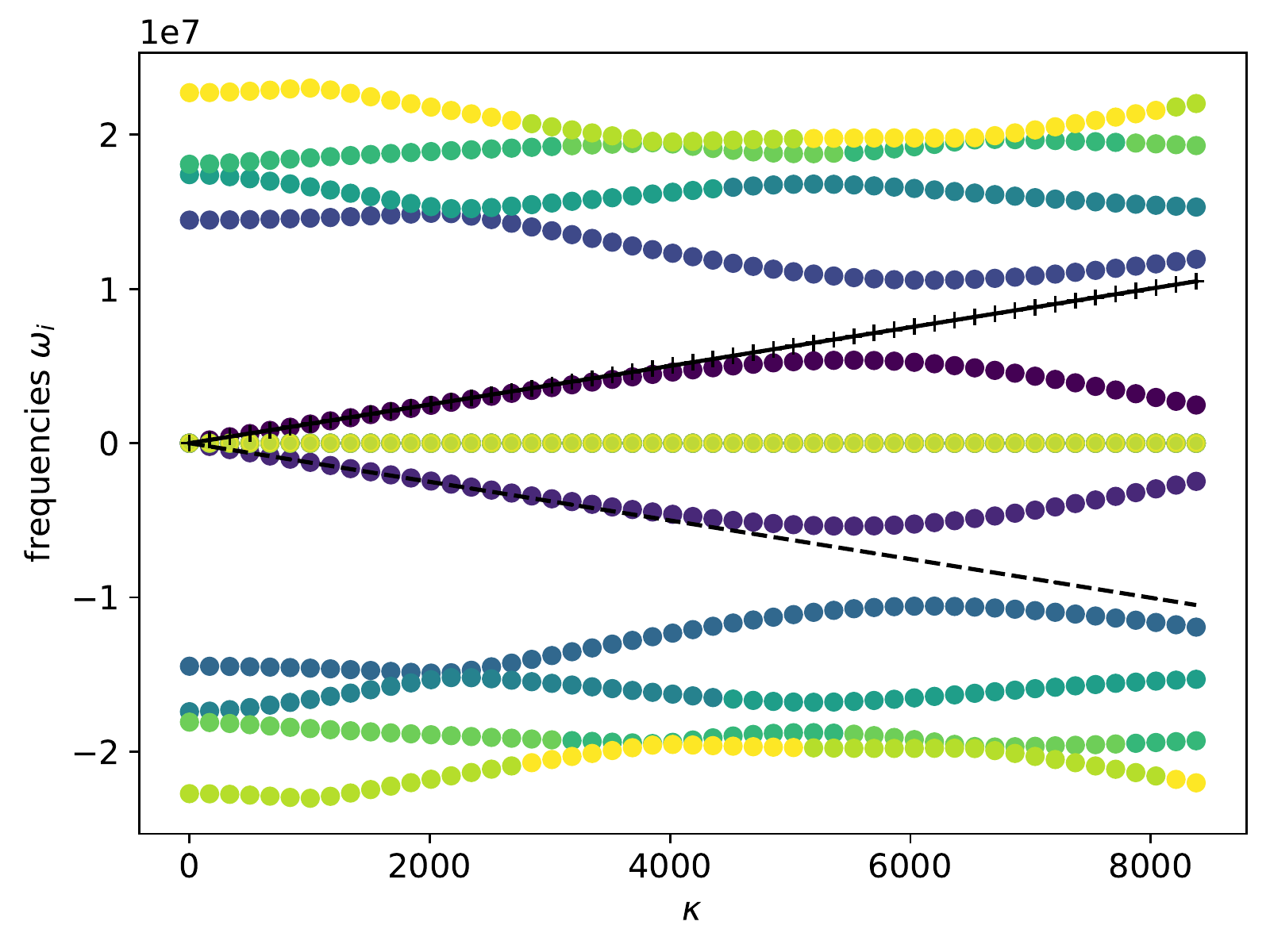} &
    \includegraphics[width=0.5\linewidth]{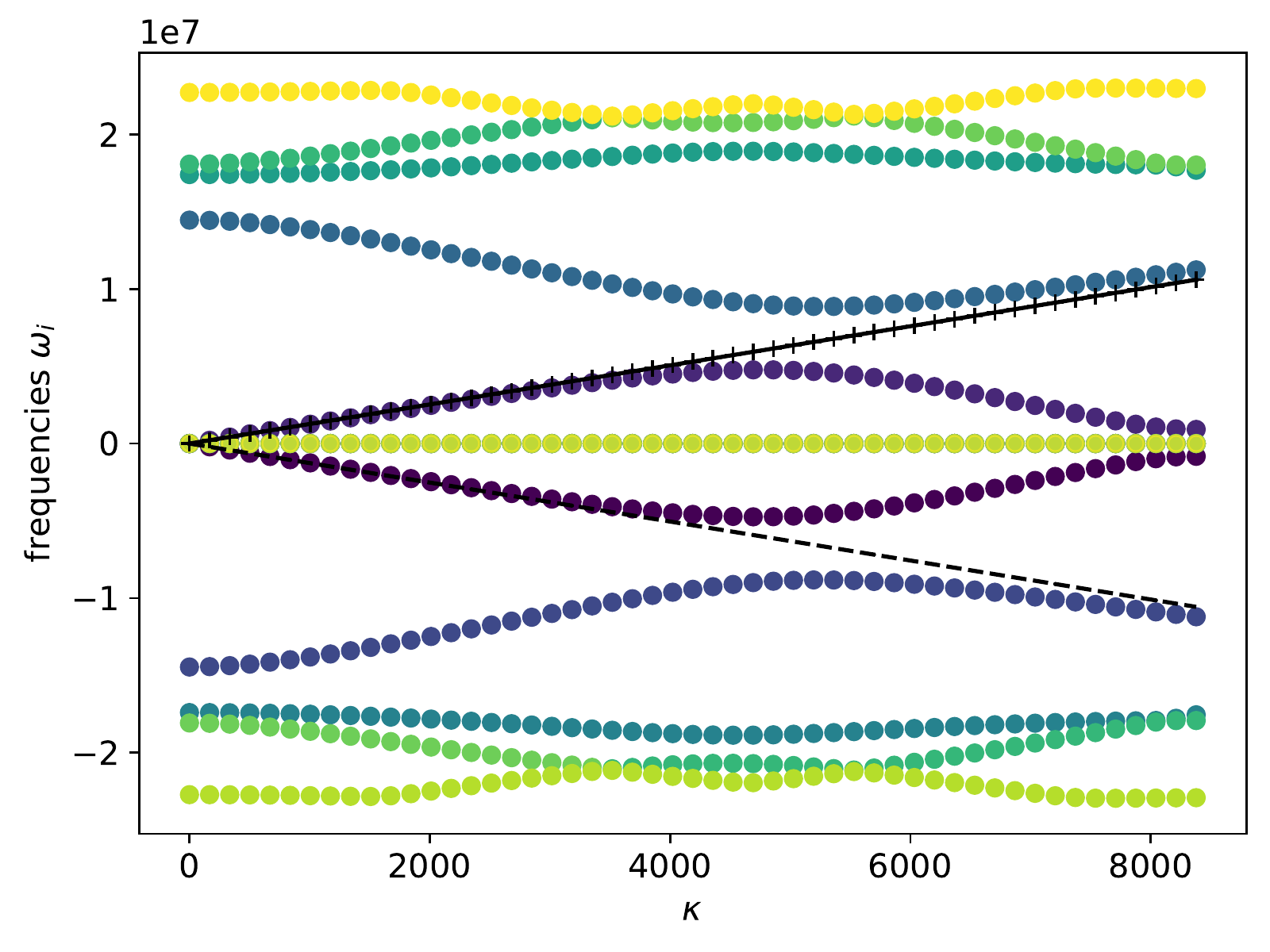} \\
    \multicolumn{2}{c}{$\mbw = 100$} \\
    \includegraphics[width=0.5\linewidth]{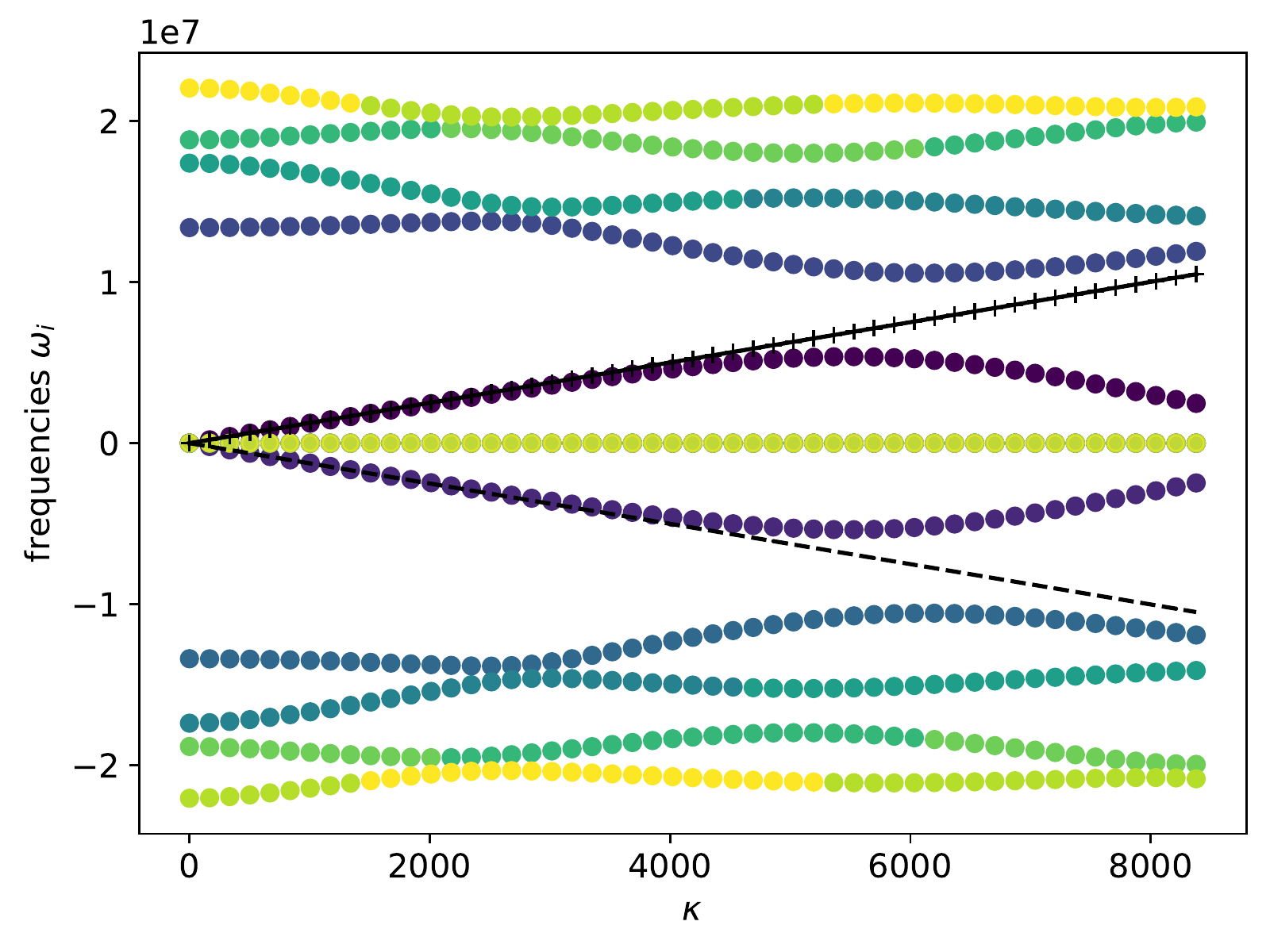} &
    \includegraphics[width=0.5\linewidth]{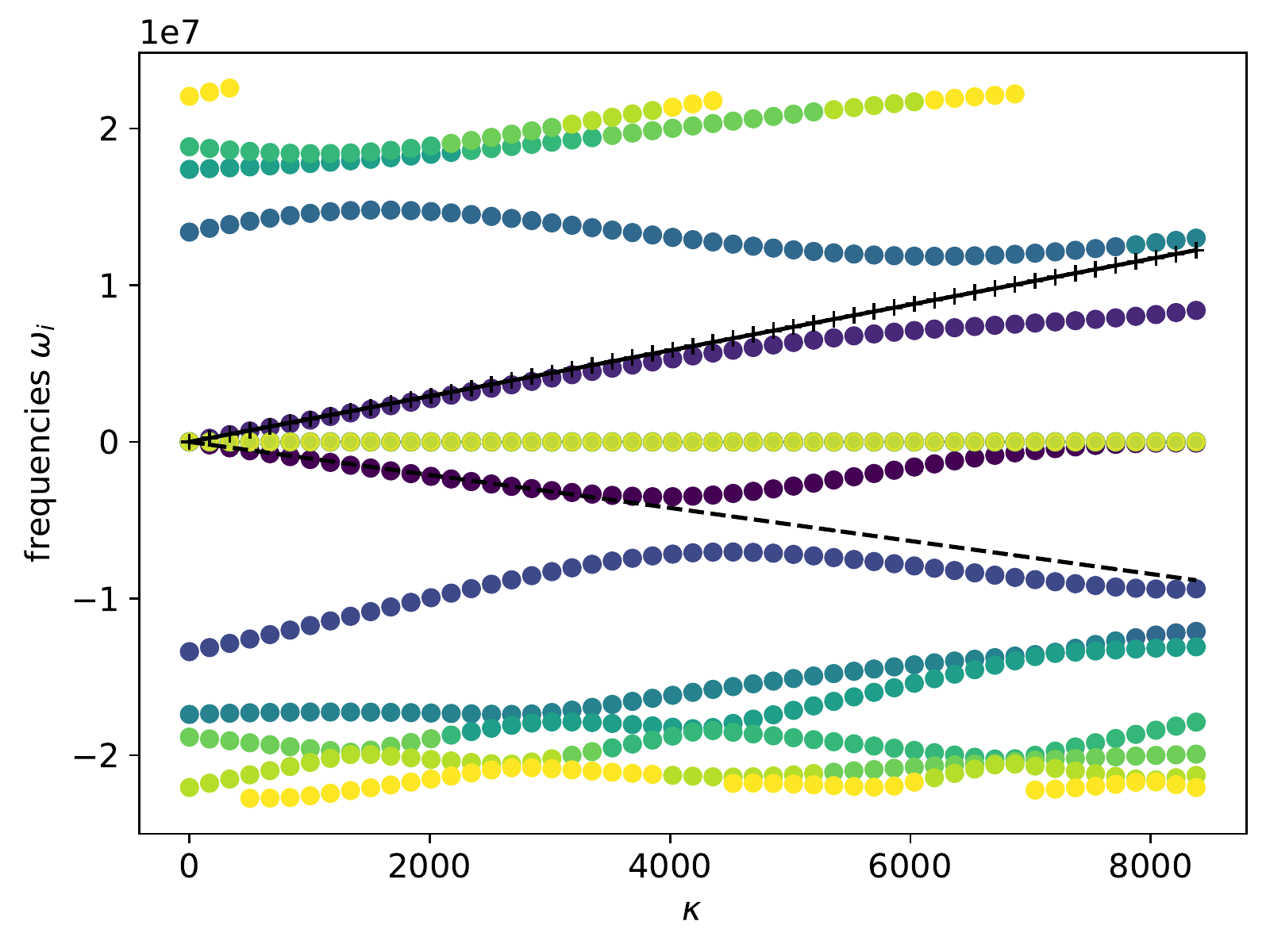} \\
    \multicolumn{2}{c}{$\mbw = 300$} \\
    \includegraphics[width=0.5\linewidth]{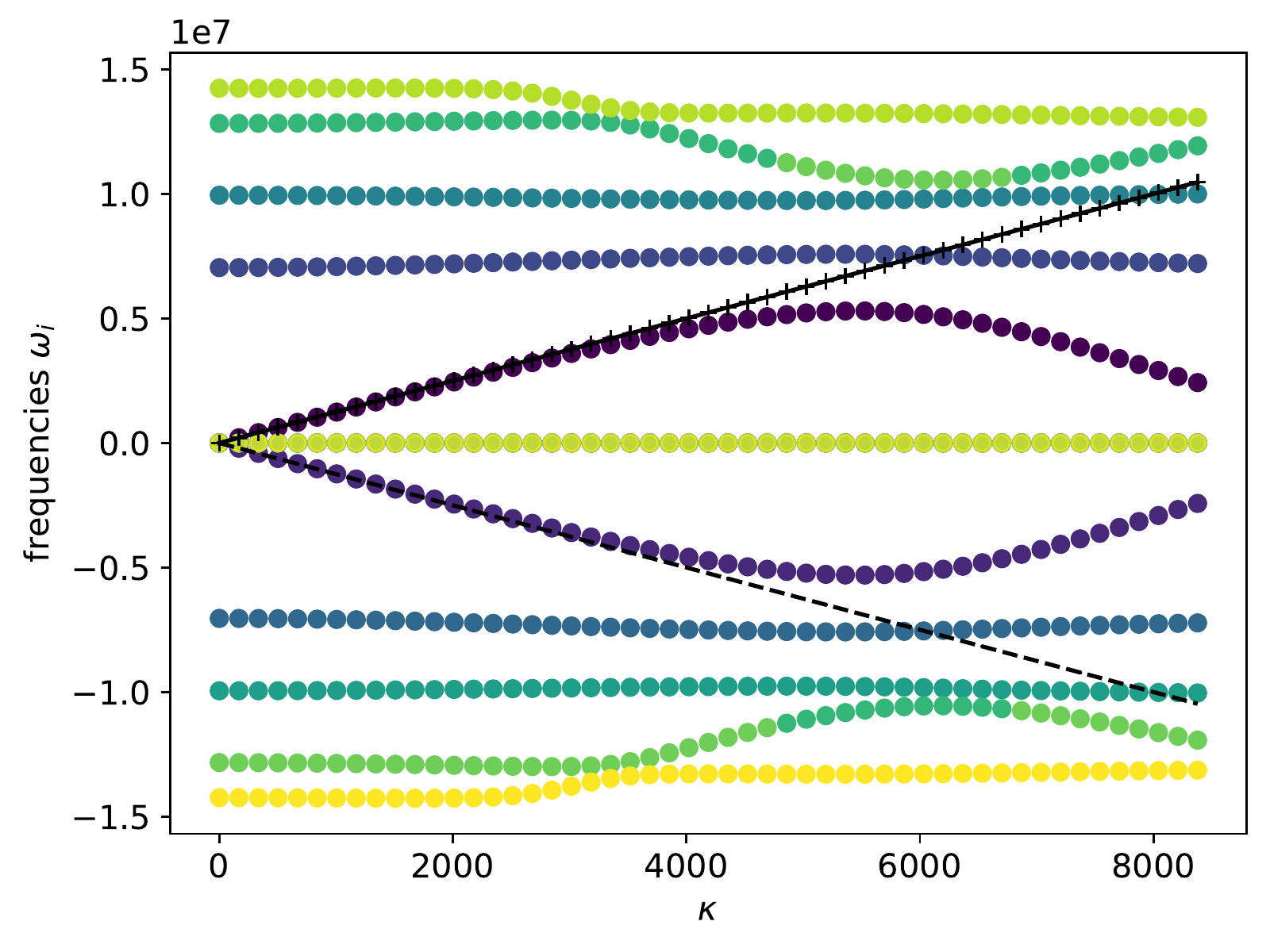} &
    \includegraphics[width=0.5\linewidth]{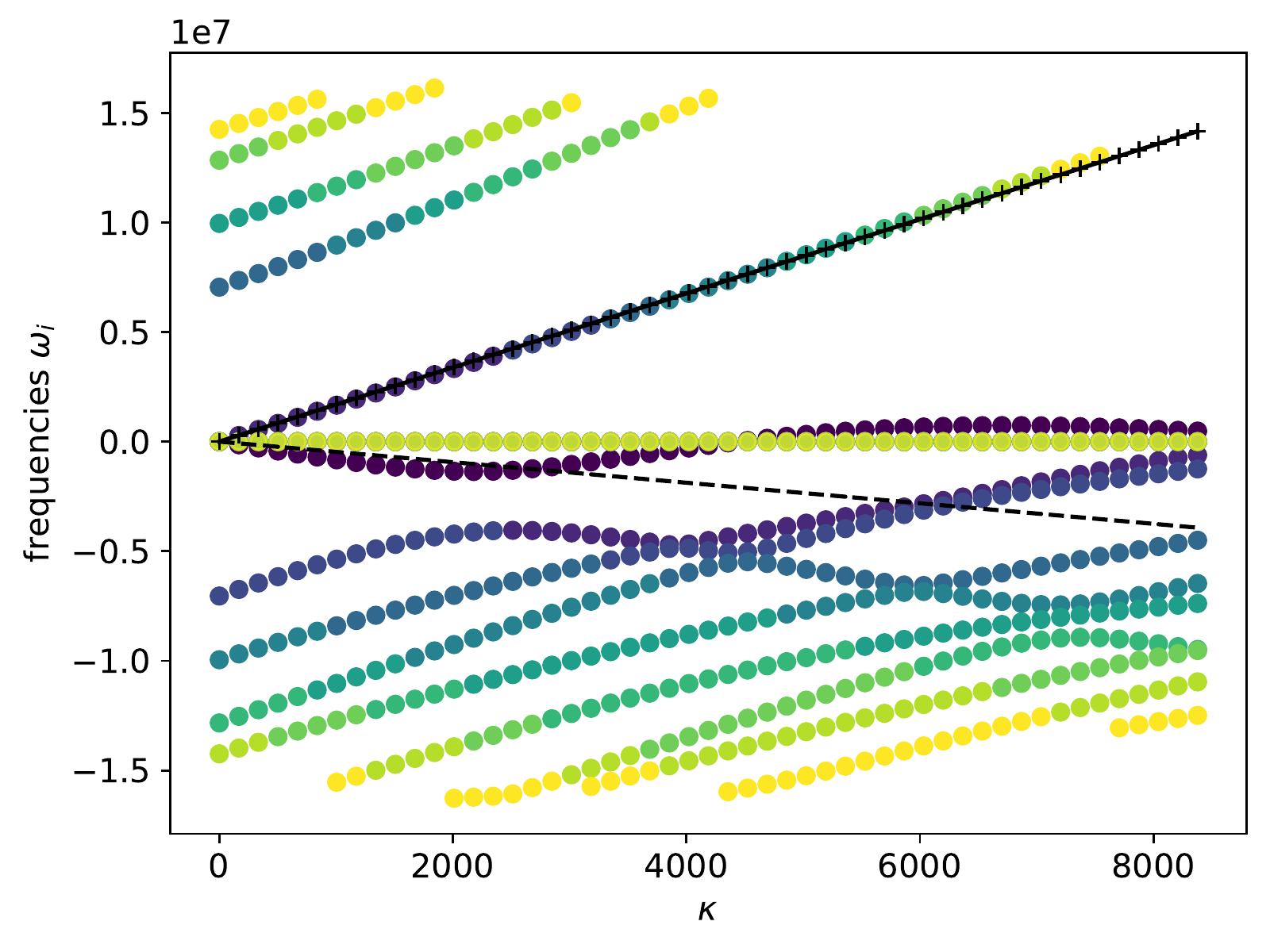}
  \end{tabular}
  \caption{The comparison of the dispersion analysis by the FB method (color
    point lines) and the homogenization-based analysis (black lines), various
    macroscopic flow velocities $\bbwM$ we considered: Left: study I-A; Right:
    study I-B.}
  \label{fig:d-overall}
\end{figure}

\begin{figure}[htp!]
  \centering
  \begin{tabular}{cc}
    \multicolumn{2}{c}{$\mbw = 100$} \\
    \includegraphics[width=0.5\linewidth]{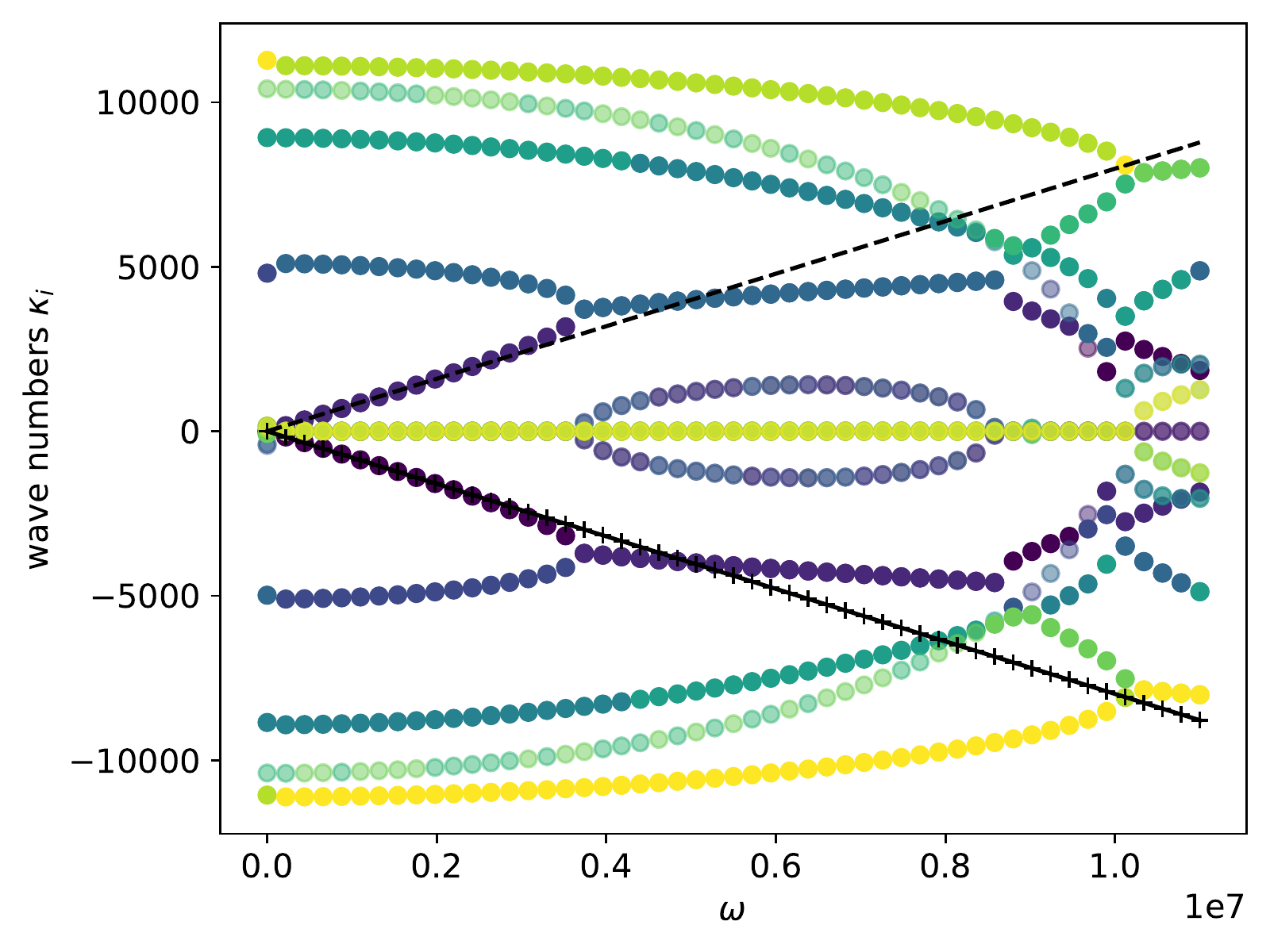} &
    \includegraphics[width=0.5\linewidth]{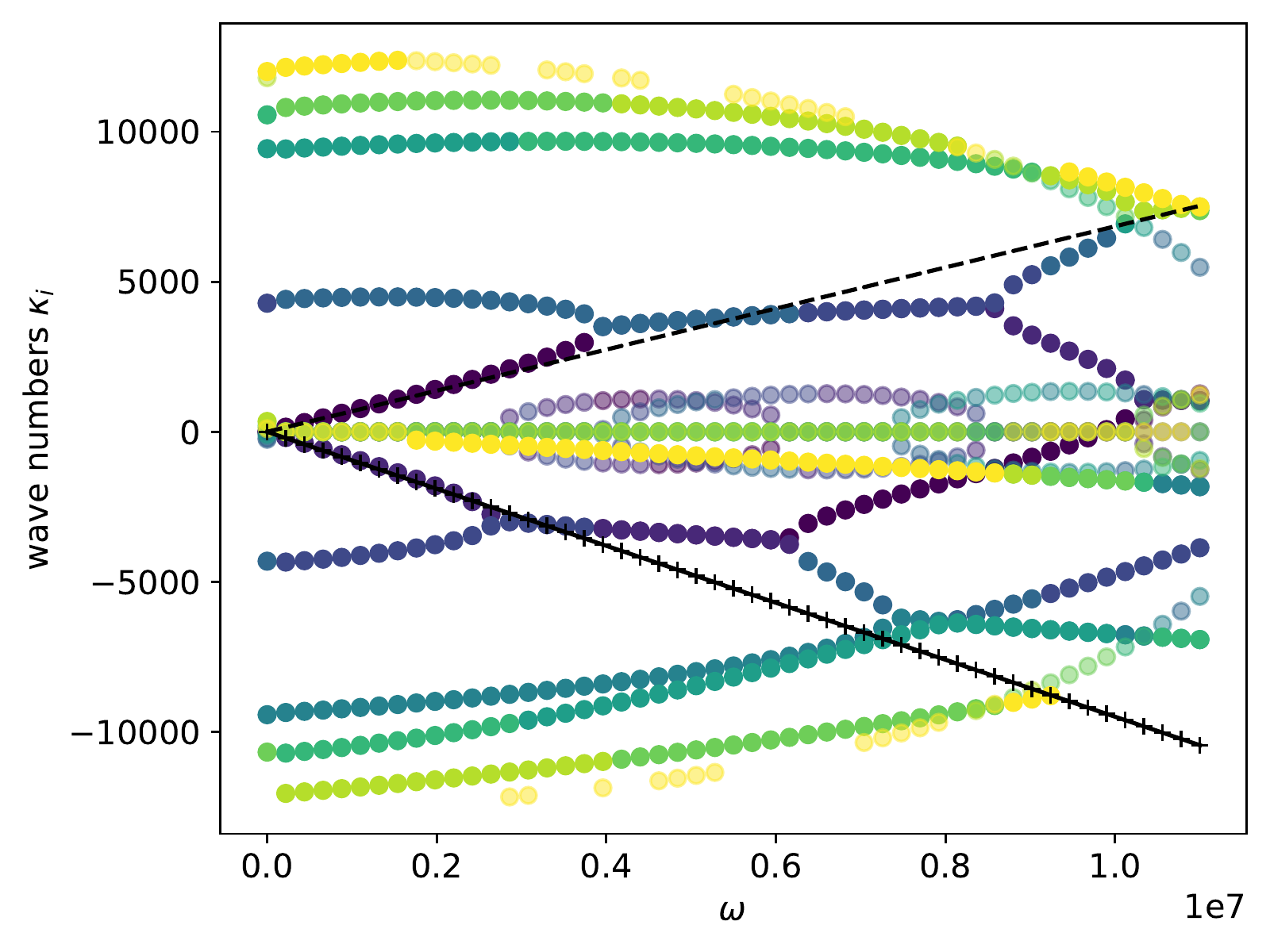} \\
  \end{tabular}
  \caption{The comparison of the dispersion analysis by the FB method (color
    point lines) and the homogenization-based analysis (black lines) for
    various inlet convective velocities $\mbw$, computed $\vkappa$ for given
    $\om$. Left: study I-A, right: study I-B.}
  \label{fig:c-overall}
\end{figure}

\begin{figure}[htp!]
  \centering
  \includegraphics[width=\linewidth]{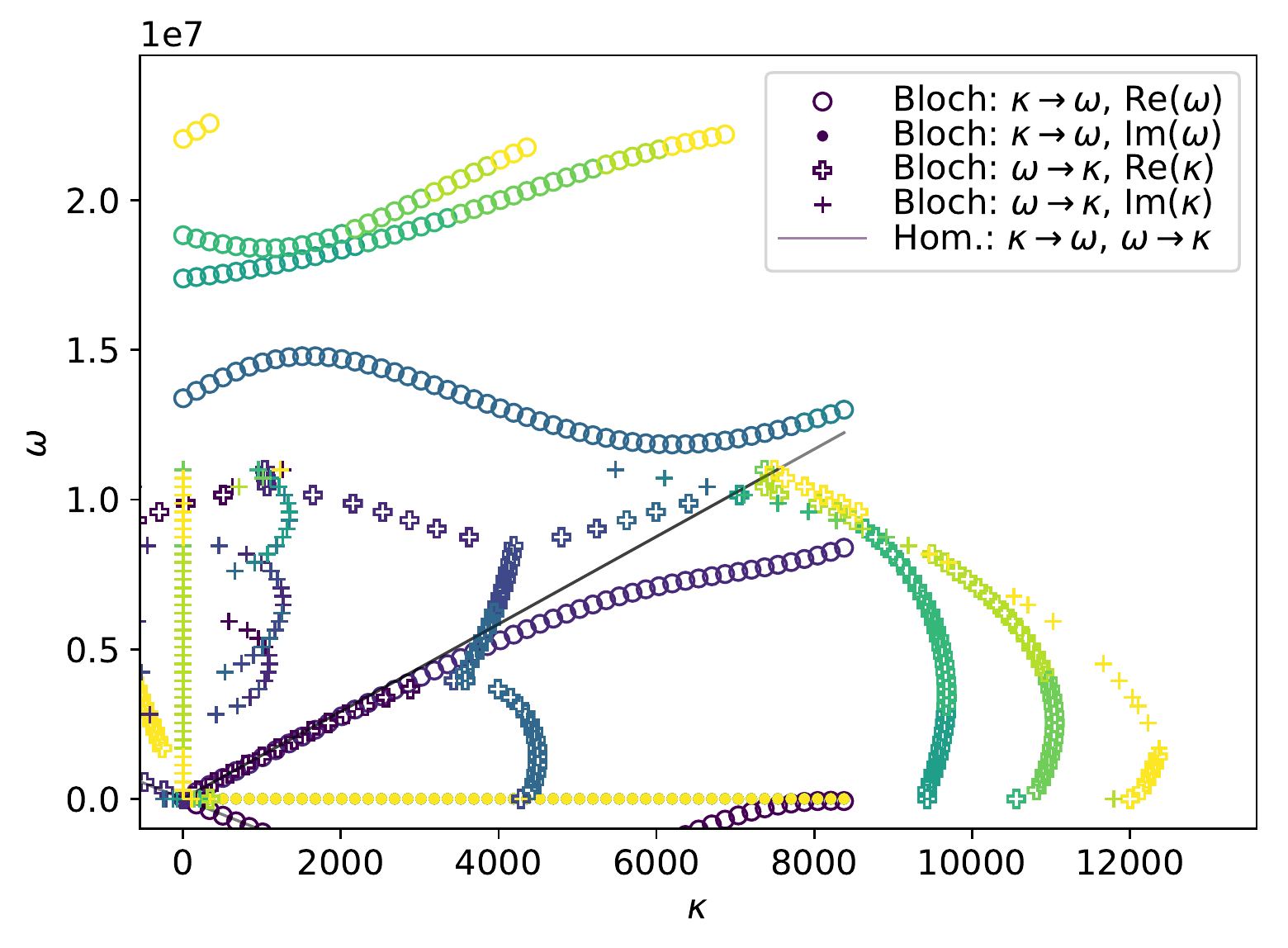}
  \caption{The comparison of $\vkappa \rightarrow \om$
    (Fig.~\ref{fig:d-overall}) and $\om \rightarrow \vkappa$
    (Fig.~\ref{fig:c-overall}) dispersion results for the study I-B,
    $\mbw = 100$. Only the first quadrant is shown.}
  \label{fig:inviscid-composite}
\end{figure}

In Fig.~\ref{fig:d-overall}, the dispersion analysis by the FB method is
compared with the homogenization-based approximation, whereby
\tododo{$\mbw \in \{1, 100, 300\}$}~m/s.
To compare the two possible methods of the FB analysis implementation, namely
the dispersion mappings $\om\mapsto\vkappa$, problem \eq{eq-NSadv24}, and
$\vkappa\mapsto\om$, problem \eq{eq-NSadv25}, in Fig.~\ref{fig:c-overall}, the
corresponding curves are displayed for one fixed macroscopic advection
velocity. For the case study I-B with the advection velocity $\mbw = 100$,
responses computed using both the FB analysis implementations, \ie
$\vkappa\mapsto\om$ and $\om\mapsto\vkappa$ are displayed in
Fig.~\ref{fig:inviscid-composite}.

\begin{figure}[htp!]
  \centering
  \begin{tabular}{cc}
    \includegraphics[width=0.5\linewidth]{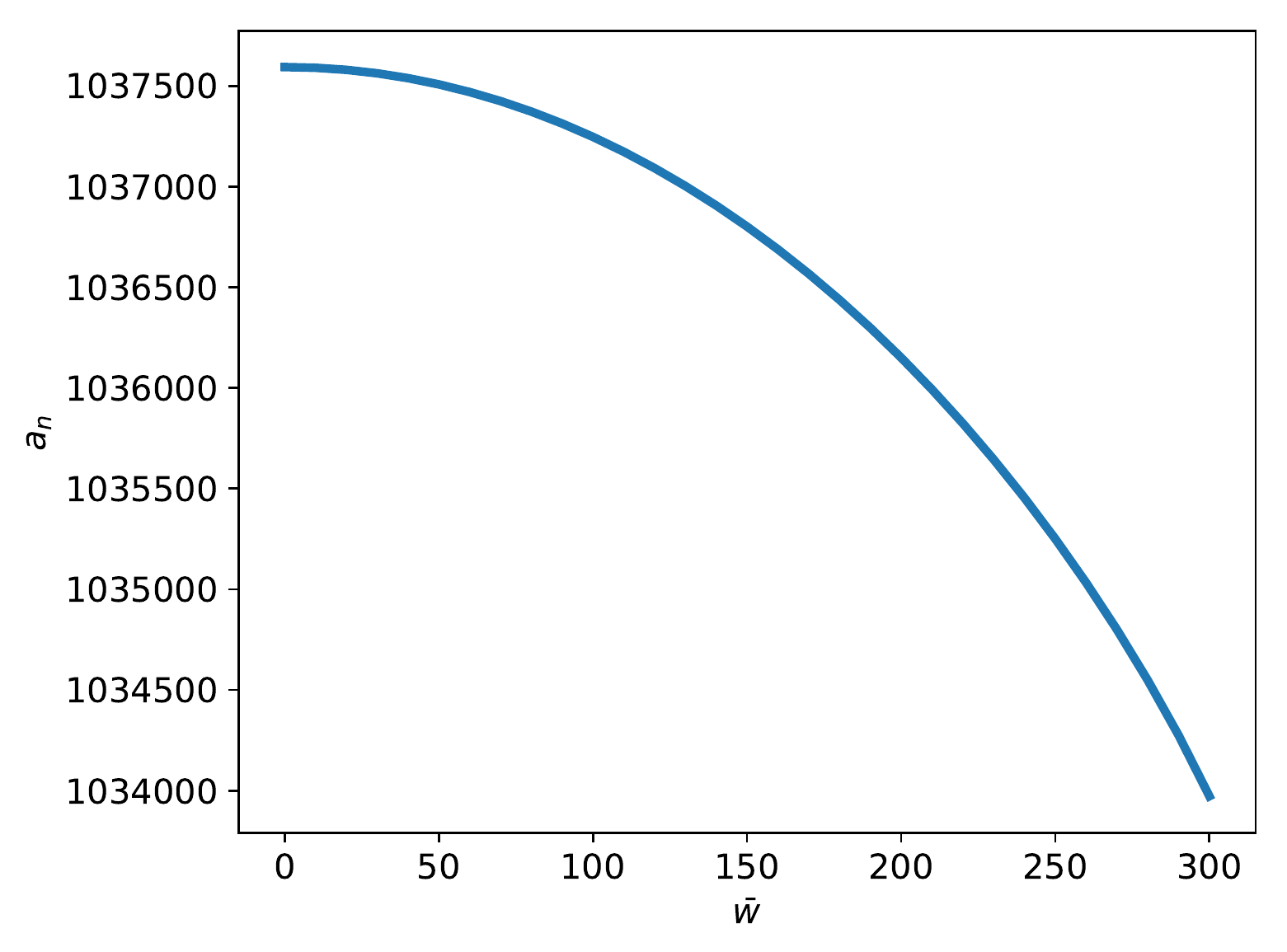} &
    \includegraphics[width=0.5\linewidth]{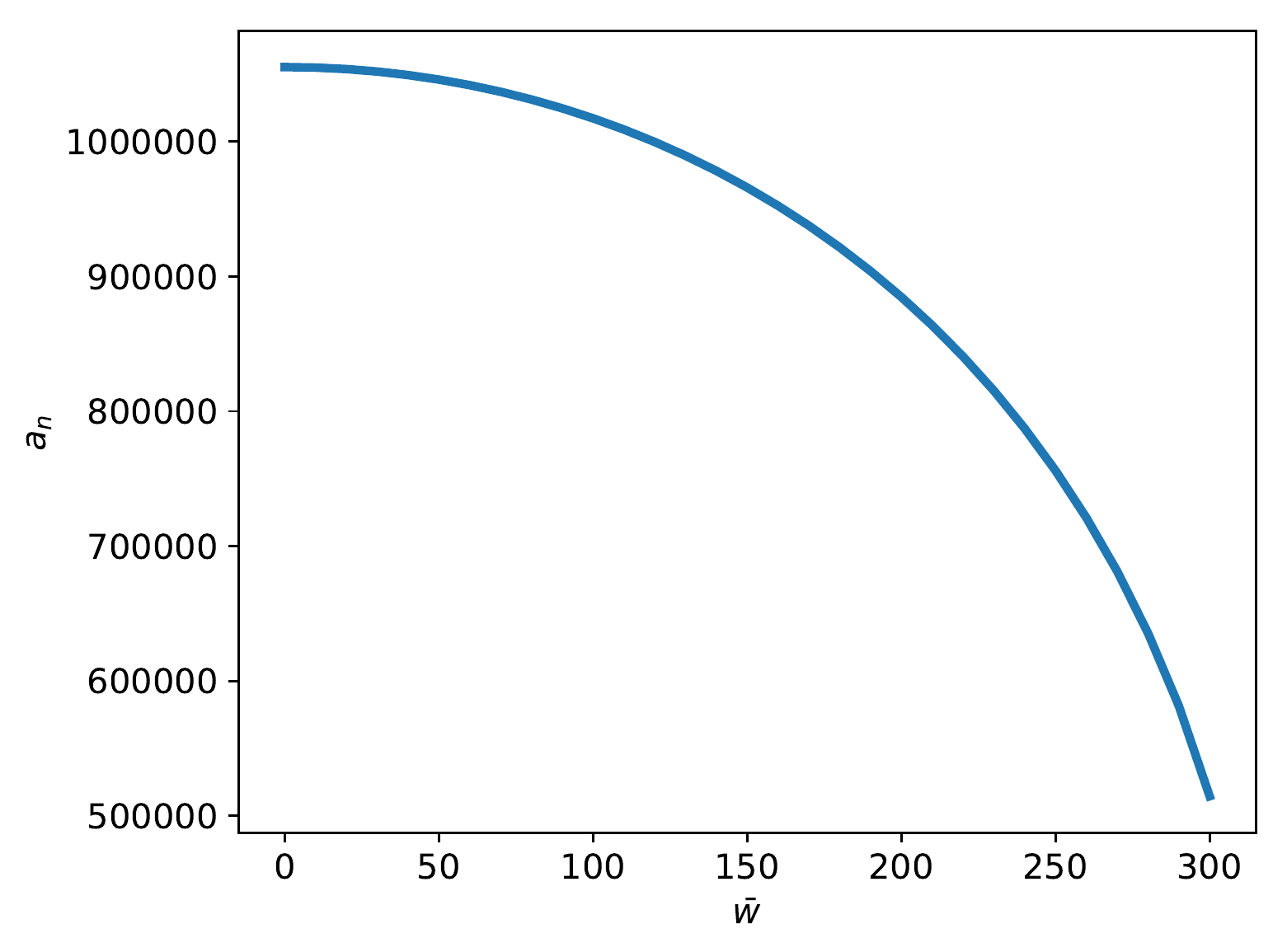} \\
    \includegraphics[width=0.5\linewidth]{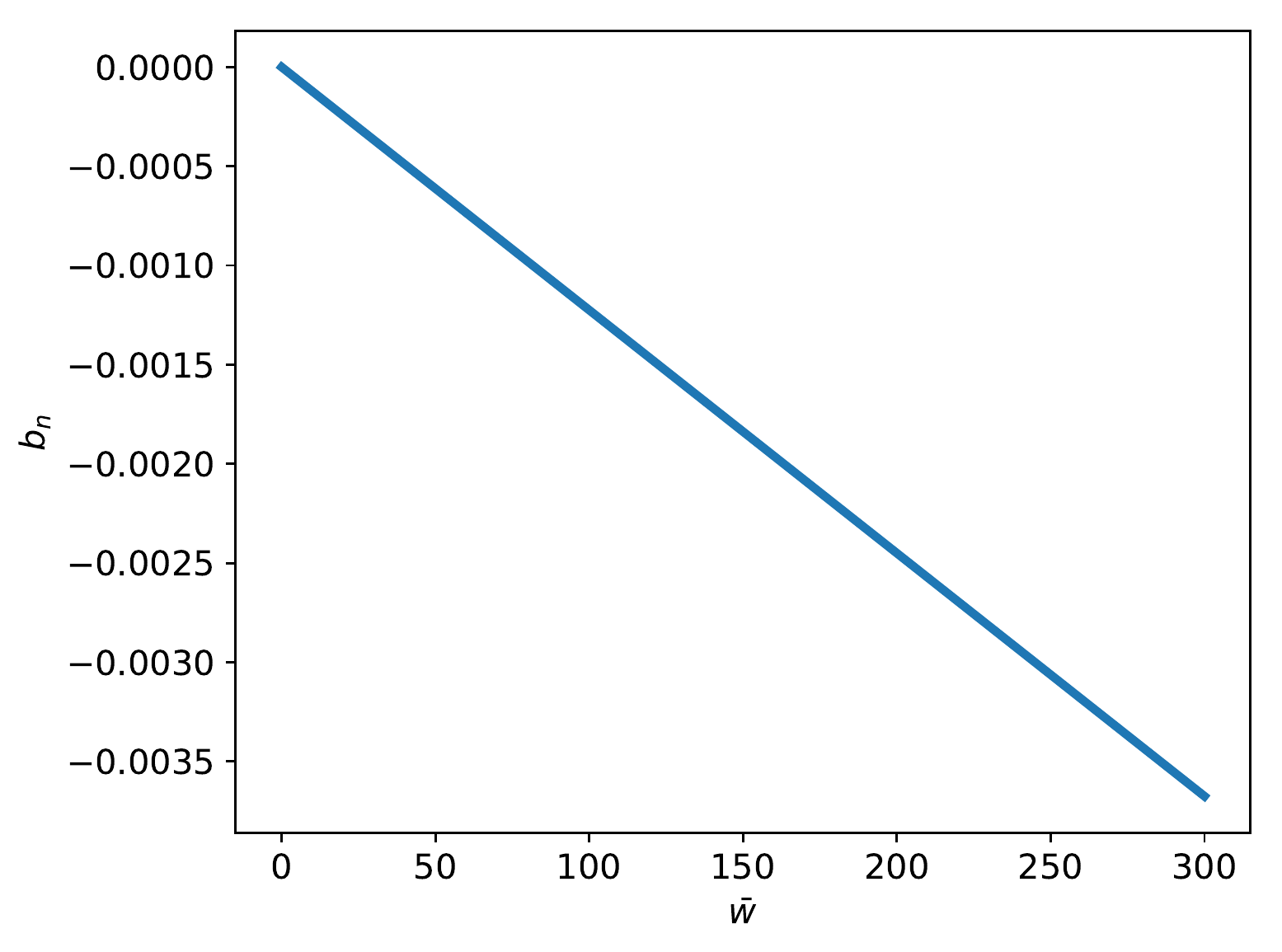} &
    \includegraphics[width=0.5\linewidth]{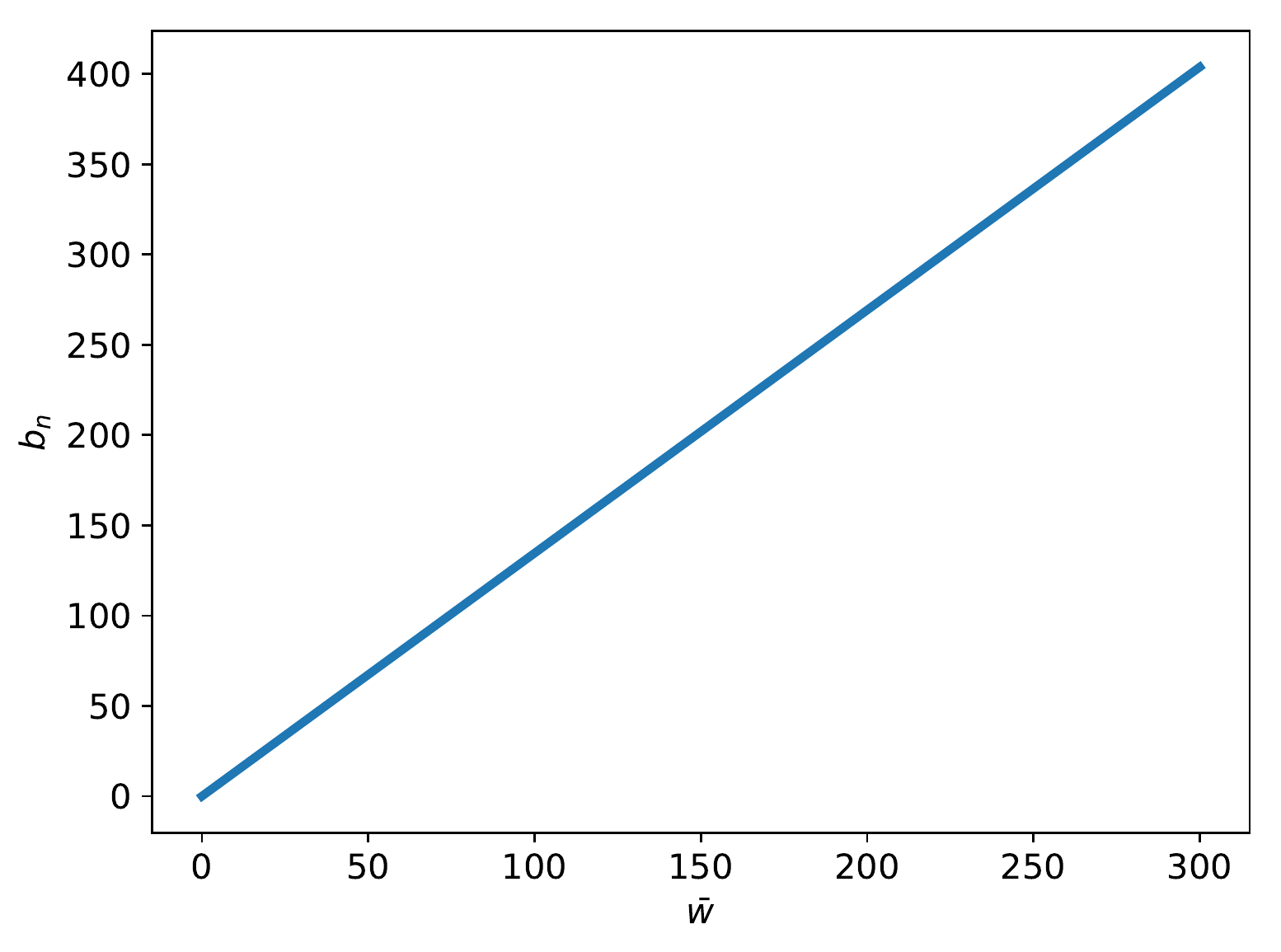} \\
  \end{tabular}
  \caption{The dependence of homogenized coefficients $a_n$, $b_n$ on inlet
    convective velocities $\mbw$. Left: study I-A, right: study I-B.}
  \label{fig:d-coefs}
\end{figure}

\begin{figure}[htp!]
  \centering
  \begin{tabular}{cc}
    \includegraphics[width=0.5\linewidth]
    {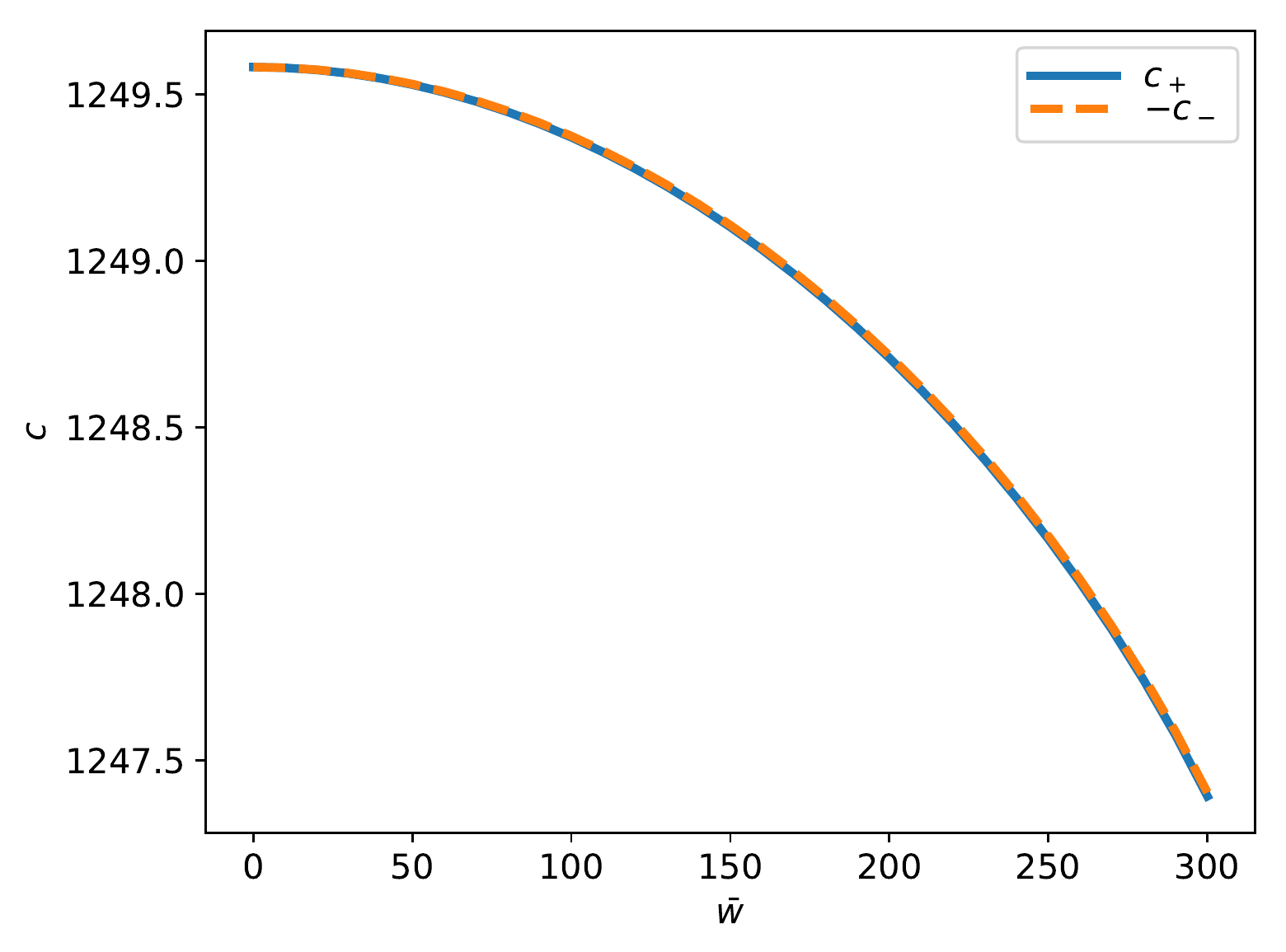} &
    \includegraphics[width=0.5\linewidth]
    {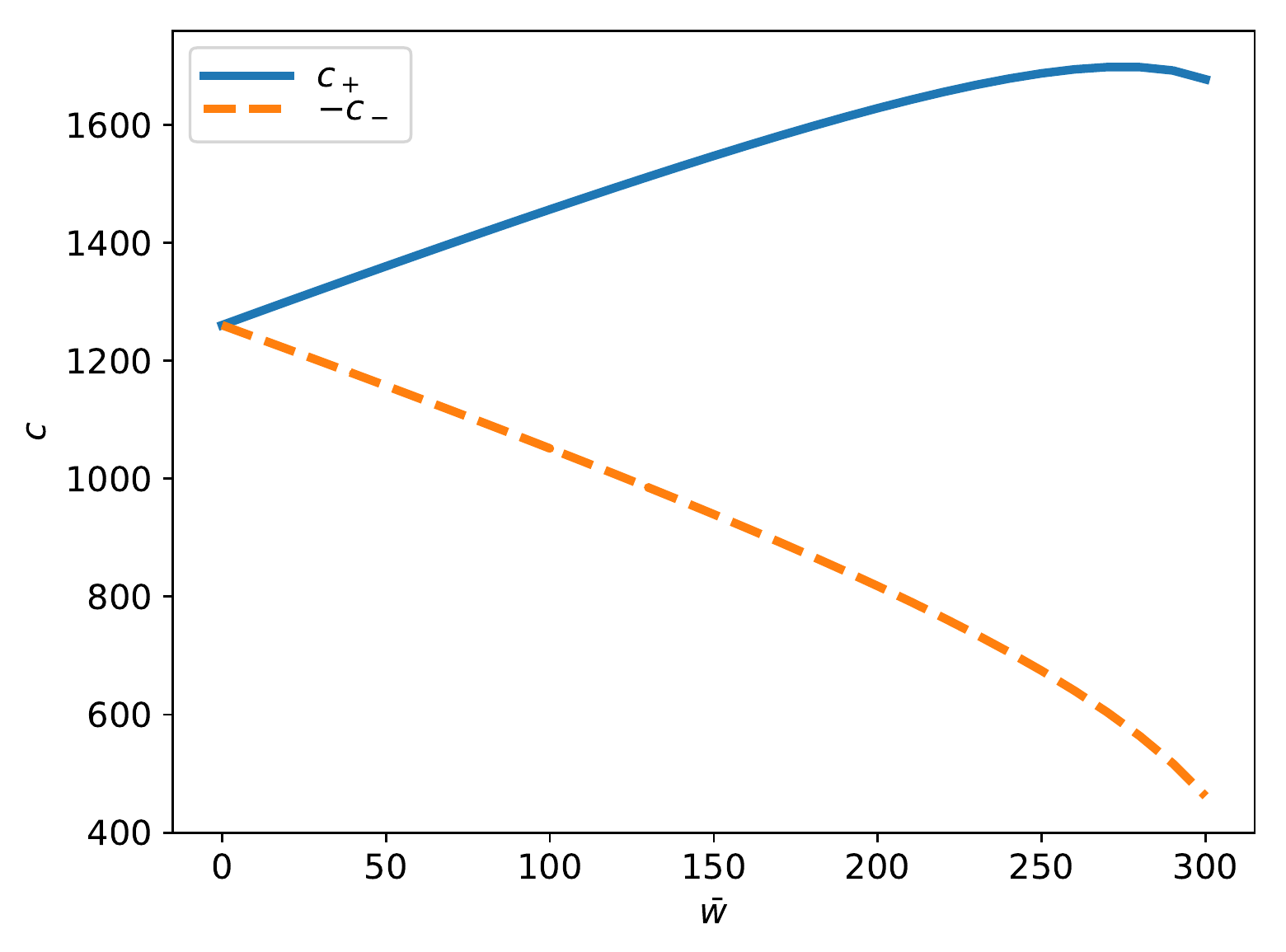} \\
    \includegraphics[width=0.5\linewidth]
    {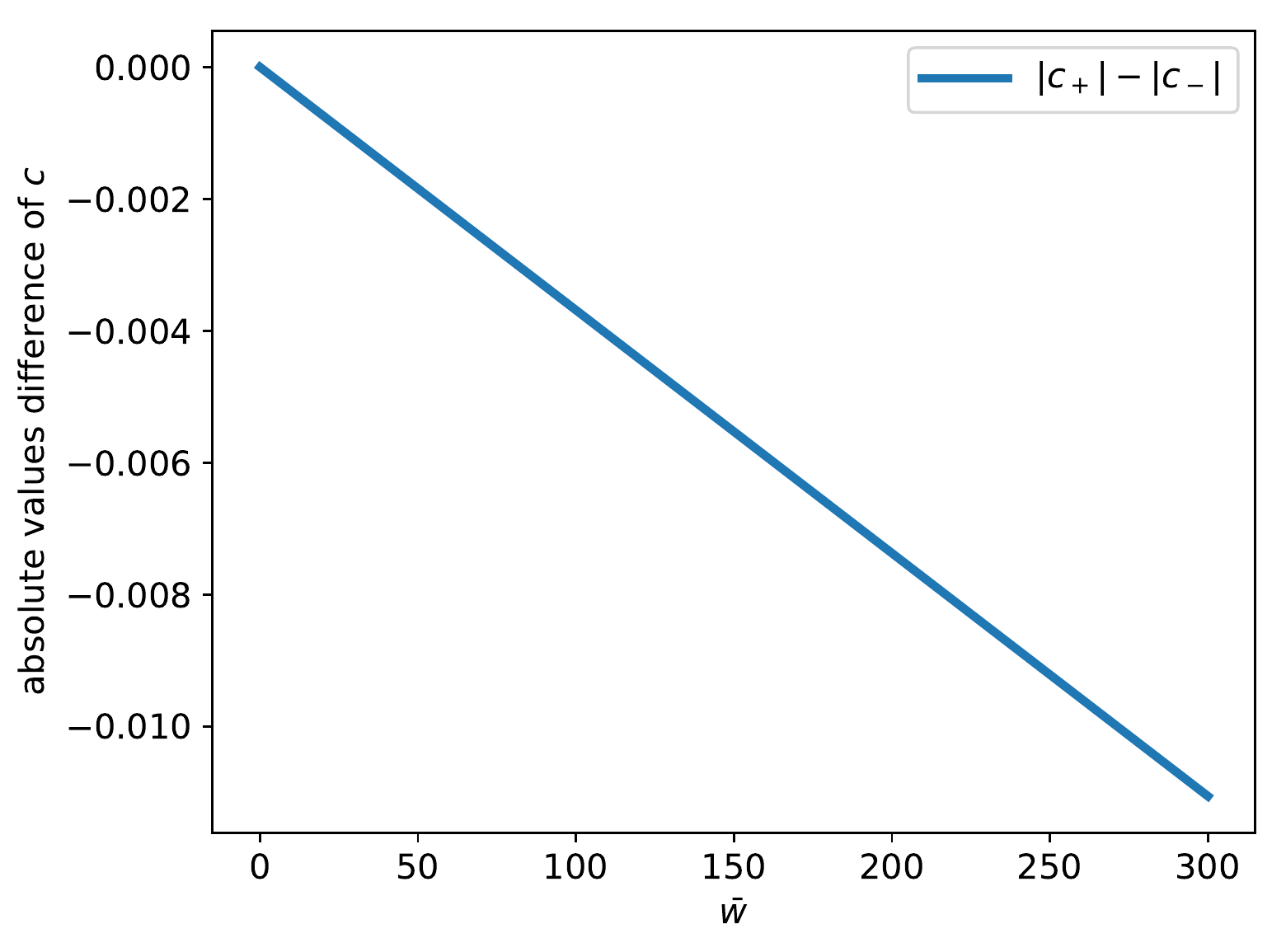} &
    \includegraphics[width=0.5\linewidth]
    {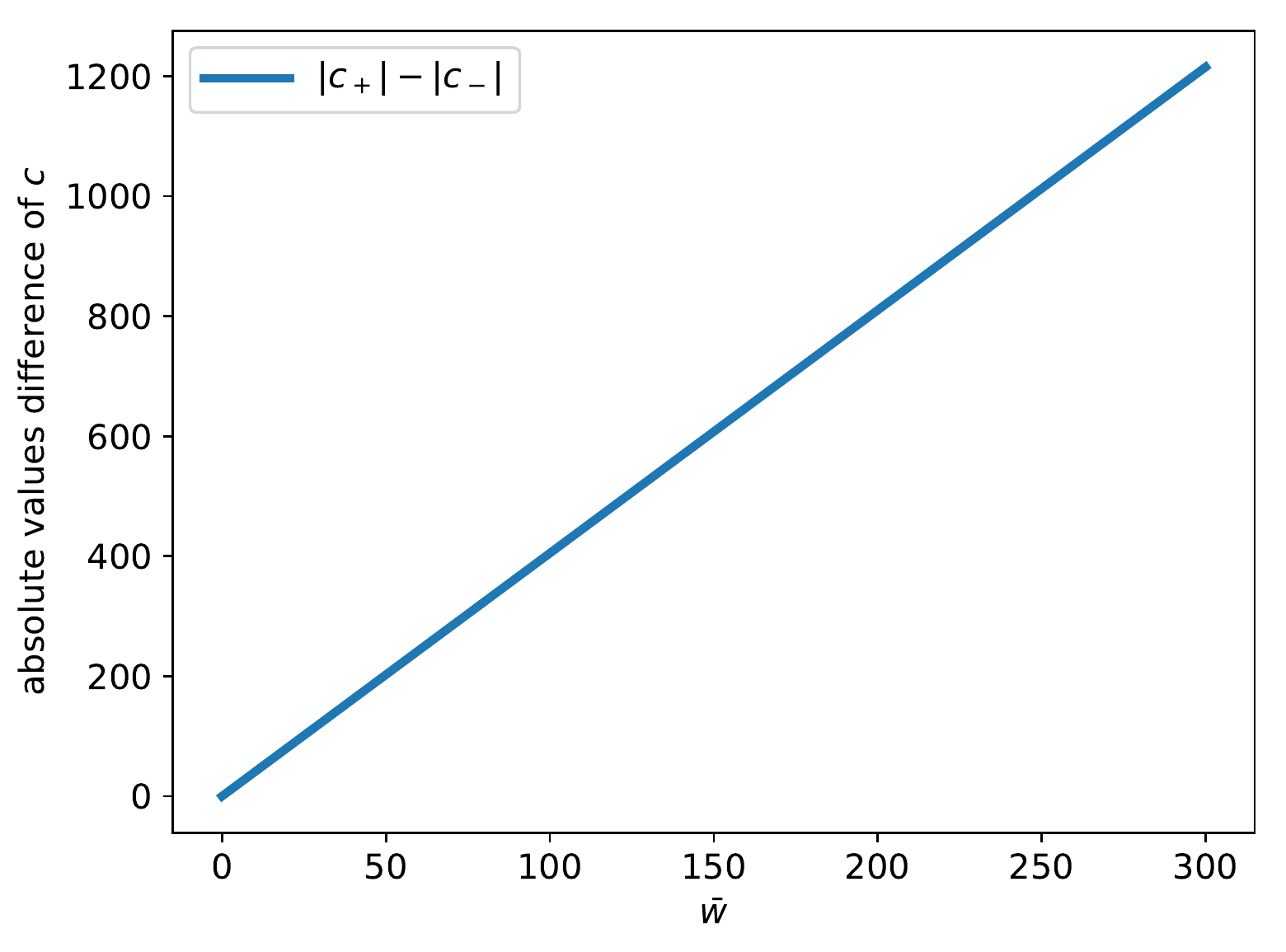}
  \end{tabular}
  \caption{The dependence of the homogenization-based phase velocities
    $c_+$, $-c_-$ on $\mbw$ (top) and the dependence of the difference
    $|c_+| - |c_-|$ on $\mbw$ (bottom). Left: study I-A, right: study I-B.}
  \label{fig:d-c-w}
\end{figure}

In Fig.~\ref{fig:d-coefs} we present the dependence of the homogenized
coefficients $a_n$, $b_n$ involved in (\ref{eq-NSadv20a}) on the macroscopic
advection velocity magnitude $\mbw$. In Fig.~\ref{fig:d-c-w} the dependence of
the homogenization-based phase velocities on $\mbw$ is shown: $c_+$, $-c_-$ in
the left subplot and $|c_+| - |c_-|$ in the right subplot. \tododo{As can be
seen in the both figures, in study I-A, the influence of $\mbw$ magnitude is
relatively small (nevertheless increasing with increasing $\mbw$): $a_n$,
$b_n$ are almost constant and $c_+$, $-c_-$ relative difference stays below
$10^{-5}$. On the other hand, in study I-B, $a_n$, $b_n$ vary significantly
and large progressively increasing differences between $c_+$ and $-c_-$ can
be observed.}

\subsection{Viscous fluid}
\label{sec:examples-viscous}

Because of computational hurdles associated with solving the quadratic
eigenvalue problems (QEP), we consider 2D structures only leading to smaller
discretized models . The viscous fluid parameters are those of water:
$\mu = 1.02\cdot10^{-3}$~Pa$\cdot$s, $\gamma = 5 \cdot 10^{-10}$~Pa$^{-1}$,
$\rho_0 = 1000$~kg/m$^3$. The P2 (for velocity variables) and P1 (for pressure
variables) finite elements on triangles were used for the discretization of
both the FB and PH analyses.

\subsubsection{Homogenization-based analysis}
\label{sec:viscous-homo}

\begin{figure}[htp!]
  \centering
  \includegraphics[width=0.5\linewidth]{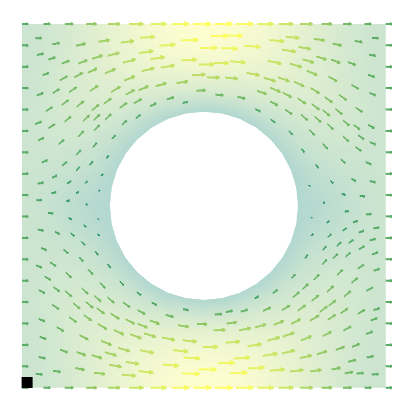}
  \caption{The finite element mesh and viscous flow velocity field.}
  \label{fig:mesh-2d-viscous-homo}
\end{figure}

By using (\ref{eq-w3A}) we studied, for a fixed finite scale parameter
$\veps = 10^{-5}$, the influence of the convective velocity magnitude $\mbw$ on
the dispersion properties. The incident wave vector direction is
$\nb = (0, 1)^T$, whereas the convective flow represented by
$\bbwM \parallel (1, 0)^T$ is aligned with the $x_1$-axis. The associated
advection field $\bar\wb$ is obtained by the standard reconstruction of the
homogenized Stokes flow, see \eg \cite{Polisevski-ZAMP1989,Allaire-nonsteady-NS-homog,CDG-Stokes-2005,Zaki-2012}; for illustration of $\bar\wb$,
the streamlines are depicted in Fig~\ref{fig:mesh-2d-viscous-homo}. Four
velocities were used: $\mbw \in \{0, 1, 5, 10\}$~m/s.

\begin{figure}[htp!]
  \centering
  \begin{tabular}{p{5mm}c|c}
    & $k=1 \quad\quad k =2$ & $k=1 \quad\quad k =2$ \\
    \vskip -50mm
    $\Im{\wb^k}$
    \vskip 25mm
    $\Re{\wb^k}$
    & \includegraphics[width=0.46\linewidth]{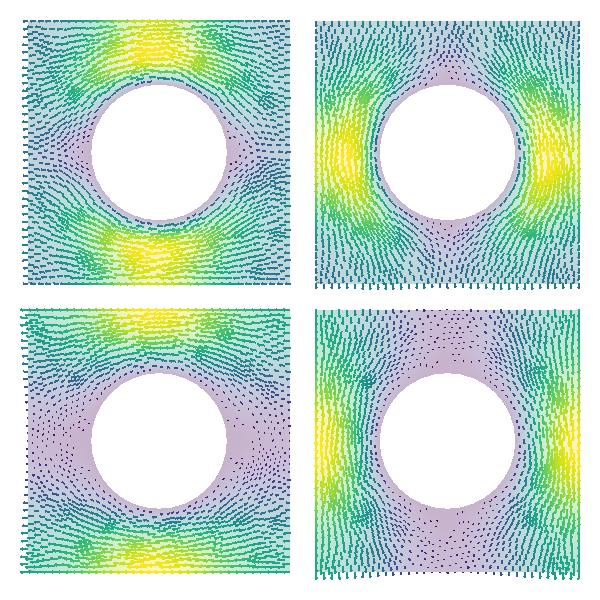}
    & \includegraphics[width=0.46\linewidth]{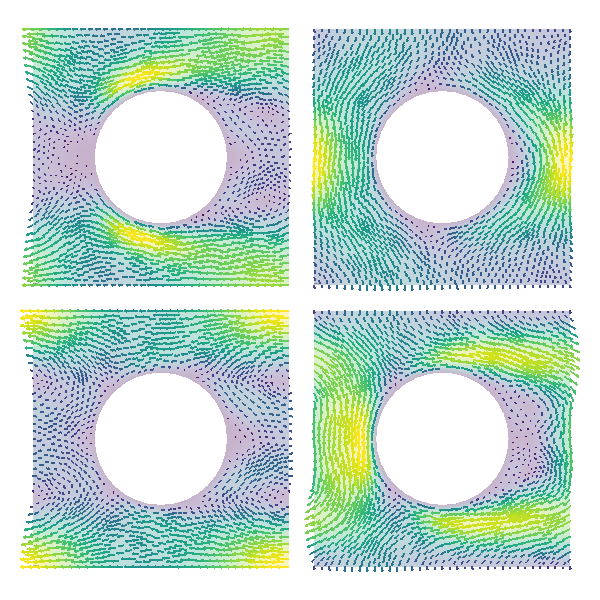} \\
    & $|\bar\wb| = 0$ [m/s] & $|\bar\wb| = 10$ [m/s]
  \end{tabular}
  \caption{Characteristic responses ${\wb^k}$ for ``horizontal'' flow $\bbw$.}
  \label{fig:viscous-char-resp-2d}
\end{figure}

\begin{figure}[htp!]
  \centering
  \begin{tabular}{cc}
    \includegraphics[width=0.45\linewidth]{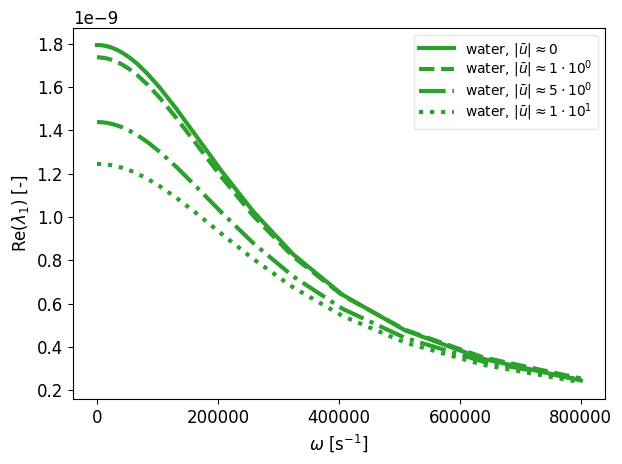}
    &
    \includegraphics[width=0.45\linewidth]{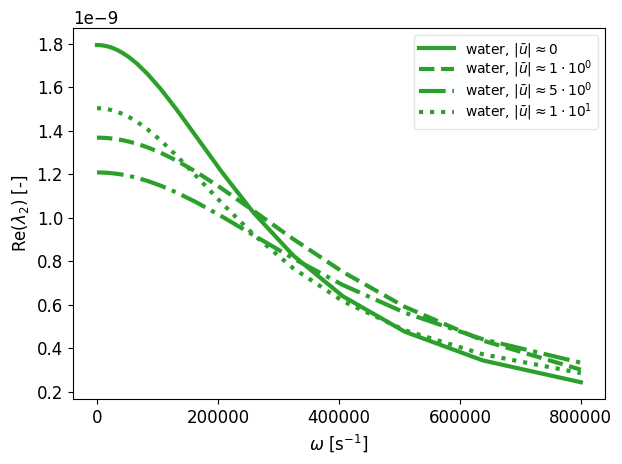}\\
    (a) 1st eig.val. of $\FT{\Kcalbf}(\imu\om)$, $\Re{\lam_1(\om)}$ &
    (b) 2nd eig.val. of $\FT{\Kcalbf}(\imu\om)$, $\Re{\lam_2(\om)}$
  \end{tabular}
  \begin{tabular}{cc}
    \includegraphics[width=0.45\linewidth]{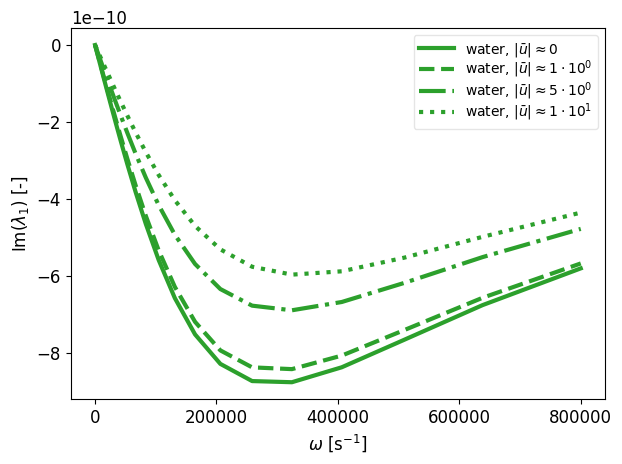}
    &
    \includegraphics[width=0.45\linewidth]{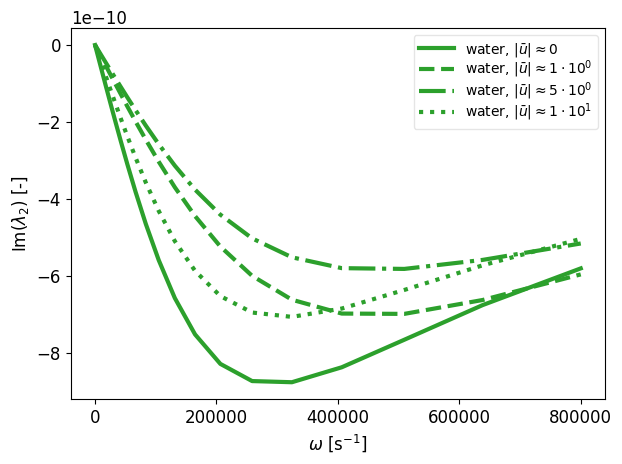}\\
    (c) 1st eig.val. of $\FT{\Kcalbf}(\imu\om)$, $\Im{\lam_1(\om)}$ &
    (d)  2nd eig.val. of  $\FT{\Kcalbf}(\imu\om)$, $\Im{\lam_2(\om)}$
  \end{tabular}
  \caption{The influence of the convective velocity magnitude $\mbw$ on the
    eigenvalues of the homogenized permeability tensor.}
  \label{fig:viscous-perm-2d}
\end{figure}

\begin{figure}[htp!]
  \centering
  \begin{tabular}{cc}
    \includegraphics[width=0.45\linewidth]{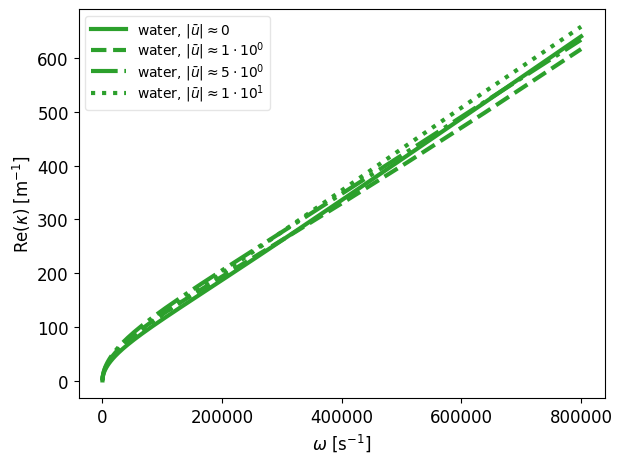}
    &
    \includegraphics[width=0.45\linewidth]{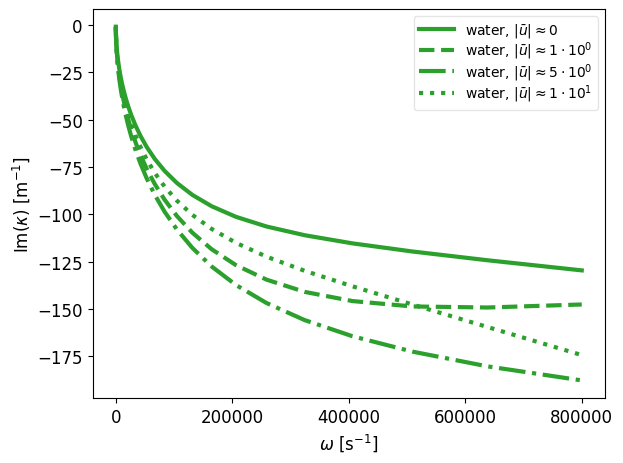} \\
    (a) $\Re{\vkappa(\om)}$ & (b) $\Im{\vkappa(\om)}$
  \end{tabular}
  \begin{tabular}{cc}
    \includegraphics[width=0.45\linewidth]{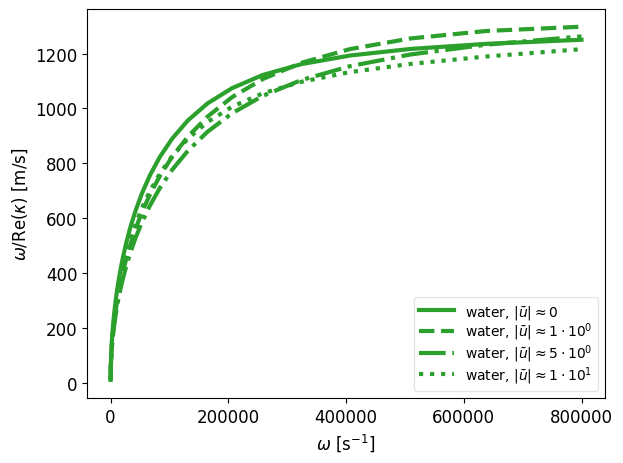}
    &
    \includegraphics[width=0.45\linewidth]{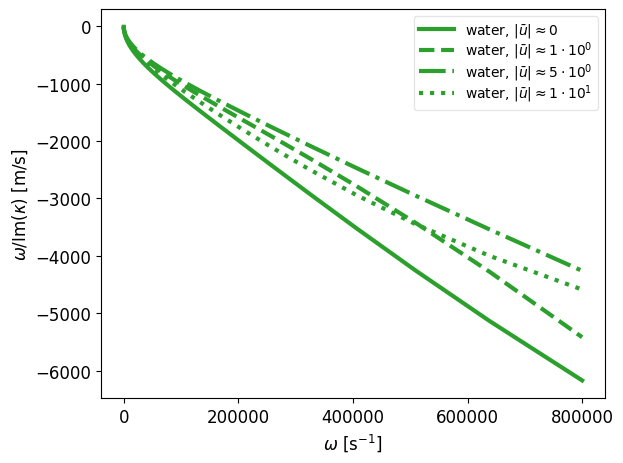}\\
    (c) phase velo.: $c_w(\om) = \om / \Re{\vkappa}$ &
    (d) attenuation $\om / \Im{\vkappa}$
  \end{tabular}
  \caption{The influence of the convective velocity magnitude $\mbw$ on the
    wave number, phase velocity and attenuation.}
  \label{fig:viscous-dispersion-2d}
\end{figure}

The characteristic responses $\LT{\chibf}^k$ defined by (\ref{eq-A10b}) are
shown in Fig.~\ref{fig:viscous-char-resp-2d} for two velocities: $\mbw = 0$ and
$\mbw = 10$~m/s. The steady advection flow modifies the so-called correctors
(the characteristic responses) associated with the macroscopic pressure
gradients, see \eq{eq-A10a}, thus, inducing a non-symmetry of the dynamic
permeability calculated according to (\ref{eq-A11}). This effect is
demonstrated in Fig.~\ref{fig:viscous-perm-2d}; while for zero advection, the
responses of the modes $k=1$ and $k=2$ are symmetric \wrt the coordinate axes
(rotation by $\pi/2$), this symmetry is lost for $\mbw \not = 0$. Finally, the
dispersion properties, as depending on $\mbw$, in terms of the $\vkappa(\om)$
plots and phase velocity and attenuation plots are shown in
Fig.~\ref{fig:viscous-dispersion-2d}.

\subsubsection{Bloch wave analysis compared to homogenization}
\label{sec:viscous-bloch-homo}

\begin{figure}[htp!]
  \centering
  \includegraphics[width=0.5\linewidth]{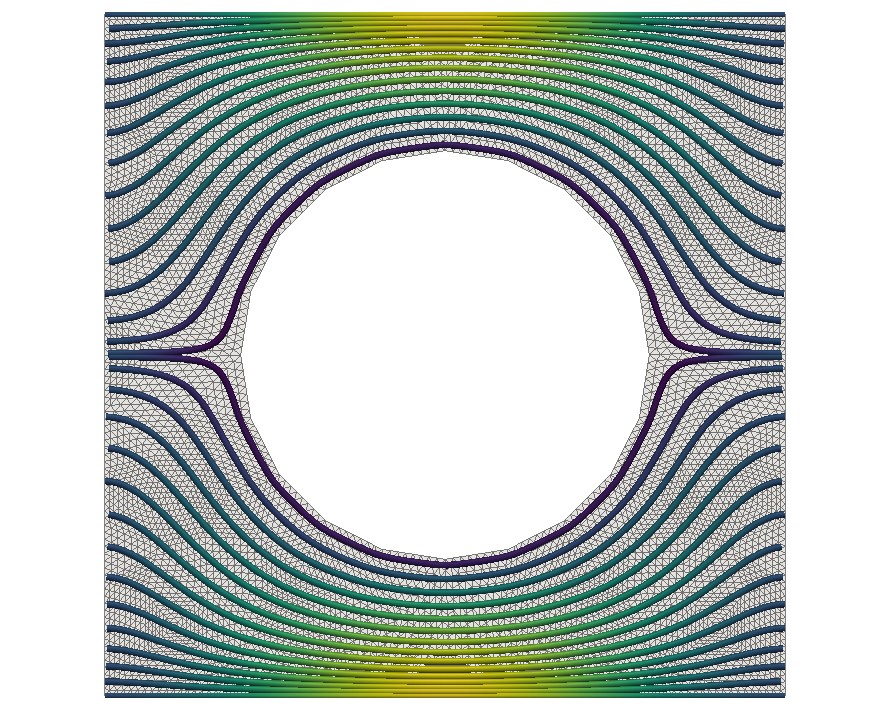}
  \caption{The finite element mesh and viscous flow streamlines.}
  \label{fig:mesh-2d-viscous}
\end{figure}

The homogenization-based dispersion analysis (\ref{eq-w3A}) results were
compared with the Bloch wave analysis (\ref{eq-vsc13b}) as follows. The
incoming wave vector direction was $\nb = (0, 1)^T$. The convective flow $\bbw$
had the overall direction along the $x_1$-axis $(1, 0)^T$, and was specified using
its magnitude $\mbw$ on the inflow (left) boundary. It is illustrated, together
with the used finite element mesh, in Fig~\ref{fig:mesh-2d-viscous} in terms of
streamlines.

Two parametric studies of the dispersion were performed:
\begin{itemize}
\item Study V-A: The finite scale
  parameter $\veps_0$ was modified in the range $[10^{-4}, 5.23 \cdot 10^{-3}]$ in
  lockstep with the finite scale advection velocity $\bar\wb^{\veps_0}$, so that $|\bbw| = 1$ in \eq{eq-A10b}, and  \eq{eq-A11}; recall the scaling \eq{eq-FS14a}.
\item Study V-B: The dispersion analyses were performed for the fixed finite
  scale parameter $\veps_0 = 10^{-3}$, while changing $\mbw$ in range
  $[1, 21.5]$.
\end{itemize}
While in the V-A study, the advection effect is being changed proportionally to the microstructure size, the V-B study is aimed to explore the advection effect for a fixed microstructure size.

The eigenvalues in the FB analysis were computed using the MATLAB\reg{}
function \textrm{eigs()} \cite{stewart2002, arpack1998} via the MATLAB Engine
API for Python. Ten eigenvalues ($\frac{1}{i \vkappa}$) with the largest real
part, i.e., $\vkappa$ with the smallest imaginary part, were requested in the
study V-A, while only three eigenvalues in the study V-B.

\begin{figure}[htp!]
  \centering
  \begin{tabular}{cc}
    \multicolumn{2}{c}{$\veps_0 = 10^{-4}$} \\
    \includegraphics[width=0.5\linewidth]{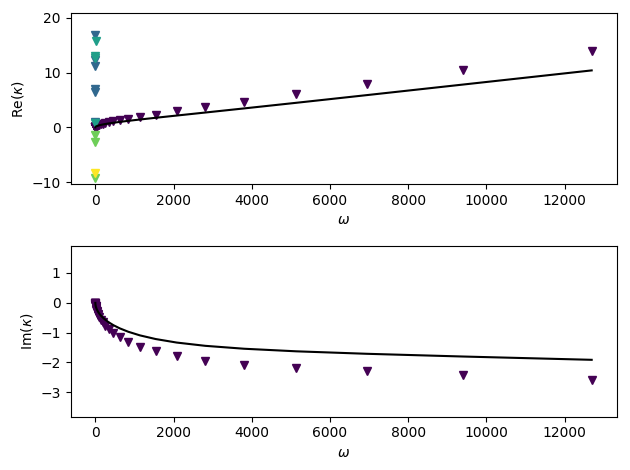}
    &
    \includegraphics[width=0.5\linewidth]{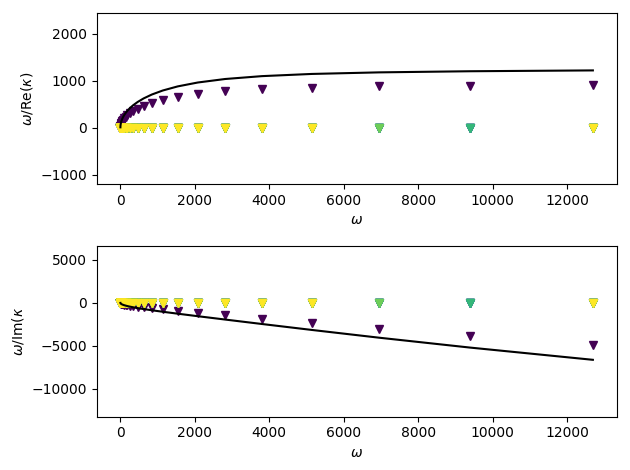}
    \\
    \multicolumn{2}{c}{$\veps_0 = 5.23 \cdot 10^{-3}$} \\
    \includegraphics[width=0.5\linewidth]{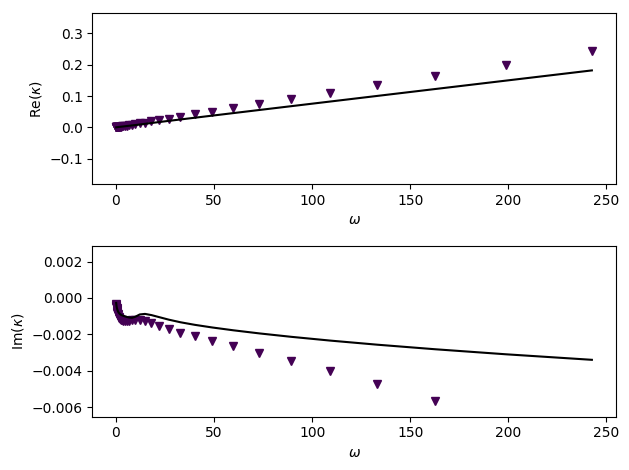}
    &
    \includegraphics[width=0.5\linewidth]{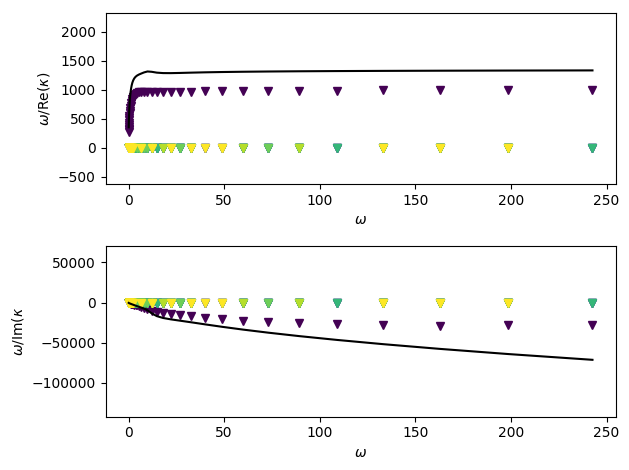}
  \end{tabular}
  \caption{Study V-A, the comparison of dispersion curves (left) and phase
    velocities (right) for homogenization (solid lines) and Bloch-based
    analyses (triangles). }
  \label{fig:a-hb-kappa}
\end{figure}

\begin{figure}[htp!]
  \centering
  \begin{tabular}{cc}
    \multicolumn{2}{c}{$\mbw = 1$} \\
    \includegraphics[width=0.5\linewidth]{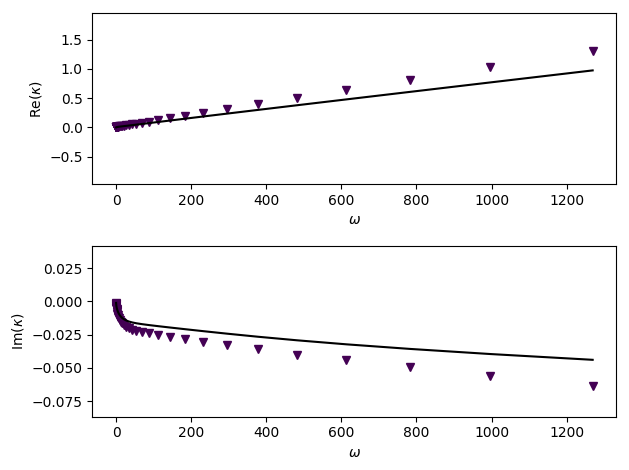}
    &
    \includegraphics[width=0.5\linewidth]{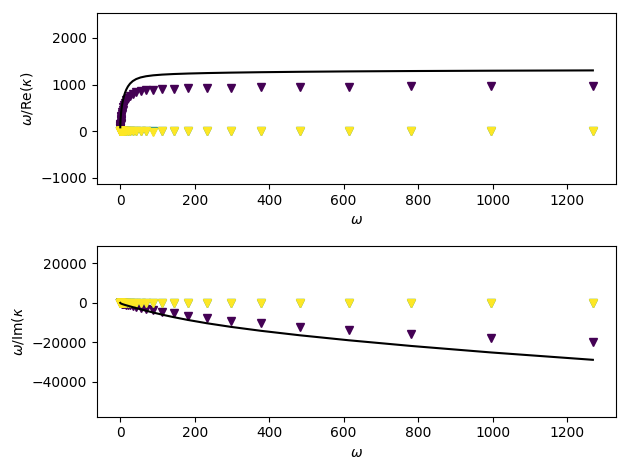}
    \\
    \multicolumn{2}{c}{$\mbw = 21.5$} \\
    \includegraphics[width=0.5\linewidth]{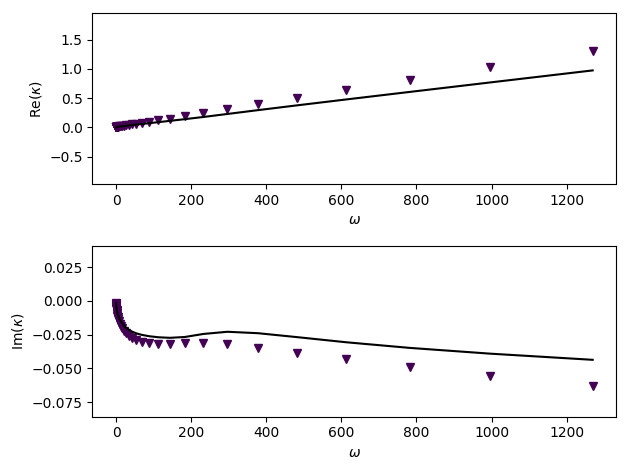}
    &
    \includegraphics[width=0.5\linewidth]{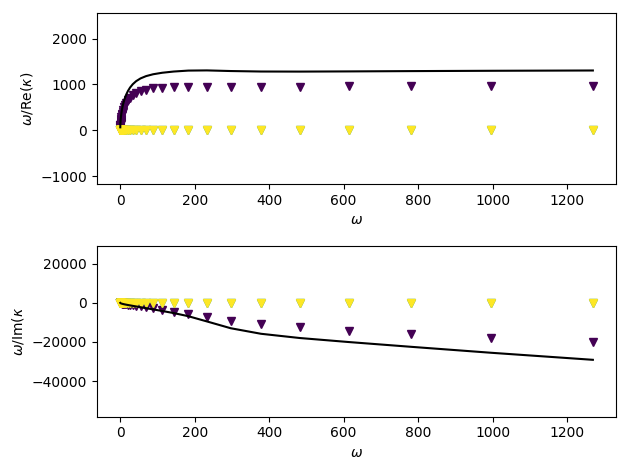}
  \end{tabular}
  \caption{Study V-B, the comparison of dispersion curves (left) and phase
    velocities (right) for homogenization (solid lines) and Bloch-based
    analyses (triangles).}
  \label{fig:b-hb-kappa}
\end{figure}

The dispersion curves and phase velocities obtained by the homogenization-based and
FB-based analyses are compared in Fig.~\ref{fig:a-hb-kappa} for the study
V-A, for two selected finite scale parameters $\veps_0$. Similarly, in
Fig.~\ref{fig:b-hb-kappa}, for the study V-B, the results for two of the $\mbw$
are shown. The dispersion curves $\om\mapsto\vkappa$ computed by the homogenization-based method were always closest to those obtained from the FB analysis with the smallest imaginary part of eigenvalues $\vkappa$, thus, to the least attenuated modes.

\begin{figure}[htp!]
  \centering
  \begin{tabular}{cc}
    \includegraphics[width=0.5\linewidth]{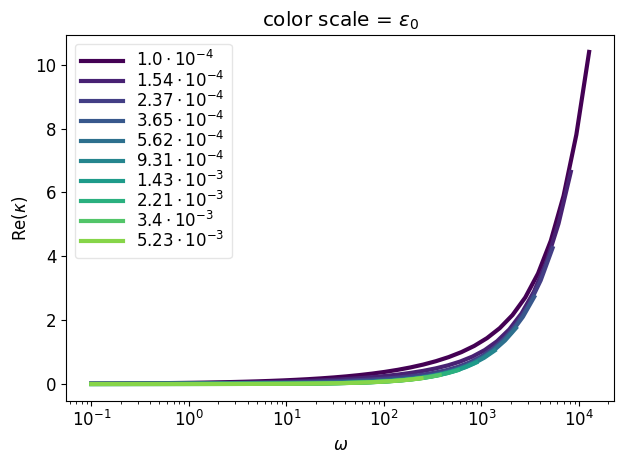}
    &
    \includegraphics[width=0.5\linewidth]{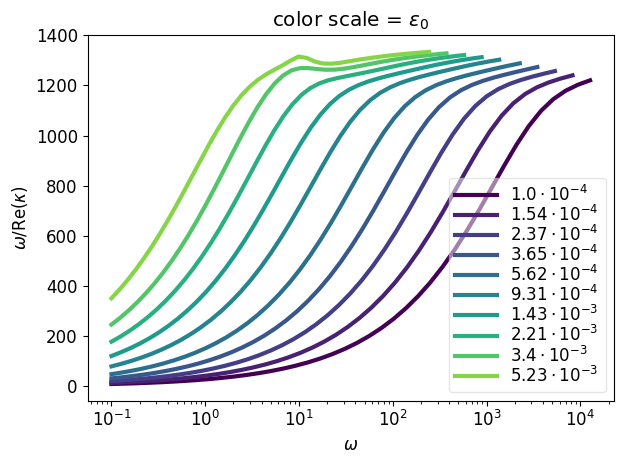}
    \\
    \includegraphics[width=0.5\linewidth]{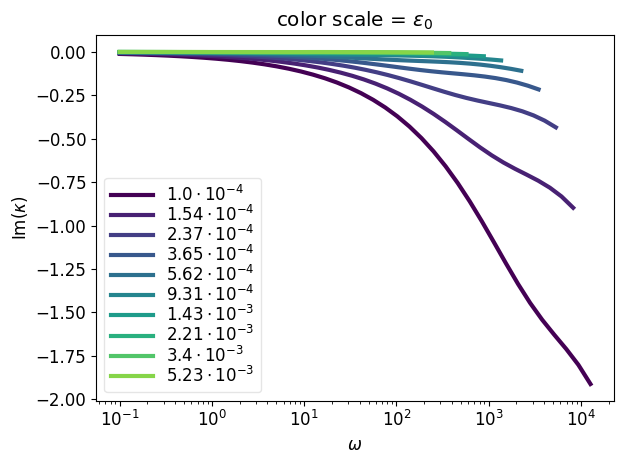}
    &
    \includegraphics[width=0.5\linewidth]{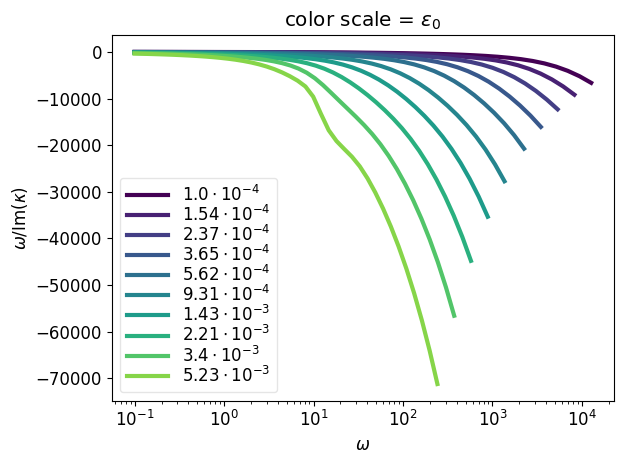}
  \end{tabular}
  \caption{Study V-A, the dependence of dispersion curves (left) and phase
    velocities (right) on $\veps_0$ obtained by the homogenization analysis.}
  \label{fig:a-h-kappas}
\end{figure}

\begin{figure}[htp!]
  \centering
  \begin{tabular}{cc}
    \includegraphics[width=0.5\linewidth]{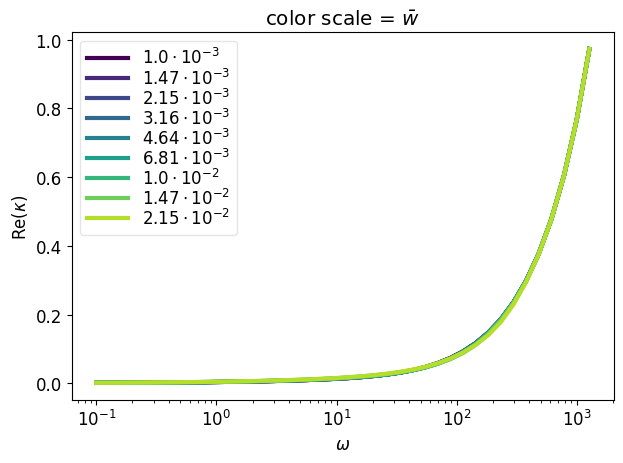}
    &
    \includegraphics[width=0.5\linewidth]{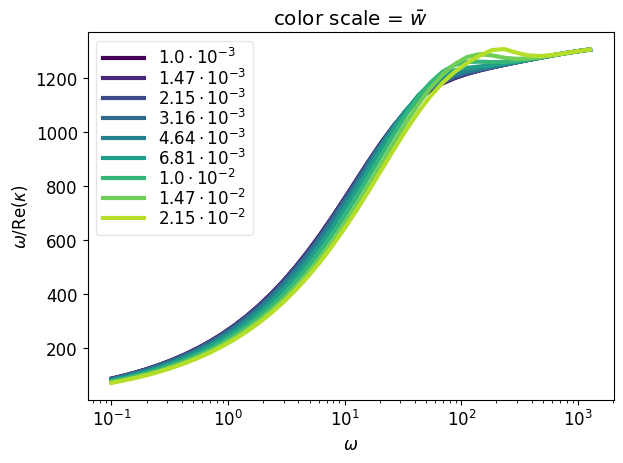}
    \\
    \includegraphics[width=0.5\linewidth]{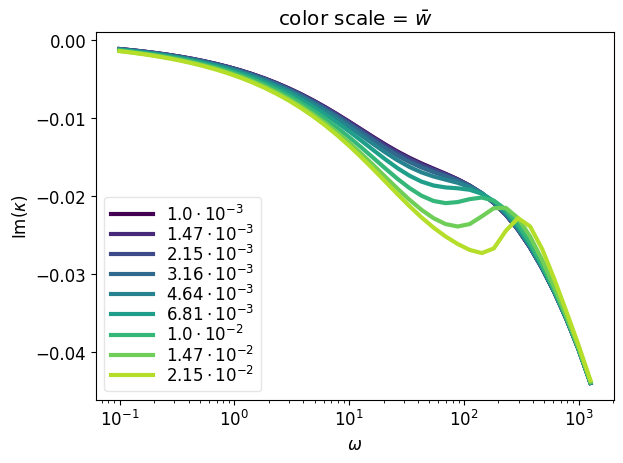}
    &
    \includegraphics[width=0.5\linewidth]{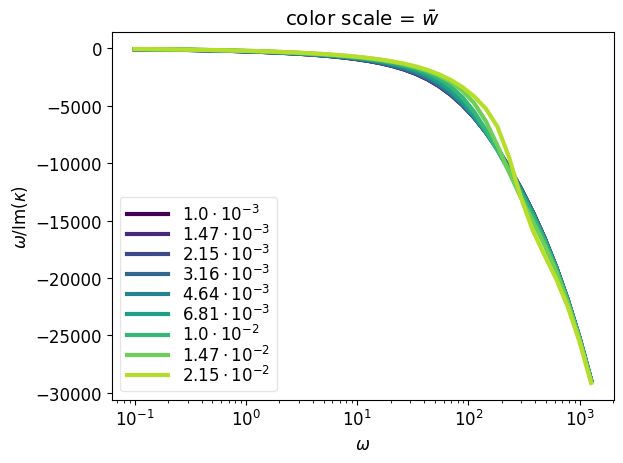}
  \end{tabular}
  \caption{Study V-B, the dependence of dispersion curves (left) and phase
    velocities (right) on $\mbw$ obtained by the homogenization analysis.}
  \label{fig:b-h-kappas}
\end{figure}

Since, in this case, the dispersion analyses by the two approaches correspond well, only those results obtained by the homogenization method are presented. For the study V-A, the dependence of
$\vkappa(\om)$, phase velocity $c_w(\om) = \om / \Re{\vkappa}$ and attenuation
$\om / \Im{\vkappa}$ curves on $\veps_0 \in [10^{-4}, 5.23 \cdot 10^{-3}]$ is
shown in Fig.~\ref{fig:a-h-kappas}. In analogy, for the study V-B, the
dependence of the dispersion curves on $\mbw \in [1, 21.5]$ for the fixed
$\veps_0 = 10^{-3}$ is presented in Fig.~\ref{fig:b-h-kappas}.

\section{Conclusion}
We derived a macroscopic model governing the acoustic waves in rigid periodically porous media  perfused by permanent flows of Newtonian, or inviscid fluids. This model extends the results reported in the literature by considering the advection phenomenon associated with the Navier-Stokes equations. The dynamic effects described by the model are limited by the linearization based on the assumptions stated in Section~\ref{sec-flow-decompose}. Besides the thermal effects which are neglected by virtue of the restriction to barotropic fluids, the acoustic linearization assumes the oscillations of the velocity field magnitudes and of its gradient being small in comparison with the permanent flow velocity field. For the inviscid fluid with considered advection permanent flow, a reduced formulation for pressure field only was proposed.

As the second contribution of the paper, the Floquet-Bloch (FB) theory was applied to study the dispersion phenomena for models featured by the advection effects for the viscous and inviscid fluids. In both considered types of fluid, the analysis leads to a general form of the quadratic eigenvalue problem (QEP). Linearizations of QEPs for all treated models were suggested.
{Numerical solutions were obtained using the finite element method implemented in the software SfePy \cite{sfepy2019}.} The numerical studies demonstrated expected relationships between the wave dispersion results obtained by the homogenization-based and the FB-based analyses. While the first approach provides a sufficiently accurate approximation for long waves only, the second approach enables to analyze planar waves propagating in infinite media. However, further research of numerical methods for the FB should be pursued to {improve robustness and efficiency of the computational algorithms, especially in the case of the QEP using techniques presented, e.g., in~\cite{Tisseur2001TheQE}.}

We believe that the modelling approach presented in this paper can be developed further towards a more detailed study of the nonlinear phenomena of acoustics in porous media, such as the acoustic streaming, with a high applicability in biomedical technologies and smart material design.



\paragraph{Acknowledgments} This research was supported by project  GACR 17-01618S of the Scientific Foundation of the Czech Republic and due to
 the European Regional Development Fund-Project ``Application of Modern Technologies in Medicine and Industry'' {(No.~CZ.02.1.01/0.0/ 0.0/17~048/0007280)}, and in part by project
 LO 1506 of the Czech Ministry of Education, Youth and Sports.

\appendix

\section{Model of acoustic fluctuations}\label{apx-ac-fl}
  Based on the Assumptions established in Section~\ref{sec-flow-decompose}, we may consider the fluctuations in \eq{eq-NS2} being proportional to a small parameter $\alpha>0$, so that, without changing the notation, \eq{eq-NS2} becomes $\wb = \bar\wb + \alpha\tilde\wb$, $p = \bar p + \alpha\tilde p$, and $\rho = \rho_0 + \alpha\tilde \rho$ where $\bar\rho = \rho_0$ is assumed to be a constant, see Assumption (A4). Upon substituting these expansions in \eq{eq-NS1}, at order $\alpha^0$, this yields the stationary flow model \eq{eq-NS-flw}. At order $\alpha^1$, the momentum equation \eq{eq-NS1}$_1$ yields \eq{eq-NS3}$_1$ straightforwardly, while from the mass conservation \eq{eq-NS1}$_2$ we obtain
  \begin{equation*}
\begin{split}
   {\pdiff{\tilde\rho}{t}} + \nabla\cdot(\rho_0\tilde\ub + {\tilde\rho\bar\wb}) & = 0\;.
\end{split}
\end{equation*}
Hence, using the incompressibility \eq{eq-NS-flw}$_2$ and the assumptions (A5) related to the barotropic approximation, \ie $\tilde p = c_f^2 \tilde \rho$, with $c_f^2 = (\gamma \rho_0)^{-1}$, the density $\tilde \rho$ can be eliminated, so that the above equation becomes
\begin{equation}\label{eq-NS3-mass}
\begin{split}
 \rho_0\gamma\left(\pdiff{\tilde p}{t} + \bar\wb\cdot\nabla\tilde p\right) + \rho_0\nabla\cdot\tilde\ub & = 0\;,
\end{split}
\end{equation}
thus,   \eq{eq-NS3}$_2$ holds. Obviously, terms of higher orders in the perturbation parameter $\alpha$ could be considered, bringing further equations involving the first-order approximation handled in this paper.

\section{Basics of the homogenization by the unfolding operator method}\label{apx-uf}
In Section~\ref{sec-homog}, the homogenization results were obtained by the \emph{periodic
  unfolding method}, 
see \cite{Cioranescu-etal-2008}. 
To introduce the periodic unfolding operator, a domain is needed, containing  the ``entire'' periods $\veps Y$ only:
\begin{equation*}\label{eq:3}
\begin{split}
\hat \Om^\veps & = \mbox{interior} \bigcup_{\zeta \in \Xi^\veps} Y_\zeta^\veps\;, \quad Y_\zeta^\veps= \veps (\ol{Y} + \zeta ) \;,\\
\mbox{ where } \Xi^\veps & = \{\zeta \in \ZZ^3\,|\; \veps (\ol{Y} + \zeta) \subset \Om\}\;.
\end{split}
\end{equation*}
For simplicity we may consider such domains $\Om$ and such subsequences $\{\veps_k\}$ for which $\Om = \hat \Om^{\veps_k}$.
For all $z \in \RR^3$,  let $[z]$ be the unique integer such that $z- [z] \in Y$.
Since $z = [z]+\{z\}$ for all $z\in \RR^3$,
for all $\veps >0$, the unique decomposition holds,
\begin{equation}\label{eq:3a}
x = \veps\left ( \left [\frac{x}{\veps}\right ] + \left \{\frac{x}{\veps}\right \}\right) =  \xi + \veps y
\quad \forall x \in \RR^3\;,\quad \xi = \veps\left [\frac{x}{\veps}\right ]\;.
\end{equation}
Based on this decomposition, the periodic unfolding operator
$\Tuftxt{}:  L^2(\Om;\RR) \rightarrow L^2(\Om \times Y;\RR)$ is defined as follows: for
 any function $v \in L^1(\Om;\RR)$, extended to $L^1(\RR^3;\RR)$ by zero outside $\Om$,
i.e. $v=0$ in $\RR^3 \setminus \Om$,
\begin{equation*}
\Tuf{v}(x,y) =
\left \{
\begin{array}{ll}
v\left( \veps \displaystyle  \left [\frac{x}{\veps}\right ] + \veps y \right)\;,
\quad &  x \in \hat\Om^\veps, y \in Y\;, \\
0 & \mbox{ otherwise }. \\
\end{array}
\right .
\end{equation*}
For product of any $u$ and $v$ the unfolding yields $\Tuf{uv}=\Tuf{u}\Tuf{v}$. The following integration formula holds:
\begin{equation*}
  \int_{\hat\Om^\veps} v\,dx = \frac{1}{|Y|}\int_{\Om \times Y} \Tuf{v}\,dy\,dx
= \int_{\Om}\intY_Y \Tuf{v}\,dy\,dx
  \quad
\forall v \in L^1(\Om)\;.
\end{equation*}


\section{Local periodic advection velocity field for the inviscid fluid}\label{appx-ivs-psi}
Here we briefly present the homogenization result of the problem \eq{eq-potflow}.
For a given ``macroscopic'' advection velocity $\wb^0$, which may depend on the macroscopic position $x \in \Om$, the fluctuation  of the velocity potential $\psi(x,\cdot) \in H_\#^1(Y_f)$ for almost all $x \in \Om$ satisfies

\begin{equation}\label{eq-appx-psi1}
\begin{split}
\int_{Y_f} \nabla_y\psi\cdot\nabla_y q = \wb^0 \cdot \int_{Y_f} \nabla_y q \quad \forall q \in H_\#^1(Y_f)\;.
\end{split}
\end{equation}
Note that the non-penetration condition holds, \ie $\nubf\cdot(\nabla_y\psi - \wb^0) = 0$ on $\Gamma$, where $\wb^1 := -\nabla_y\psi$ is the fluctuating part of the velocity.
Then, the local advection field $\bar\wb$ is computed, as follows:
\begin{equation}\label{eq-appx-psi2}
\begin{split}
\bar\wb(x,y) = \wb^0(x) - \nabla_y \psi(x,y)\;.
\end{split}
\end{equation}
In the present paper, we assume periodic advection $\bar\wb^\veps$, so that $\wb^0$ is a constant vector.

\section{Limit seepage velocity}\label{appx-bc-u}

We prove the convergence of $\ub^{\pd,\veps}$ to trace of the limit seepage velocity $\ub^0$ on $\pd\Om$, see \eq{eq-FS29}.
Let $\vphi^\dlt \in C^\infty(\Om)$. 
 We compute limits of the identity,
\begin{equation}\label{eq-FS29a}
  \begin{split}
\int_{\pd \Om_f^\veps}\vphi^\dlt \ub^{\pd,\veps}\cdot \nb - \int_{\Om_f^\veps} \nabla\cdot\ub^\veps \vphi^\dlt = \int_{\Om_f^\veps}\ub^\veps\cdot\nabla \vphi^\dlt\;.
  \end{split}
\end{equation}
Clearly, the \rhs integral converges, as follows
\begin{equation}\label{eq-FS29b}
  \begin{split}
    \int_{\Om_f^\veps}\ub^\veps\cdot\nabla \vphi^\dlt & \rightarrow \int_\Om \intY_{Y_f} \hat\ub\cdot \nabla_x \vphi^\dlt = \int_\Om\nabla_x \vphi^\dlt\cdot(\phi_f\ub^0)\\
    &  = -\int_\Om\vphi^\dlt\nabla_x\cdot(\phi_f\ub^0) + \int_{\pd\Om}(\phi_f\ub^0)\cdot \nb\;.
  \end{split}
\end{equation}
The first \lhs integral in \eq{eq-FS29a} converges by virtue of the given data $\ub^{\pd,\veps}$, thus
\begin{equation}\label{eq-FS29c}
  \begin{split}
 \int_{\pd \Om_f^\veps}\vphi^\dlt \ub^{\pd,\veps}\cdot \nb \rightarrow  \int_{\pd\Om}\vphi^\dlt\bar\phi_f\ub^\pd\cdot \nb\;.
  \end{split}
\end{equation}
The volume integral on the \lhs in \eq{eq-FS29a} converges, as follows
\begin{equation}\label{eq-FS29d}
  \begin{split}
- \int_{\Om_f^\veps}\nabla\cdot\ub^\veps \vphi^\dlt \rightarrow 
 -\int_\Om \nabla_x\vphi^\dlt\cdot\intY_{Y_f}\hat y \nabla_y\cdot\hat\ub = 0\;,
  \end{split}
\end{equation}
since $\nabla_y\cdot\hat\ub = 0$ in $Y_f$. Above $\hat y$ is the relative position of $y$ \wrt the \correction{barycenter} of $Y$. Hence \eq{eq-FS29b} and \eq{eq-FS29c} yield
\begin{equation}\label{eq-FS29e}
  \begin{split}
\int_{\pd\Om}\vphi^\dlt\bar\phi_f\ub^\pd\cdot \nb = -\int_\Om\vphi^\dlt\nabla_x\cdot(\phi_f\ub^0) + \int_{\pd\Om}\vphi^\dlt(\phi_f\ub^0)\cdot \nb\;.
  \end{split}
\end{equation}
The assertion \eq{eq-FS29} now follows by $\dlt\rightarrow 0$.


\RELEASE{
\section{Modified formulation -- inviscid fluid...}\label{appx-eig-vsc-inv}
\ToDo{???}{to remove?}

Let us consider $\bar\wb = 0$ and inviscid fluid, so that $\Abm, \Sbm$ and $\Xbm$ vanish. Then $\Pop, \Nop$ and $\vop$ attain the following form
\begin{equation}\label{eq-vsc12e}
  \begin{split}
    \Pop := \left (
    \begin{array}{ll}
      \imu\om\rho_0\Mbm\;, & -\Bbm^T \\
      \Bbm\;,& \imu\om \gamma\Qbm
    \end{array}\right) \;,\\
    \Nop := \left (
    \begin{array}{lll}
      \textbf{0}\;, & \Nbm_n^T\\
      \Nbm_n\;,  & \textbf{0}
    \end{array}\right) \;,\quad\quad
    \vop = \left (
    \begin{array}{l}
      \ubm \\ \pbm
    \end{array}\right)\;.
\end{split}
\end{equation}
Now \eq{eq-vsc13b} is replaced by problem
\begin{equation}\label{eq-vsc13f}
  \begin{split}
    \frac{1}{\imu\vkappa}\Pop\vop = \Nop \vop\;,
\end{split}
\end{equation}
} 
\bibliographystyle{plain}
\bibliography{references,Biblio-Acoustics-Porous-Advection-2019}

\begin{thebibliography}{10}

\bibitem{Allaire-nonsteady-NS-homog}
G.~Allaire.
\newblock Homogenization of the unsteady {S}tokes equations in porous media.
\newblock In C.~Bandle, J.~Bemelmans, M.~Chipot, M.~Gr\"{u}ter, and J.~Saint
  Jean~Paulin, editors, {\em Progress in Partial Differential Equations:
  Calculus of Variations, Applications}, volume 296 of {\em Pitman Research
  Notes in Mathematics Series}, pages 109--123. Longman Scientific \&
  Technical, 1992.

\bibitem{Bensoussan1978book}
A.~Bensoussan, J.L. Lions, and G.~Papanicolaou.
\newblock {\em Asymptotic methods in periodic media}.
\newblock North-Holland, 1978.

\bibitem{ThierryCoussyZinszner1987acoustics-porous}
Thierry Bourbi{\'e}, O.~Coussy, and B.~Zinszner.
\newblock {\em Acoustics of Porous Media}.
\newblock Institut fran{\c{c}}ais du p{\'e}trole publications. Editions
  Technip, 1987.

\bibitem{CarcioneBook2014}
J.M. Carcione.
\newblock {\em Wave fields in real media. Wave propagation in anisotropic,
  anelastic, porous and electromagnetic media}, volume~38 of {\em Handbook of
  geophysical exploration. Section {I.} Seismic exploration}.
\newblock Elsevier, third edition, extended and revised edition, 2014.

\bibitem{Chafin2016WaveFlowIA}
Clifford~E. Chafin.
\newblock Wave-flow interactions and acoustic streaming.
\newblock 2016.
\newblock arXiv:1602.04893.

\bibitem{Chen-homog-NS-Forchheimer2001}
Z.~Chen, L.~Lyons, and G.~Qin.
\newblock Derivation of the {F}orchheimer law via homogenization.
\newblock {\em Transport in Porous Media}, 44:325--335, 2001.

\bibitem{sfepy2019}
Robert Cimrman, Vladimír Lukeš, and Eduard Rohan.
\newblock Multiscale finite element calculations in python using sfepy.
\newblock {\em Advances in Computational Mathematics}, 45(4):1897–1921, Aug
  2019.

\bibitem{Cioranescu2010}
D.~Cioranescu, A.~Damlamian, P.~Donato, G.~Griso, and R.~Zaki.
\newblock The periodic unfolding method in domains with holes.
\newblock {\em SIAM Journal on Mathematical Analysis}, 44(2):718--760, 2012.

\bibitem{Cioranescu-etal-2008}
D.~Cioranescu, A.~Damlamian, and G.~Griso.
\newblock The periodic unfolding method in homogenization.
\newblock {\em SIAM Journal on Mathematical Analysis}, 40(4):1585--1620, 2008.

\bibitem{CDG-Stokes-2005}
D.~Cioranescu, A.~Damlamian, G.~Griso, and et~al.
\newblock The {S}tokes problem in perforated domains by the periodic unfolding
  method. new trends in continuum mechanics.
\newblock {\em Theta Series in Advanced Mathematics}, 3:67--80, 2005.

\bibitem{Collet2011}
M.~Collet, M.~Ouisse, M.~Ruzzene, and M.N. Ichchou.
\newblock {Floquet-Bloch decomposition for the computation of dispersion of
  two-dimensional periodic, damped mechanical systems}.
\newblock {\em International Journal of Solids and Structures},
  48(20):2837--2848, 2011.

\bibitem{Diaz-Alban-CPDE2014}
Jose Diaz-Alban and Nader Masmoudi.
\newblock Asymptotic analysis of acoustic waves in a porous medium:
  Microincompressible flow.
\newblock {\em Communications in Partial Differential Equations},
  39(11):2125--2167, 2014.

\bibitem{Gilbert-Panchenko2004}
R.P. Gilbert and A.~Panchenko.
\newblock Effective acoustic equations for a two-phase medium with
  microstructure.
\newblock {\em Mathematical and Computer Modelling}, 39:1431--1448, 2004.

\bibitem{Wu-ac-stream-2018}
Wu~J.
\newblock Acoustic streaming and its applications.
\newblock {\em Fluids}, 3(4):108--125.

\bibitem{Kandel-long-wave-Forchheimer-2019}
Hom~N. Kandel and Dong Liang.
\newblock The long wave fluid flows on inclined porous media with nonlinear
  {F}orchheimer’s law.
\newblock {\em AIP Advances}, 9(9):095302, 2019.

\bibitem{kruisova2018}
Alena Kruisová, Martin Ševčík, Hanuš Seiner, Petr Sedlák, Benito
  Román-Manso, Pilar Miranzo, Manuel Belmonte, and Michal Landa.
\newblock Ultrasonic bandgaps in 3d-printed periodic ceramic microlattices.
\newblock {\em Ultrasonics}, 82:91–100, Jan 2018.

\bibitem{Laschet-2008-Forchheimer-homog}
G.~Laschet.
\newblock Forchheimer law derived by homogenization of gas flow in
  turbomachines.
\newblock {\em Jour. of Comput. and Appl Math.}, 215:467--476, 2008.

\bibitem{arpack1998}
R.~B. Lehoucq, D.~C. Sorensen, and C.~Yang.
\newblock {\em ARPACK Users’ Guide}.
\newblock Software, Environments and Tools. Society for Industrial and Applied
  Mathematics, Jan 1998.

\bibitem{Masmoudi-compressNS2002}
N.~Masmoudi.
\newblock Homogenization of the compressible {N}avier--{S}tokes equations in a
  porous medium.
\newblock {\em ESAIM: Control, Optimisation and Calculus of Variations},
  8:885--906, 2002.

\bibitem{Mikel91}
A.~Mikelic.
\newblock Homogenization of nonstationary {N}avier-{S}tokes equations in a
  domain with a grained boundary.
\newblock {\em Annali di Matematica Pura ed Applicata}, 158:167--179, 1991.

\bibitem{Mikel-Paoli99}
A.~Mikelic and L.~Paoli.
\newblock Homogenization of the inviscid incompressible fluid flow through a
  {2D} porous medium.
\newblock {\em Proceedings of the American Mathematical Society Volume 127,
  Num}, 127(7):2019--2028, 1999.

\bibitem{Miroshnikova-2016}
E.~Miroshnikova.
\newblock {\em Some new results in homogenization of flow in porous media with
  mixed boundary condition}.
\newblock PhD thesis, Lulea University of Technology, 2016.
\newblock Graphic Production.

\bibitem{Norris1986}
A.N. Norris.
\newblock On the viscodynamic operator in {Biot}'s equations of poroelasticity.
\newblock {\em J. Wave-Material Interaction}, 1:365--380, 1986.

\bibitem{Peszynska2010-proc}
M.~Peszy\'nska and A.~Trykozko.
\newblock Forchheimer law in computational and experimental studies of flow
  through porous media at porescale and mesoscale.
\newblock In {\em Current Advances in Nonlinear Analysis and Related Topics},
  volume~32 of {\em GAKUTO International Series}, pages 463--482, 2010.

\bibitem{Polisevski-ZAMP1989}
D.~Poli\v{s}evski.
\newblock Homogenization of {N}avier-{S}tokes model: the dependence upon
  parameters.
\newblock {\em Z. angew. Math. Phys.}, 40:387--394, 1989.

\bibitem{Raghavan-acoustic-streaming-2018}
Raghu Raghavan.
\newblock Theory for acoustic streaming in soft porous matter and its
  applications to ultrasound-enhanced convective delivery.
\newblock {\em Journal of Therapeutic Ultrasound}, 6:6, 2018.

\bibitem{Sanchez1980Book}
E.~Sanchez-Palencia.
\newblock {\em Non-homogeneous media and vibration theory}.
\newblock Number 127 in Lecture Notes in Physics. Springer, Berlin, 1980.

\bibitem{stewart2002}
G.~W. Stewart.
\newblock A krylov--schur algorithm for large eigenproblems.
\newblock {\em SIAM Journal on Matrix Analysis and Applications},
  23(3):601–614, Jan 2002.

\bibitem{Tisseur2001TheQE}
F.~Tisseur and K.~Meerbergen.
\newblock The quadratic eigenvalue problem.
\newblock {\em SIAM Review}, 43:235--286, 2001.

\bibitem{scipy2019}
Pauli {Virtanen}, Ralf {Gommers}, Travis~E. {Oliphant}, Matt {Haberland}, Tyler
  {Reddy}, David {Cournapeau}, Evgeni {Burovski}, Pearu {Peterson}, Warren
  {Weckesser}, Jonathan {Bright}, St{\'e}fan~J. {van der Walt}, Matthew
  {Brett}, Joshua {Wilson}, K.~{Jarrod Millman}, Nikolay {Mayorov}, Andrew
  R.~J. {Nelson}, Eric {Jones}, Robert {Kern}, Eric {Larson}, CJ~{Carey},
  {\.I}lhan {Polat}, Yu~{Feng}, Eric~W. {Moore}, Jake {Vand erPlas}, Denis
  {Laxalde}, Josef {Perktold}, Robert {Cimrman}, Ian {Henriksen}, E.~A.
  {Quintero}, Charles~R {Harris}, Anne~M. {Archibald}, Ant{\^o}nio~H.
  {Ribeiro}, Fabian {Pedregosa}, Paul {van Mulbregt}, and SciPy 1.~0
  {Contributors}.
\newblock {SciPy} 1.0: fundamental algorithms for scientific computing in
  {Python}.
\newblock {\em Nature Methods}, page 1–12, Feb 2020.

\bibitem{PhysRevB.99.134304}
Yan-Feng Wang, Jun-Wei Liang, A-Li Chen, Yue-Sheng Wang, and Vincent Laude.
\newblock Wave propagation in one-dimensional fluid-saturated porous
  metamaterials.
\newblock {\em Phys. Rev. B}, 99:134304, Apr 2019.

\bibitem{Zaki-2012}
R.~Zaki.
\newblock Homogenization of a {S}tokes problem in a porous medium by the
  periodic unfolding method.
\newblock {\em Asymptotic Analysis}, 79(3--4):229--250, 2012.

\bibitem{Zhengan-Hongxing-2008}
Y.~Zhengan and Z.~Hongxing.
\newblock Homogenization of a stationary {N}avier--{S}tokes flow in porous
  medium with thin film.
\newblock {\em Acta Mathematica Scientia}, 28B(4):963--974, 2008.

\end{thebibliography}

\end{document}